\documentclass[3p]{elsarticle} 
\usepackage{amsmath}
\usepackage{amsfonts} 
\usepackage{graphicx}
\usepackage{lineno,hyperref}
\usepackage[T1]{fontenc}
\usepackage{ae,aecompl}
\usepackage{amssymb}
\usepackage{listings} 
\usepackage{float} 
\usepackage{color}
\usepackage{multirow}

\usepackage{graphicx}
\usepackage{caption}
\usepackage{subcaption}

\modulolinenumbers[5]

\begin{document}
\begin{frontmatter}

\title{Mass reconstruction in disk like galaxies using strong lensing and rotation curves: The Gallenspy package}
\tnotetext[code]{\url{https://github.com/ialopezt/GalLenspy}}

\author[astro_unal]{Itamar A. López Trilleras\corref{correspondingauthor}}
\cortext[correspondingauthor]{Corresponding author}
\ead{ialopezt@unal.edu.co}

\author[astro_unal]{Leonardo Casta\~neda}
\ead{lcastanedac@unal.edu.co}

\address[astro_unal]{Observatorio Astron\'omico Nacional, Universidad Nacional de Colombia, Carrera 30 Calle 45-03, P.A. 111321 Bogot\'a, Colombia}

\begin{abstract}
    Two methods for mass profile reconstruction in disk-like galaxies  are used in this work, the first one is done with the rotation curve's fit based on the data of circular velocity obtained observationally in a stellar system, while the other method is focused on the Gravitational Lensing Effect (GLE). For these mass reconstructions, two routines developed in the programming language python were used: one of them is \textbf{Galrotpy} \cite{galrotpy}, which was built by members of the Galaxies, Gravitation and Cosmology group from the Observatorio Astronómico Nacional of the Universidad Nacional de Colombia  and whose funtionality is applied in the rotation curves,   the second routine  is \textbf{Gallenspy} which was created in the development of this work \cite{tesis}, and is focused in the GLE. It should be noted that both routines perform a parametric estimation from Bayesian statistics, which allow to obtain the uncertainties of the estimated values. Finally the great power of combining galactic dynamics and GLE is shown. For this purpose the mass profiles of the galaxies SDSSJ2141-001 and SDSSJ1331+3628 were reconstructed with \textbf{Galrotpy} and \textbf{Gallenspy}, the results obtained are compared with those reported by other authors regarding these systems.
\end{abstract}
\end{frontmatter}
\section{Introduction} 

The study of mass distribution in galaxies allows to obtain valuable information about the universe structure on a large scale  and the process of stellar evolution. For this reason, the rotation curves and the GLE present in galaxies are relevant, because they allow the analysis of the distribution of baryonic and dark matter in these systems. With this approach we can obtain significant restrictions on values such as cosmological densities, the Hubble constant, and the cosmological constant among others. 

The analysis of the rotation curves is done based on the Newtonian gravitation theory, in this case is important to point out that the flatness in the contours of these curves is the cause why dark matter is included by other authors in different mass reconstructions \cite{galrotpy, Dutt, Dutt2, Curv}. From this perspective, dark matter reconciles the keplerian decrease with the observations done along the astronomy history, which forces to include the superposition of different mass components in the fitting of the rotation curves with observational data. 

In addition to the rotation present in the galaxies, GLE has been evidenced in many of them, an effect related to the deflection that the light coming from a background source presents due to gravitational potential of these stellar systems. Thanks to the achievements of the General Relativity Theory, it is possible to estimate the mass distribution of these galaxies based on the deflector angle of the light beams \cite{Cast,Sch}. For this reason this effect can be complemented with the dynamical analysis of mass reconstruction profiles in galaxies.  

Between the types of GLE observations present in galaxies and clusters, the formation of Einstein rings and giant arcs is very common. This is evidenced in this work by the mass reconstructions of the galaxies SDSSJ2141-001 and SDSSJ1331+3628, regarding to the rotational velocity data of this galaxies it is important to clarify that these were obtained from Dutton et.al. \cite{Dutt,Curv} hence the mass profiles reconstruction was done with both methods.

Finally it is important to point out that in this work, the advantages of combining GLE and galactic dynamics are presented, showing that both methods are complementary and powerful in mass reconstructions. 

\section{Galactic Dynamics}\label{sec: Galactic Dynamics}

In the past years, has been evidenced that the disk-like galaxies have different mass components, these can be classified into four kinds: dark matter halo, stellar halo, disk, and bulge. These components  interact in concordance with Newtonian dynamics, where each mass distribution is essential in the understanding of the functional form of the gravitational potential for these stellar systems.   

\begin{figure}[h]
	\centering
	\includegraphics[scale=0.40]{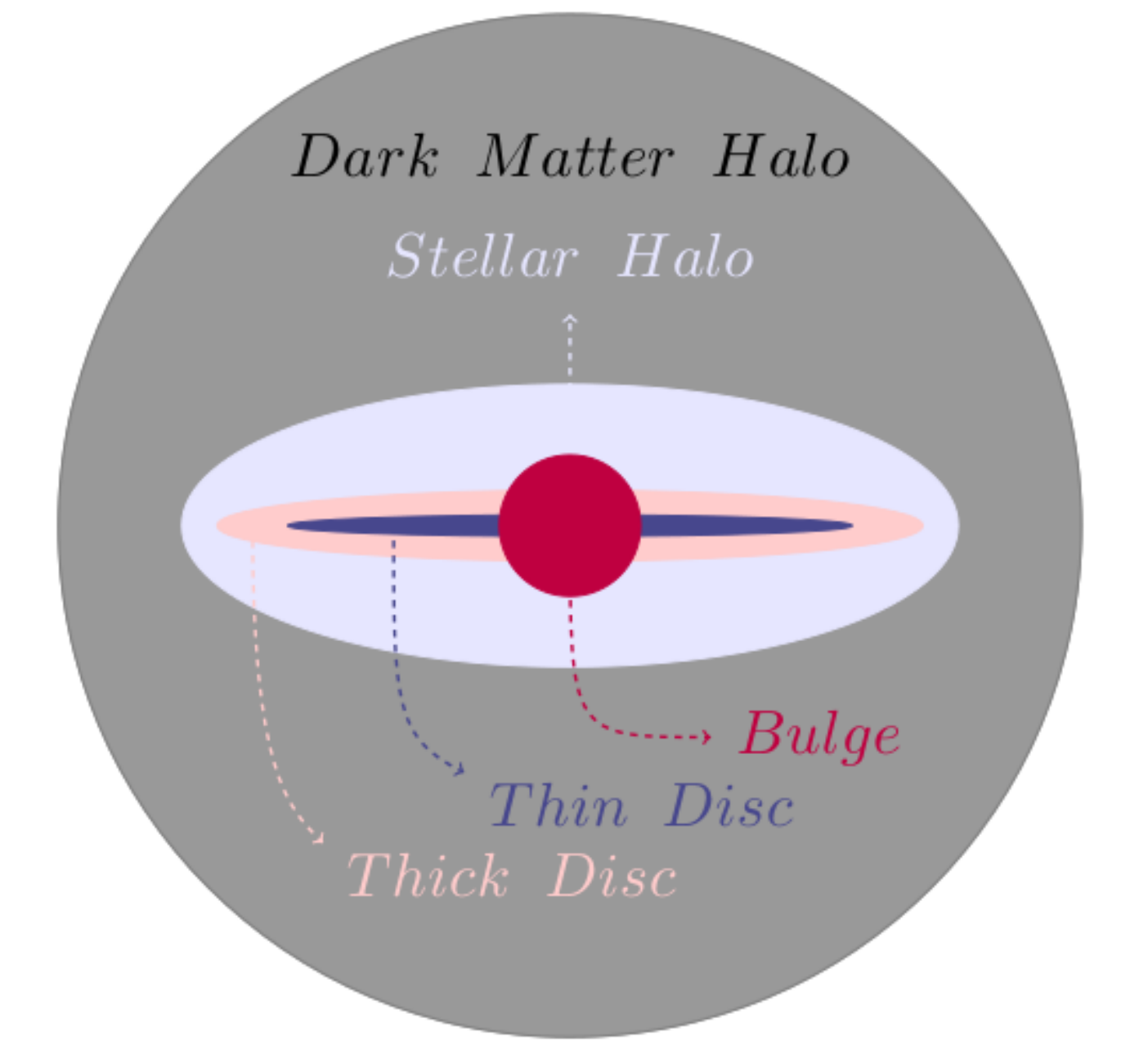}
	\caption{ Scheme of the main mass components of disk-like galaxies.\\}
	\label{fig:masscomp}
\end{figure}

Since the gravitational force ($\vec{F}$) present in galaxies is conservative \cite{Binney}, it can be related to the gravitational potential ($\Phi$) like:

 \begin{equation} \label{force_grav}
\vec{F}=-\nabla\Phi,
\end{equation}

This implies that to the mass volumetric density ($\rho$) and  $\Phi$ are related by means of the Poisson equation \cite{galrotpy}:

\begin{equation} \label{poisson}
\nabla^2\Phi(\vec{r},t)= 4\pi G \rho(\vec{r},t),
\end{equation}

where $G$ is the universal gravitation constant.

Due to the linearity of the Poisson equation \cite{Galactic, galrotpy}, in the case of a galaxy with N components and respective volumetric mass densities $\rho_{1}$, $\rho_{2}$, ...$\rho_{N}$,  the total density of this system is: 

\begin{equation} \label{densidad}
\rho=\displaystyle\sum_{i=1}^{N} \rho_i,
\end{equation}

Therefore the total gravitational potential is expressed like $\Phi=\displaystyle\sum_{i=1}^{N} \Phi_i$.

\begin{figure}[h]
	\centering
	\includegraphics[width=3.0in]{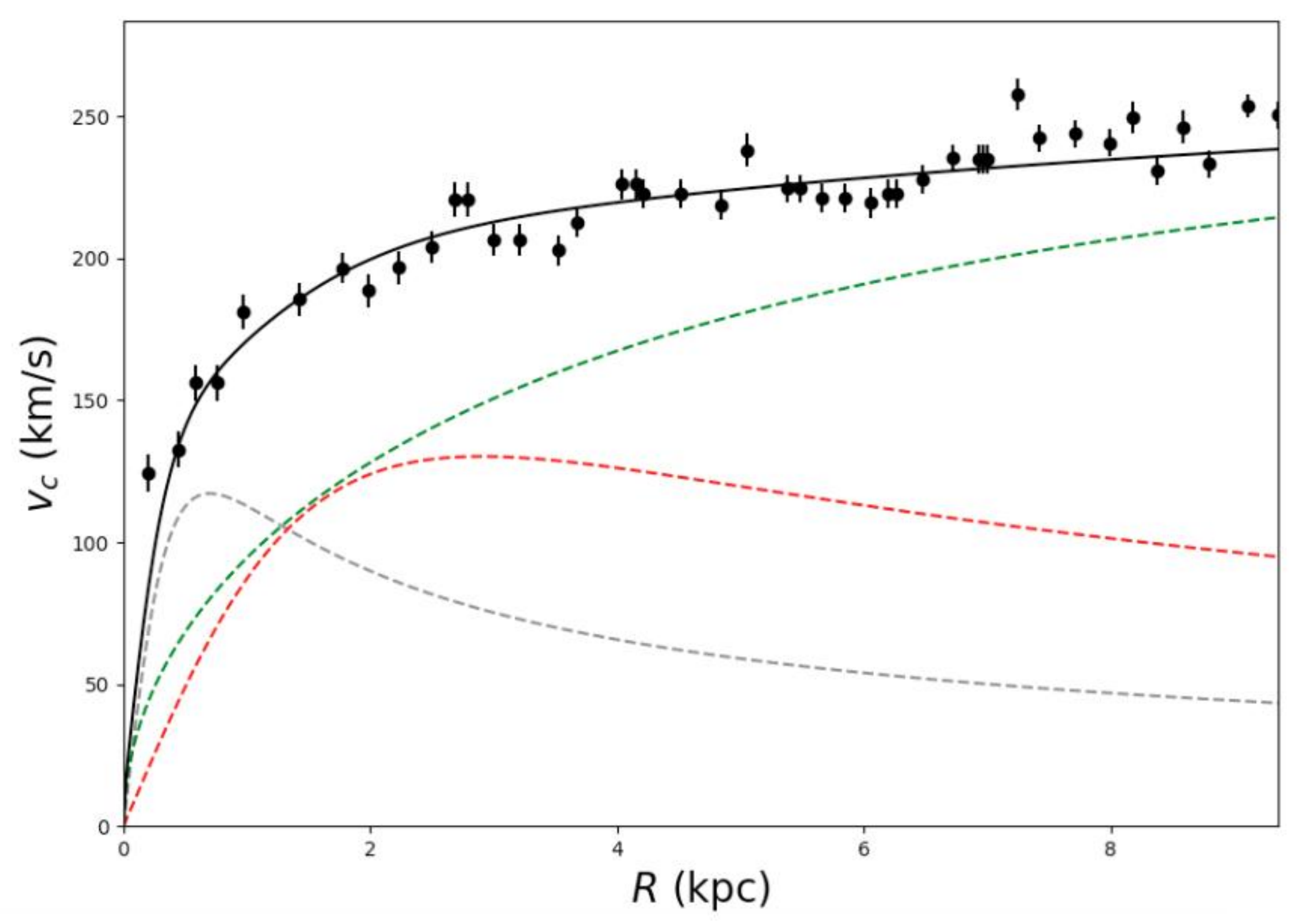}
	\caption{Fitting of rotational velocities with the rotation curve belonging to the galaxy NGC6361 (the observational data are presented with black dots and the fitting curve is the continuous line), in this case the Miyamoto-Nagai profile was used for the bulge (gray dotted line) and the stellar disk (red dotted line), while the Navarro-Frenk-White (green dotted line) belongs to the dark matter halo.\\}
	\label{fig:ej_rotcurv}
\end{figure}

As the circular velocity (in the equatorial plane) associated to a gravitational potential $\Phi(R,z=0)$ in disc-like galaxies is described by:

\begin{equation} \label{vel_pot}
V^{2}_{c}(R)=R\dfrac{\partial\Phi}{\partial r}|_{r=R},
\end{equation}

the total circular velocity of this kind of galaxies is expressed as a superposition of the velocities belonging to each gravitational potential, which is evidenced in the equation \ref{vel}:

\begin{equation} \label{vel}
V_c^2=\displaystyle\sum_{i=1}^{N} V_{c(i)}^2, 
\end{equation}

and therefore it is possible to make reconstructions of mass profiles in disk-like galaxies through the fitting of rotation curves with the observational values of circular velocity, as shown in figure \ref{fig:ej_rotcurv}.  

\section{Gravitational Lensing Effect}\label{sec:Gravitational Lensing Effect}

In GLE, the relation between the coordinates of the images $\vec{\theta}$ and the source $\vec{\beta}$ is given by the equation:

\begin{equation} \label{eq_len}
\vec{\beta}=\vec{\theta}-\nabla_{\vec{\theta}} \psi(\vec{\theta}),
\end{equation}

where $\psi$ is the deflector potential, which has the information of the lens and the cosmological distances. For the case of mass profiles with spherical symmetry, it is possible to assume that  \cite{Rog}

\begin{equation}\label{potential_deflec}
    \psi(\vec{\theta})=2\int_{0}^{|{\vec{\theta}}|} \theta^{'} \kappa(\theta^{'})ln\bigg(\dfrac{|{\vec{\theta}|}}{\theta^{'}}\bigg) d\theta^{'},
\end{equation}

where $\kappa$ is the convergence, defined as

\begin{equation}\label{convergence}
    \kappa=\dfrac{\Sigma}{\Sigma_{crit}}
\end{equation}

where $\Sigma$ is the superficial mass density and $\Sigma_{crit}$ is given by the relation

\begin{equation}\label{sigma_crit}
    \Sigma_{crit}=\dfrac{c^2}{4\pi G} \dfrac{D_s}{D_{d}D_{ds}},
\end{equation}

in this case, $c$ is the speed of light, $D_d$, $D_s$ and $D_{ds}$ are the angular diameter distances between observer-lens, observer-source and lens-source respectively. 

Additionally, the mapping between the planes of the lens and the source is done through the jacobian transformation matrix

\begin{equation}
    A_{i,j}=\dfrac{\partial\beta_{i}}{\partial\theta_{j}},
\end{equation}

for $i,j = 1,2$.

With this transformation matrix, the magnification of the images can be defined as  

\begin{equation}\label{magnification}
\mu=\dfrac{1}{\vert detA\vert},
\end{equation}

therefore the critical curve is the set of points in the lens plane where ${\vert detA\vert}=0$.

\textbf{Note:} The set of points in the source plane, which through the equation \ref{eq_len} belongs to the critical points is denominated caustic curve.   

\section{Combining GLE and Galactic Dynamics}\label{sec:Combining GLE and Galactic Dynamics}

Due to the superposition between the different mass distributions in galaxies, there are mass profiles reconstructions done with galactic dynamics and other with lensing where the obtention of parameters does not occur within acceptable reliability regions \cite{Dutt}.

A solution proposal for this problem has been proposed by other authors \cite{Koopmans,Dutt}, regarding the combination of these mass reconstructions methods, which means taking advantage of the geometry used for each one of them.

For this purpose it must be taken into account, that while the mass projection in the galactic dynamics is done in the equatorial plane of the galaxy, in the case of the GLE the operator $\Sigma$ is projected in the plane ($\theta_1$, $\theta_2$) where the deflected images are formed.  

\begin{figure}[h]
	\centering
	\includegraphics[width=3.0in]{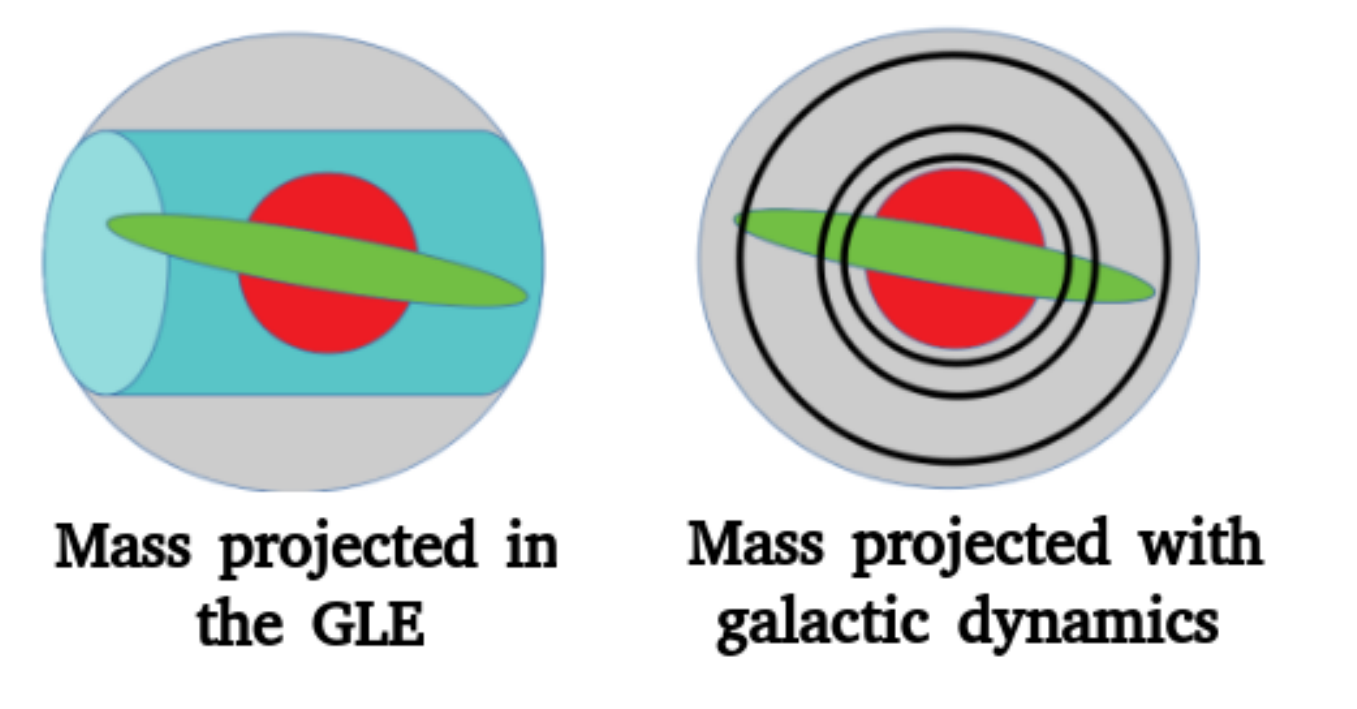}
	\caption{Illustration of the geometries used in galactic dynamics and GLE for the mass projection in 2-D based on what was stated by Dutton et al \cite{Dutt2}.\\}
	\label{fig:geometry}
\end{figure}

In figure \ref{fig:geometry}, the geometries belonging to the methods of mass reconstruction used in this work are illustrated, wherein the GLE the projection of mass in a cylinder through the line of sight $z$ and within a radius $R_{eins}$ restricted by the position of the deflected images is done, which from the formalism of strong lensing is related to the formation of the Einstein ring \cite{Dutt} and is described by the equation \ref{radio_eins}.

\begin{equation}\label{radio_eins}
R_{eins}\approx \Bigg(\dfrac{M_{eins}}{\pi \Sigma_{crit}}\Bigg)^{1/2},
\end{equation}

with $M_{eins}$ the mass projected in the cylinder of figure \ref{fig:geometry}. 

In the case of mass projection from galactic dynamics, it corresponds to enclosed spheres of differents radius due to the estimated circular velocities in the galaxy. This geometrical approach is optimal, as long as the disc has an inclination, so that the observer can see it from its edges. 

Based on the theory previously shown, the combination of GLE and galactic dynamics for this work is proposed around the restrictions in the parameter space that both methods can provide, in such a way that it is possible to distinguish more clearly the gravitational contribution of each mass component in disc-like galaxies.

\section{Gallenspy}

\textbf{Gallenspy} is an open source code created in python, designed for the mass profiles reconstruction in disk-like galaxies using GLE. It is important to note that this algorithm allows to invert numerically the lens equation (equation \ref{eq_len}) for gravitational potentials with spherical symmetry, in addition to the estimation in the position of the source ($\beta_1$,$\beta_2$), given the positions ($\theta_1$, $\theta_2$) of the images produced by the lens.

The main libraries used in this routine are: \textbf{numpy} \cite{Numpy} for the data handling, \textbf{matplotlib} \cite{Matplotlib} regarding the generation of graphic interfaces, \textbf{galpy} \cite{galpy} to obtain mass superficial densities, as to the parametric adjust with Markov-Montecarlo chains (MCMC) it was used \textbf{emcee} \cite{emcee} and for the graphics of reliability regions \textbf{corner} \cite{corner} is used.

The deflector potential is obtained numerically in \textbf{Gallenspy} by means of the equation \ref{potential_deflec}, which is very helpful because some mass distribution models do not have analytical solutions for the equations \ref{densidad}, \ref{eq_len} and \ref{potential_deflec}.

Also it is important to note others tasks of \textbf{Gallenspy} like computing of critical and caustic curves and obtaining the Einstein ring. For a more detailed description, it is recommended to see the repository page where the source code \footnote{https://github.com/ialopezt/GalLenspy} is available together with its requirements and instructions for its use.

\subsection{Tested case with Isotherm Singular Sphere (SIS)}

\begin{figure}[h]
	\centering
	\includegraphics[width=3.5in]{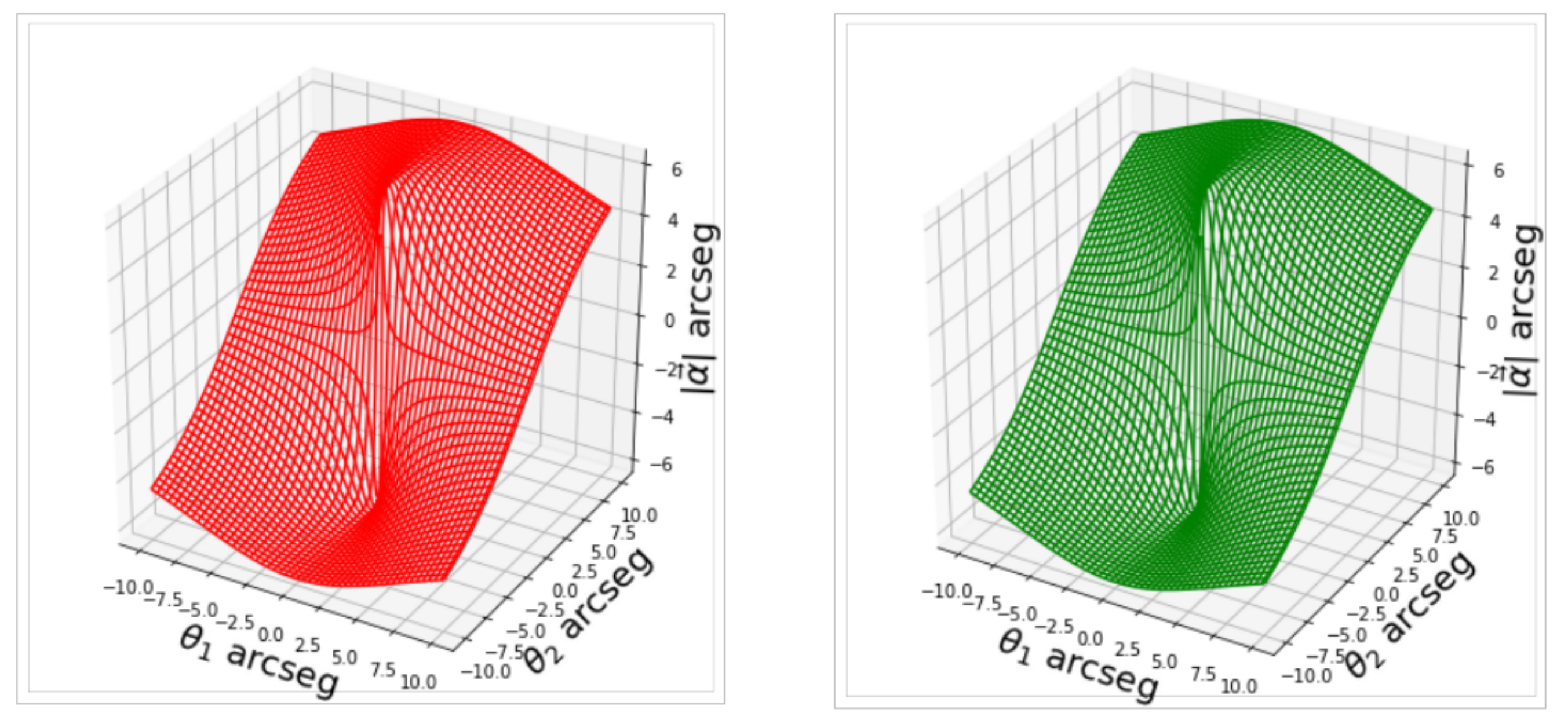}
	\includegraphics[width=3.5in]{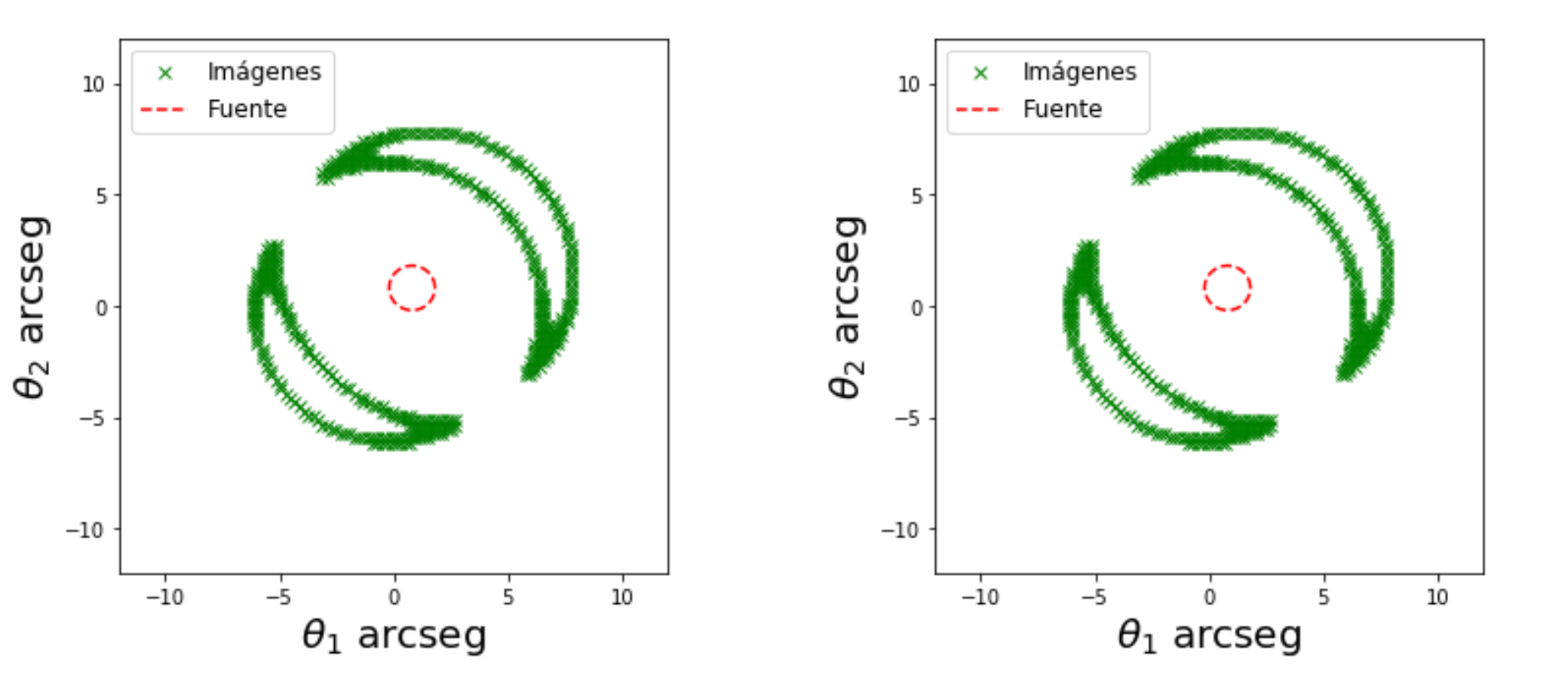}
	\includegraphics[width=3.5in]{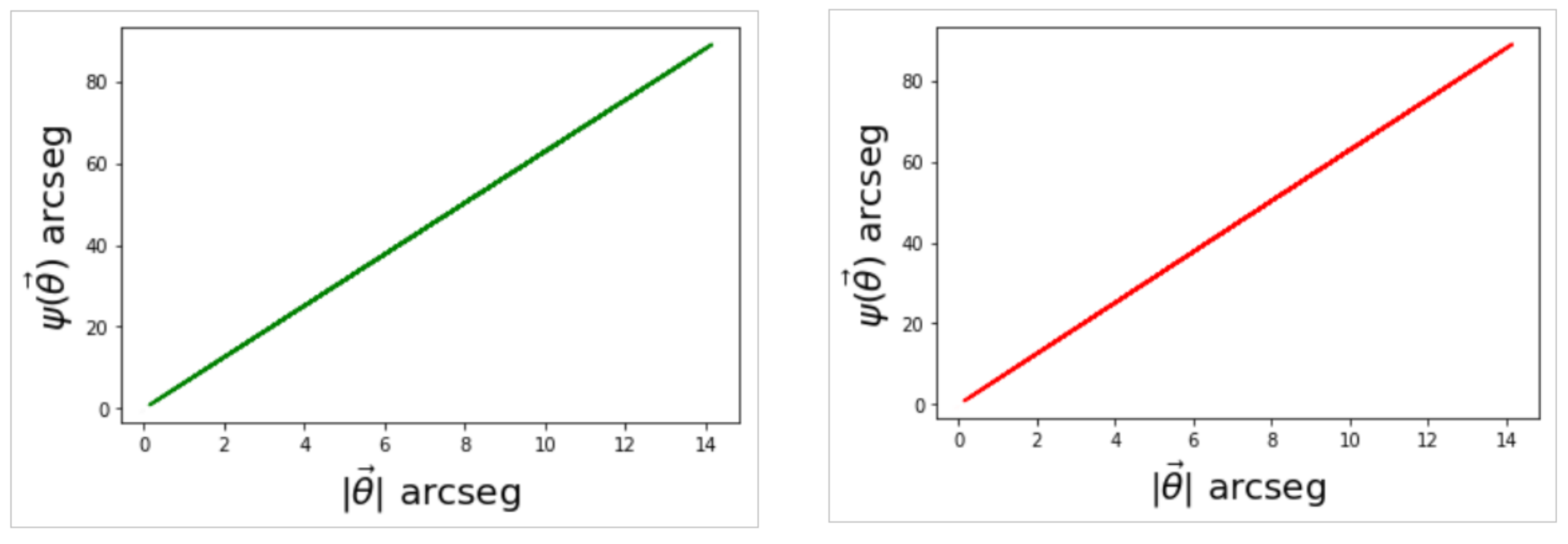}
	\caption{Comparison between the results obtained by the analytical solution of Hurtado (left side) and \textbf{Gallenspy} (right side). These results of deflection angle, images formation and deflection angle were evaluated in a grid of $100X100$ points.\\}
	\label{fig:Comparacion1_SIS_galenspy}
\end{figure}

A way by which \textbf{Gallenspy}  was tested, was the comparison with the analytical solutions given by Hurtado \cite{Rog} of the isotherm singular sphere (SIS) under certain specific conditions. In figure \ref{fig:Comparacion1_SIS_galenspy} some of these comparisons are shown for the obtention of deflection angle, the images formation, and deflector potential in the case of a circular source of radius $1kpc$ to a distance of $2kpc$ with respect to the observer, this mass profile was modeled with a dispersion velocity of $100 km/s$.    

As it is possible to observe in each comparative graphic, the results obtained numerically with \textbf{Gallenspy} present high reliability where the percentage error is of $0.1$ due to the grid used in this case.

\subsection{Gallenspy input}

To use \textbf{Gallenspy}, it is important to give the values of cosmological distances in $Kpc$ and critical density in $M_{\odot}/Kpc^2$, which are introduced by means of a file named \verb+Cosmological_distances.txt+. On the other hand, the user must introduced the coordinates of the observational images (in radians) in the file \verb+coordinates.txt+.(\textbf{Note:} for the case of a circular source it is present the file \verb+alpha.txt+, where the user must introduce angles value belonging to each point of the observational images.) 

\subsection{Visual fitting with Gallenspy}

Gallenspy presents an interactive fitting of parameters through a routine developed in Jupyter Notebook named \verb+Interactive_data+, in this case the user has the possibility of choose freely the parametric range for each value, however it is suggested a parametric space illustrated in  table  \ref{tabla:Param_Gallenspy} which is based on the values used by other authors and with the ones used to model dwarf and milky-way type galaxies \cite{galrotpy}.  

\begin{table}[h]
\centering
\begin{tabular}{|c|c|c|}
\hline
\multicolumn{3}{|c|}{\textbf{Range of values with Gallenspy}} \\
\hline
\textbf{Component} & \textbf{Range of parameters} & \textbf{Units} \\
\hline \hline
\multirow{3}{2cm}{Bulge I} & $a=0$ & $kpc$ \\ \cline{2-3}
& 0.0 $< b <$ 0.5 & $kpc$ \\ \cline{2-3}
& 0.1 $< M <$ 1.0 & $10^{10}M_{\odot}$ \\ \hline
\multirow{3}{2cm}{Bulge II} & 0.01 $< a <$ 0.05 & $kpc$ \\ \cline{2-3}
& 0.5 $< b <$ 1.5 & $kpc$ \\ \cline{2-3}
& 1 $< M <$ 5 & $10^{10}M_{\odot}$ \\ \hline
\multirow{3}{2cm}{Disk thin} & 1 $< a <$ 10 & $kpc$ \\ \cline{2-3}
& 0.1 $< b <$ 1.0 & $kpc$ \\ \cline{2-3}
& 0.5 $< M <$ 1.5 & $10^{11}M_{\odot}$ \\ \hline
\multirow{3}{2cm}{Disk thick} & 1 $< a <$ 10 & $kpc$ \\ \cline{2-3}
& 0.1 $< b <$ 15.0 & $kpc$ \\ \cline{2-3}
& 0.5 $< M <$ 1.5 & $10^{11}M_{\odot}$ \\ \hline
\multirow{2}{2cm}{Exponential Disk} & 2 $< h_{r} <$ 6 & $kpc$ \\ \cline{2-3}
& 1 $< \Sigma_0 <$ 15 & $10^{2}M_{\odot}/pc^{2}$\\ \hline
\multirow{2}{2cm}{Halo NFW} & 0.1 $< a <$ 30 & $kpc$ \\ \cline{2-3}
& 0.1 $< M_0 <$ 10 & $10^{11}M_{\odot}$\\ \hline
\multirow{2}{2cm}{Halo Burket} & 2 $< a <$ 38 & $kpc$ \\ \cline{2-3}
& 0.1 $< \rho_0 <$ 10 & $10^{6}M_{\odot}/kpc^{3}$\\ \hline
\end{tabular}
\caption{Parametric space used in \textbf{Gallenspy}}
\label{tabla:Param_Gallenspy}
\end{table}

\begin{figure}[h]
	\centering
	\includegraphics[width=3.0in]{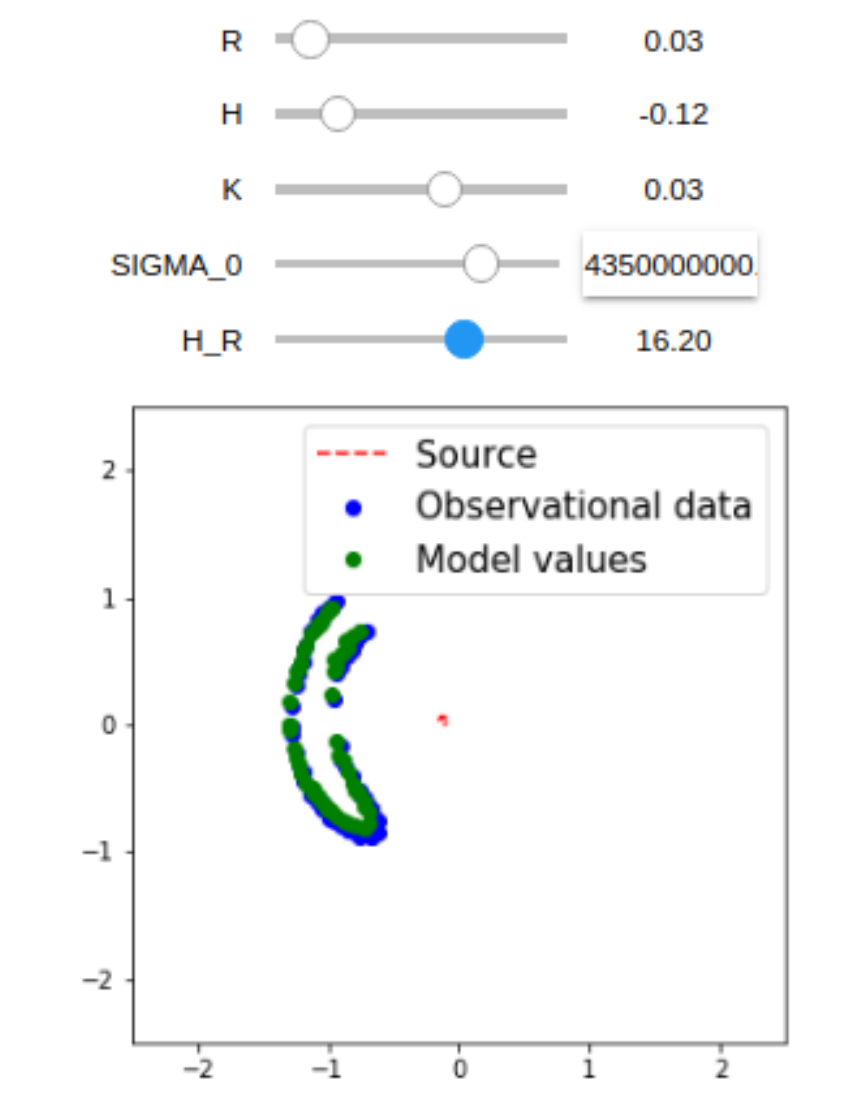}
	\caption{Interactive panel of \textbf{Gallenspy}, for the Exponential disk lens model and the fitting of a circular source.\\}
	\label{fig:interactive_gallenspy}
\end{figure}

Figure \ref{fig:interactive_gallenspy} shows the interactive panel, for the fitting of an Exponential Disk lens model and a circular source, this observational data belong to the galaxy J2141 which is analyzed later.

\subsection{Bayesian statistics with Gallenspy}

In mass reconstructions, \textbf{Gallenspy} allows assigning a mass profile to each component of the lens galaxy, where each free parameter has a range of possible values for the obtention of a set of initial values in the parametric exploration. 

Also, it is important to point out that for the mass reconstruction of each component of the galaxy, it is possible to choose between different profiles for the parametric fitting with \textbf{Gallenspy}: for example, in the case of galactic disk it is possible choose between the options of Miyamoto-Nagai  and Exponential Disk \cite{galrotpy} \cite{Binney}, for the dark matter halo between Navarro-Frenk-White (NFW) and Burket \cite{galrotpy} \cite{Binney}, while in the bulge is used the Miyamoto-Nagai even though there are two possible ranges of data in this profile.  

When the positions ($\theta_1$, $\theta_2$) of the GLE are known, the work with \textbf{Gallenspy} is to find the model and parameters set which can reproduce these provided data, for this reason in this routine the bayesian statistics is not only based on the exploration of all possible positions of the source. 

For this parametric exploration, \textbf{Gallenspy} implements the Metropolis-Hasting algorithm through MCMC \cite{Bayesians}, where it is obtained a posterior probability distribution $P(p|D,M)$ for each parameter set of the lens model selected of the table \ref{tabla:Param_Gallenspy}, which is given by the relation: 

\begin{equation}\label{mcmc}
P(p|D,M)=\dfrac{P(D|p,M)P(p|M)}{P(D|M)},
\end{equation}

with:

\begin{itemize}
    \item  $P(p|D,M)$ the probability that this parameter set $p$ is appropriate for the model $M$ and the data $D$.
   \item $P(D|p, M)$ the probability that the data $D$ are obtained with the model $M$ and the parameter set $p$, is known as \textbf{likelihood} \cite{Bayesians} and it is denoted with $L$.
   \item$P(p|M)$ the \textbf{prior} which is the reliability that the parameter set is correct for the model.
   \item $P(D|M)$ is denoted as $Z$, this is the normalization factor and is the probability of obtaining the data $D$ with the model $M$.
\end{itemize}

It is important to note that in \textbf{Gallenspy} the normalization factor is not considered, hence the fundamental work is the compute of the \textbf{likelihood} (this is because the \textbf{prior} for this parameter set has the same value). From this perspective the method for obtaining the initial values of $P(D|p, M)$ is through a visual fitting in the \verb+Interactive_data+ routine from which it is possible to make a first approximation between the data set $D$ and the model values.

Later in this process, the user must execute the code created in the respective file.py (for the estimation of the source \verb+source_lens.py+ and in the case of mass reconstruction \verb+parameters_estimation.py+) where \textbf{Gallenspy} requests to introduce the initial values obtained in the visual fitting. Next, \textbf{Gallenspy} lets  the user choose the number of steps and walkers, which is enough for the MCMC of this computational routine. 

In \textbf{Gallenspy} the minimization function $\chi^2$, for a source with a number $n_i$ of images in the GLE is given by:

\begin{equation}
\chi^2_i =\sum_{j=1}^{n_i} \dfrac{\vert \theta_{obs}^{j} -\theta^{j}(p) \vert^2}{\sigma_{ij}^2},
\label{chi_cuadrado}
\end{equation} 

where in this equation $\theta_{obs}^{j}$ is the position of observed image j in the data set $D$, $\sigma_{ij}$ the error in position $\theta_{obs}^{j}$ due to the noise in the image and $\theta^{j}(p)$ the image j predicted by the mass model used with the parameter set.

The index $i$ appears in the function $\chi^2$, this because in the GLE the formation of images with various sources is possible, for this reason the \textbf{likelihood} for each explored parameter set is expressed through the Gaussian distribution  

\begin{equation} 
L = P(D|p, M) = \prod_{i=1}^{N} \dfrac{1}{\prod_{j}\sigma_{ij}\sqrt{2\pi}}exp\bigg(-\dfrac{\chi_i^2}{2}\bigg),
\label{likelihood_GLE}
\end{equation} 

where N is the number of sources.

\begin{figure}[h]
	\centering
	\includegraphics[width=3.5in]{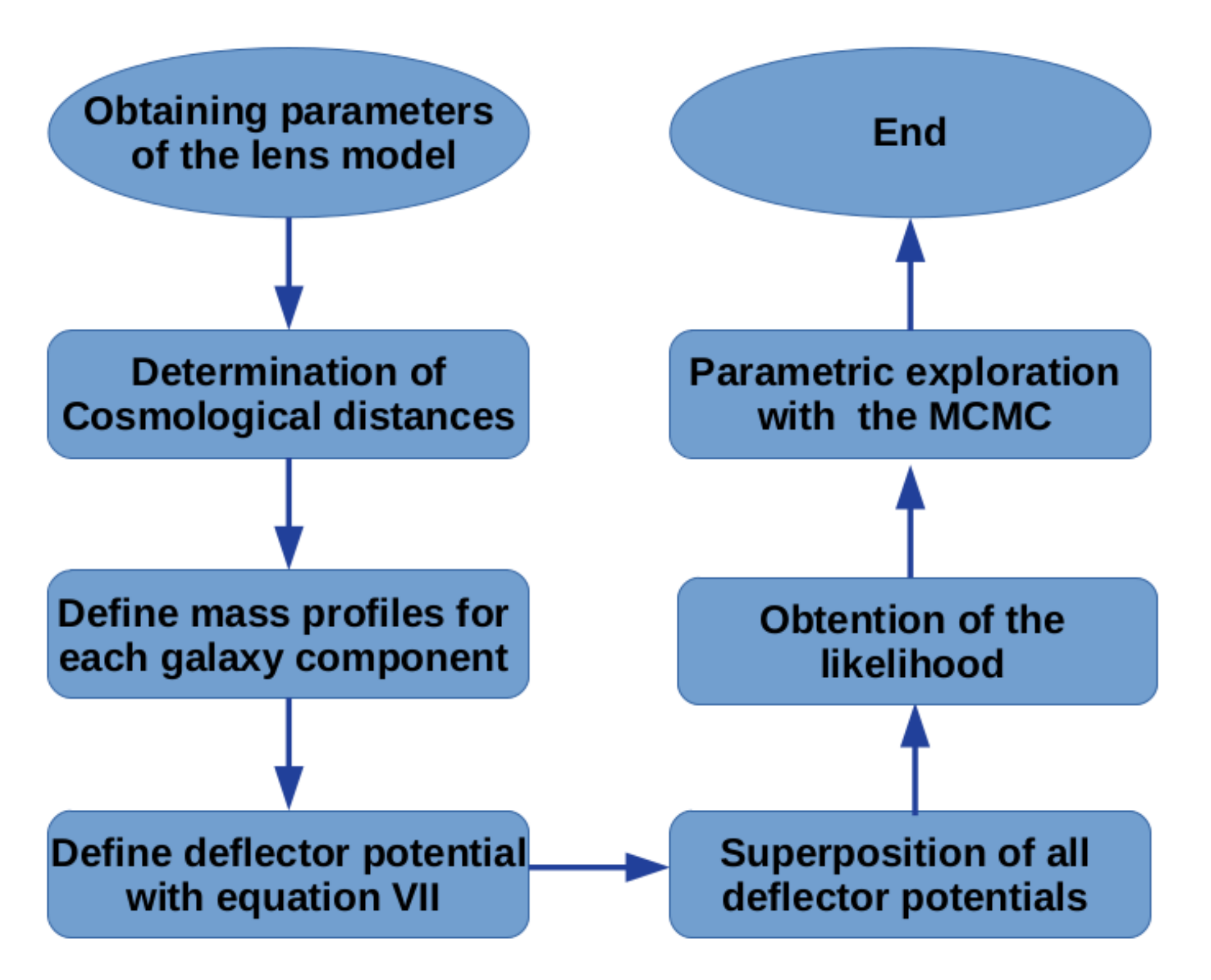}
	\caption{Process flow diagram of the parametric exploration with \textbf{Gallenspy}.\\}
	\label{fig:diagrama_flujo1}
\end{figure}

In figure \ref{fig:diagrama_flujo1}, the algorithm of \textbf{Gallenspy} is shown whereas in the next section an example of this with the SIS profile is illustrated. 

Finally it is important to note that \textbf{Gallenspy} generates a file.txt with the final parameters obtained: in the case of source estimation \verb+parameters_lens_source.txt+ and \verb+parameters_MCMC.txt+, these files are then request by \textbf{Gallenspy} for other tasks as compute of Einstein ring and mass estimations.  

\subsection{Illustrative example with the SIS profile}

Although the SIS is not included in the profiles of \textbf{Gallenspy}, it is possible to show an illustration of the mode which this routine performs the parameters exploration with this mass distribution.

Because of the analytical solutions shown by Hurtado \cite{Rog} to the lens equation in the SIS profile, the formation of images in the GLE are given by the relations:

\begin{equation}\label{image_sis1}
\vert{\vec{\theta_{p}}}\vert=\vert{\vec{\beta}}\vert+\dfrac{4\pi\sigma^{2}D_{ds}}{c^{2}D_{s}},
\end{equation}

\begin{equation}\label{image_sis2}
\vert{\vec{\theta_{n}}}\vert=-\vert{\vec{\beta}}\vert+\dfrac{4\pi\sigma^{2}D_{ds}}{c^{2}D_{s}},
\end{equation}

with $\sigma$ the dispersion velocity, $\theta_{p}$ and $\theta_{n}$ are the positive and negative solutions respectively where the images are formed.

\begin{figure}[h]
	\centering
	\includegraphics[width=2.0in]{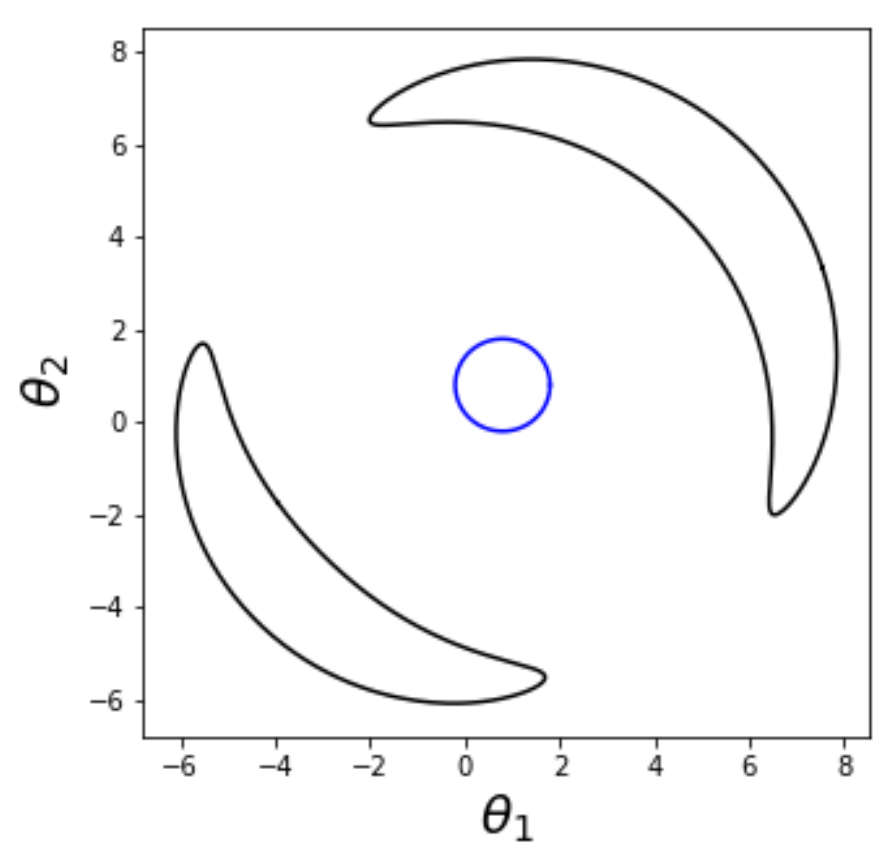}
	\caption{Images formed with the SIS profile for $\vert{\vec{\theta_{n}}}\vert$ (left-down image) and $\vert{\vec{\theta_{p}}}\vert$ (right-up image) in the case of a circular source (blue image).\\}
	\label{fig:images_sis}
\end{figure}

Figure \ref{fig:images_sis} evidences the images formation, when $\sigma=1X10^5km/s$ for a circular source of radius $r=1$arcs whose center has coordinates (h = 0.8arcs, k = 0.8arc) and where the cosmological distances are $D_{ds} = 1Kpc$ and $D_{s} = 2Kpc$.

If for this specific case the parametric exploration with \textbf{Gallenspy} is applied, then it is necessary to assume that the values of $\sigma$, $r$, h y k are not known and that the objective is the obtention of the parameters family which through of GLE can reproduce the images illustrated in figure VI, where in this example only values of $\vert{\vec{\theta_{p}}}\vert$ are supposed to be known. This time, the ranges established for the parametric exploration were $10^4km/s<\sigma<2X10^5km/s$, $0.1$arcs$<r<2$arcs, -8arcs<h<8arcs and 8arcs<k<8arcs.  

\begin{figure}[!tbp]
  \begin{subfigure}[b]{0.42\textwidth}
    \includegraphics[width=\textwidth, height=\textwidth]{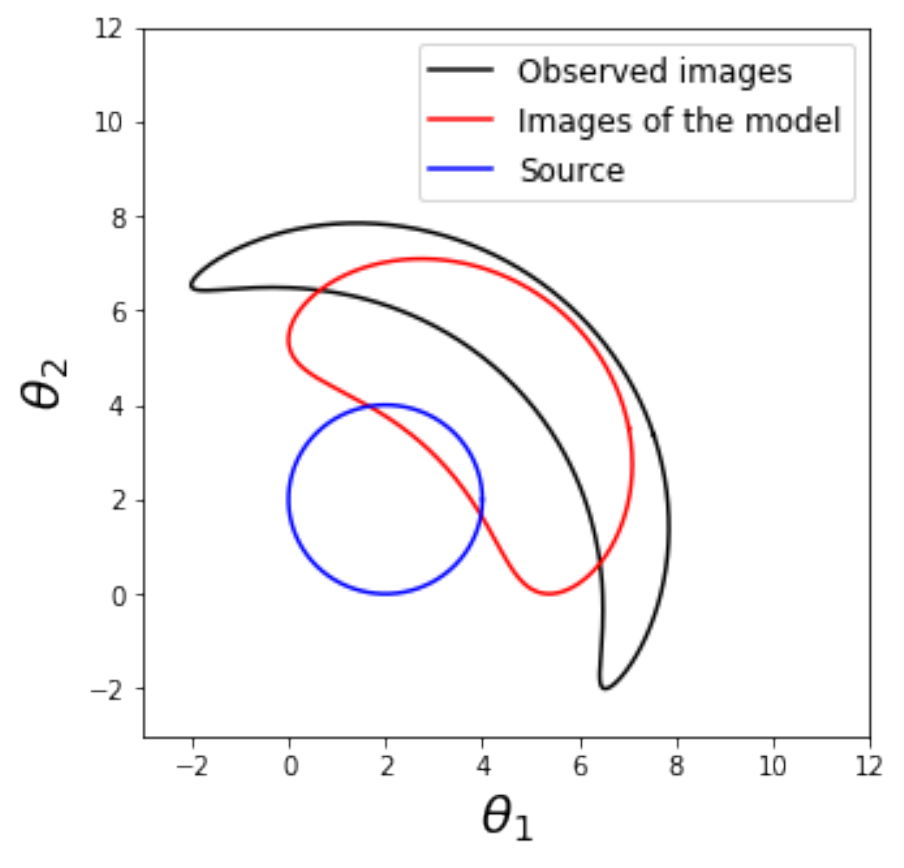}
    \caption{\textbf{Red} Image formed in the GLE for a SIS with dispersion velocity $0.38X10^5km/s$ and a circular velocity of radius $2$arc, the center of this source is in $(2,2)arcseg$. \textbf{Black} Image to reproduce by means of \textbf{Gallenspy}.\\}
    \label{fig:chi_sis}
  \end{subfigure}
  \hfill
  \begin{subfigure}[b]{0.49\textwidth}
    \includegraphics[width=\textwidth, height=\textwidth]{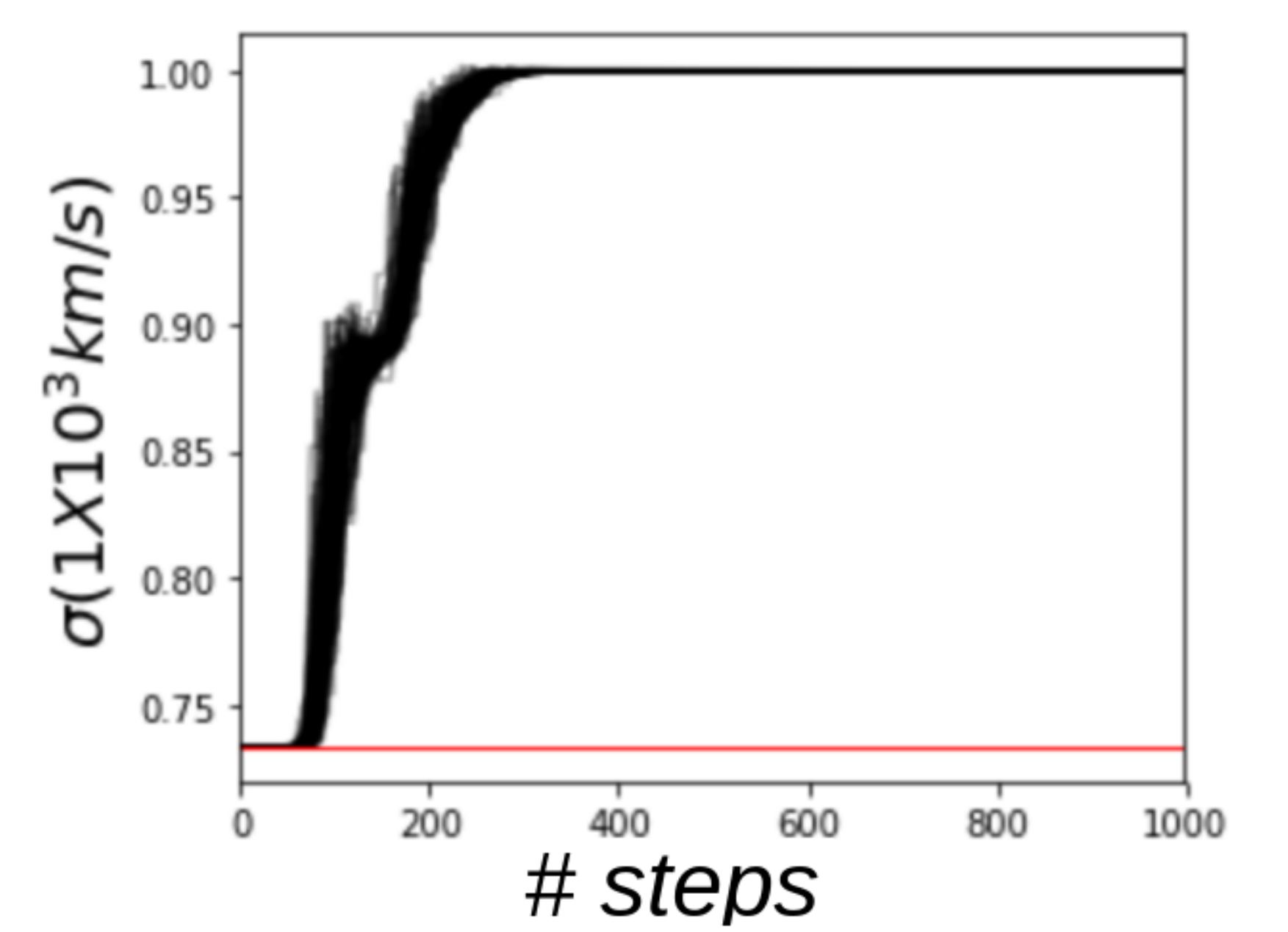}
    \caption{MCMC in the exploration of the $\sigma$ parameter.\\}
    \label{fig:mcmc_sis}
  \end{subfigure}
  \caption{Illustration of the arc belonging to the initial guess and evolution of the MCMC for the $\sigma$ parameter, in this case the values converge around the 300th step.}
\end{figure}

With the application of the function $\chi^2$ based on the equation \ref{chi_cuadrado}, and with an error of $0.1$ in the position of the images, the initial parameters set of the \textbf{likelihood} for the MCMC was obtained. In figure \ref{fig:chi_sis}, the comparison between the images of  $\vert{\vec{\theta_{p}}}\vert$ with those produced by the parameters obtained of $\chi^2$ is evidenced.

From these values obtained with the multiparametric minimization, the MCMC was executed with \textbf{Gallenspy}. The best result was obtained for a number of 100 walkers and 1000 steps. In figure \ref{fig:mcmc_sis} the chain convergence in the obtention of parameter $\sigma$ is shown, where it is possible to observe that from the obtained data in step 400 the parameter estimation can be done.

\begin{figure}[!tbp]
  \begin{subfigure}[b]{0.49\textwidth}
    \includegraphics[width=\textwidth, height=\textwidth]{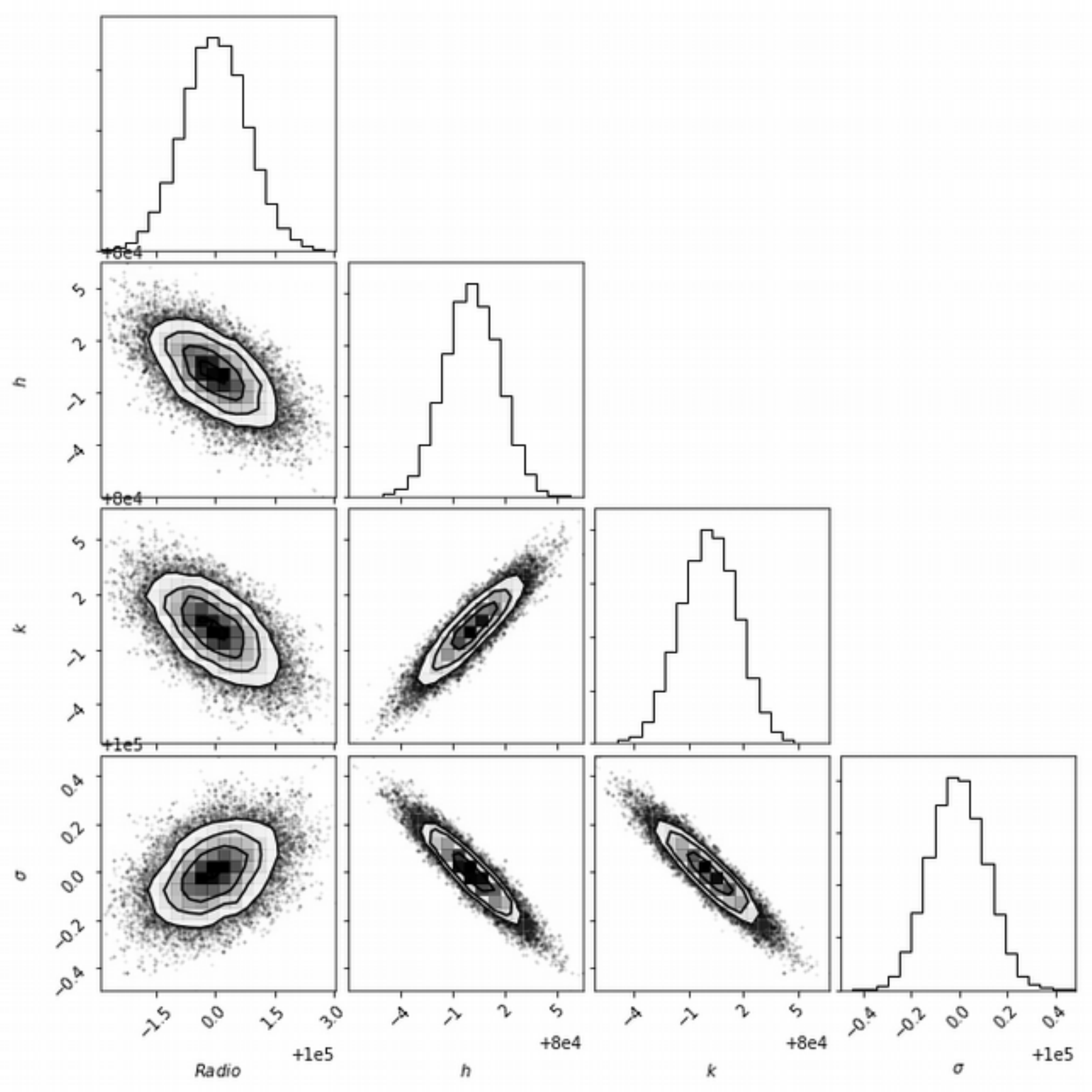}
    \caption{Results obtained with \textbf{Gallenspy}, in the exploration of parameters for the SIS profile.\\}
    \label{fig:graph_mcmc}
  \end{subfigure}
  \hfill
  \begin{subfigure}[b]{0.49\textwidth}
    \includegraphics[width=\textwidth, height=\textwidth]{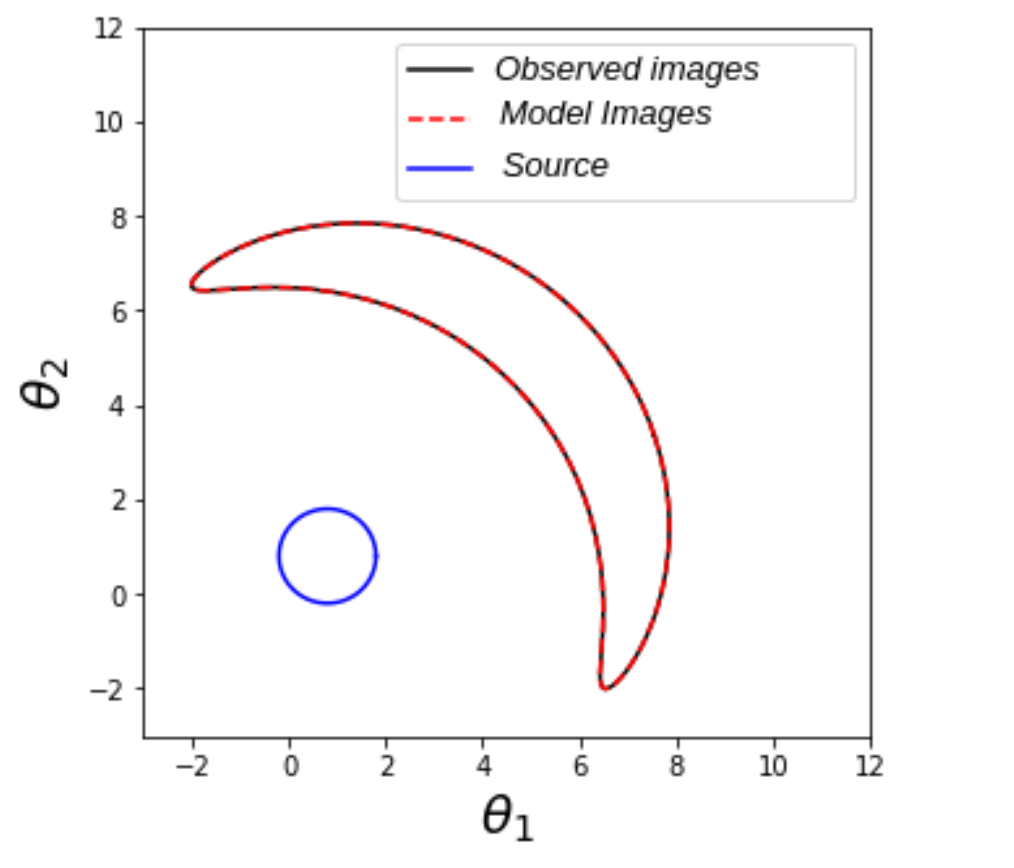}
    \caption{Comparative graph  between the produced images by the SIS model and the images of  $\vert{\vec{\theta_{p}}}\vert$.\\}
    \label{fig:mcmc_sis_fin}
  \end{subfigure}
  \caption{Final results obtained with \textbf{Gallenspy} for the fitting of the arc generated with a SIS model. }
\end{figure}

The graphs illustrated in figure \ref{fig:graph_mcmc} show the reliability regions, under which the parameters family was obtained for the reproduction of the images belonging to $\vert{\vec{\theta_{p}}}\vert$.

\begin{table}[h]
\begin{center}
\begin{tabular}{ |c ||| c|  }
\hline
\hline
\textbf{Parameter} & $\textbf{95\%}$ \\
\hline
\hline
\multicolumn{2}{ |c| }{\textbf{SIS}}\\
\hline
\hline
$\sigma \, \left(10^5 \text{km/s} \right)$ & $0.999_{-0.266}^{+4.456X10^{-6}}$\\
\hline
\hline
\multicolumn{2}{ |c| }{\textbf{Source}}\\
\hline
\hline
$Radio \, \left(\text{arcseg} \right)$ & $1.000_{-0.039}^{+0.999}$ \\
$h \, \left(\text{arcseg} \right)$ & $0.800_{8.393X10^{-5}}^{+0.199}$  \\
$k \, \left(\text{arcseg} \right)$ & $0.800_{5.267X10^{-5}}^{+0.199}$  \\
\hline
\end{tabular}
\vspace{0.5cm}
\\
\end{center}
\caption{Parameters obtained with \textbf{Gallenspy} for the 
SIS model.}
\label{tabla:Param_SIS}
\end{table}

In table \ref{tabla:Param_SIS} the parameters obtained are shown with its uncertainties for the quantile of 95\%, these values are consistent with those established for the obtention of the images of $\vert{\vec{\theta_{p}}}\vert$, for this reason with this illustrative example it was possible to observe the way in which \textbf{Gallenspy} proceeds and its efficiency in the estimation of the parameters from the GLE.

Finally the figure \ref{fig:mcmc_sis_fin} shows the superposition of images, where the reliability of results obtained with \textbf{Gallenspy} becomes evident.

\subsection{Critical and caustic curves with Gallenspy}

Another process that \textbf{Gallenspy} performs is the obtention of critical and caustic curves for distinct lens model. To understand this method, it is important to remember that $detA$ depends on the convergence $\kappa$ and the shear $\gamma$, which are relation to the deflector potential computed by \textbf{Gallenspy} in the fitting of images $(\theta_1, \theta_2)$ \cite{Sch,Rog}.  

In this way, the first step given by \textbf{Gallenspy} is the obtention of $\gamma_1$ and $\gamma_2$ with the equation \ref{shear1} and \ref{shear2}

\begin{equation}\label{shear1}
\gamma_{1}\Big(\vec{\theta}\Big)= \dfrac{1}{2}\Bigg(\dfrac{\partial^{2}\psi\Big(\vec{\theta}\Big)}{\partial\theta^{2}_{1}}-\dfrac{\partial^{2}\psi\Big(\vec{\theta}\Big)}{\partial\theta^{2}_{2}}\Bigg),
\end{equation}

\begin{equation}\label{shear2}
\gamma_{2}\Big(\vec{\theta}\Big)=\dfrac{\partial^{2}\psi\Big(\vec{\theta}\Big)}{\partial\theta_2 \partial\theta_1},
\end{equation}

and from these results, \textbf{Gallenspy} computes the points of the lens plane where $detA=0$ based on the given relation by Hurtado \cite{Rog}

\begin{equation}\label{matriz_A}
A=
\begin{pmatrix}
1-\kappa-\gamma_{1} & -\gamma_{2} \\
-\gamma_{2} & 1-\kappa+\gamma_{1} 
\end{pmatrix}
\end{equation}

\begin{figure}[h]
	\centering
	\includegraphics[width=2.5in]{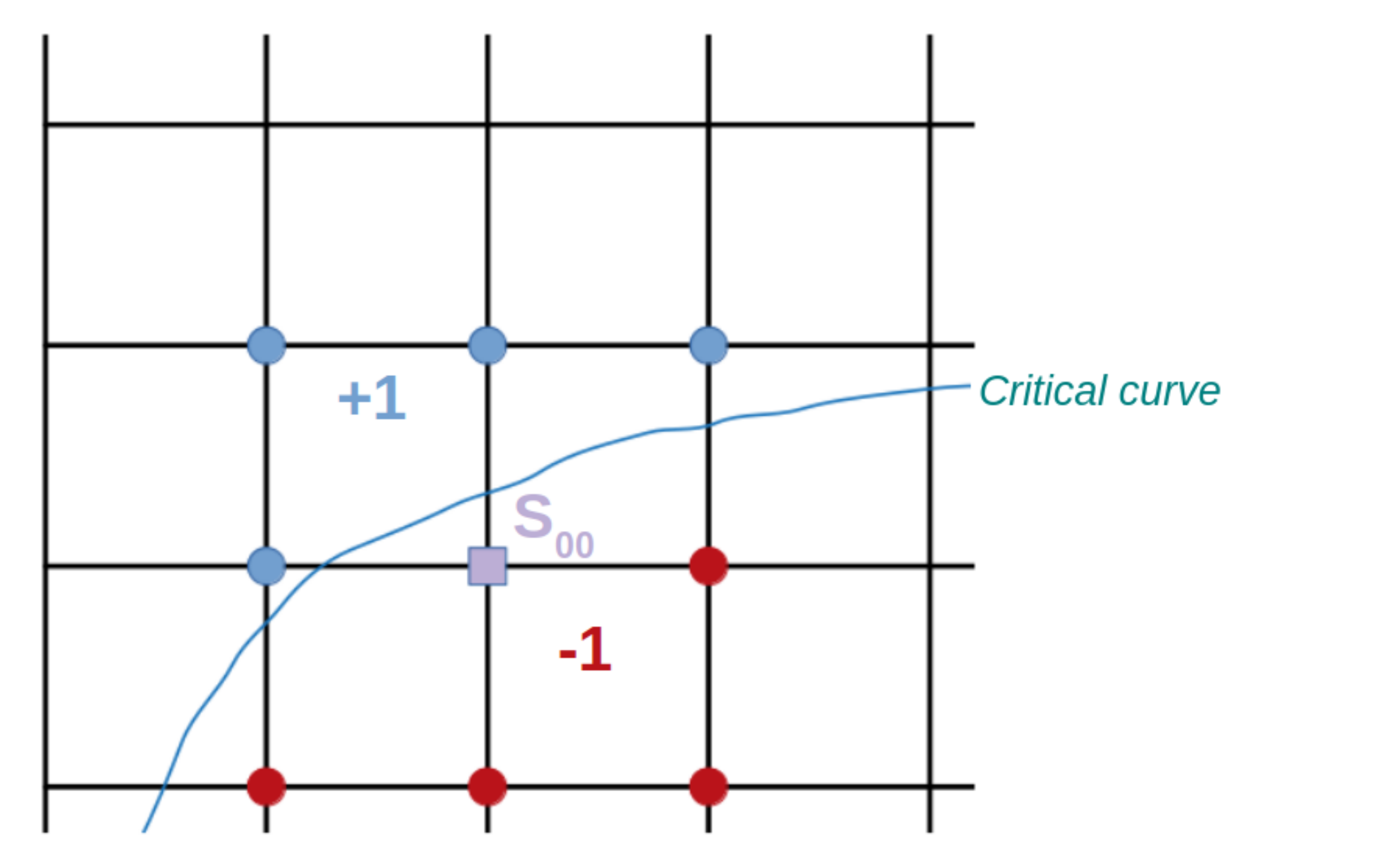}
	\caption{Illustration of the Bartelmann method for the obtention of critical points in \textbf{Gallenspy}, in this case the change in the sign of $detA$ determinant from the critical curve it is taken into account.\\}
	\label{fig:metodo_critic}
\end{figure}

As the obtention of this critical points is not a trivial process, \textbf{Gallenspy} makes this process with the Bartelmann method \cite{Bartelmann}, in which the critical curve in the lens plane is a border that changes the  sign of the determinant ($detA$) as illustred the figure \ref{fig:metodo_critic}.

As it is shown in this image, there are points of the plane $(\theta_{1},\theta_{2})$ even though do not belong to the critical curve, can be considered close to it: from this perspective, if $S=Sign(detA)$ then a point $S_{00}$ is considered adjacent to the curve in a grid when $S$ changes between $S_{00}$ and any of these neighboring points.   

In figure \ref{fig:metodo_critic} presents the estimation of each point for a grid with dimension $NXN$ based on the sign $S$, this means that for positive values $S_{i,j}=1$ and in the case of negative values $S_{i,j}=-1$, where $i$ and $j$ are in a range of $0$ to $N-1$. From this perspective, the established restriction in \textbf{Gallenspy} to know if a point $S_{i,j}$ is adjacent to the critical curve in the grid is based on the condition:  

\begin{equation} \label{restricc_curv_crit}
S_{i,j}(S_{i-1,j}+S_{i+1,j}+S_{i,j-1}+S_{i,j+1})<4,    
\end{equation}

This method is very effective to compute the critical curve as the grid is refined since it allows to reduce the range in which each critical point can be. 

\begin{figure}[h]
	\centering
	\includegraphics[height=3.2in,width=3.6in]{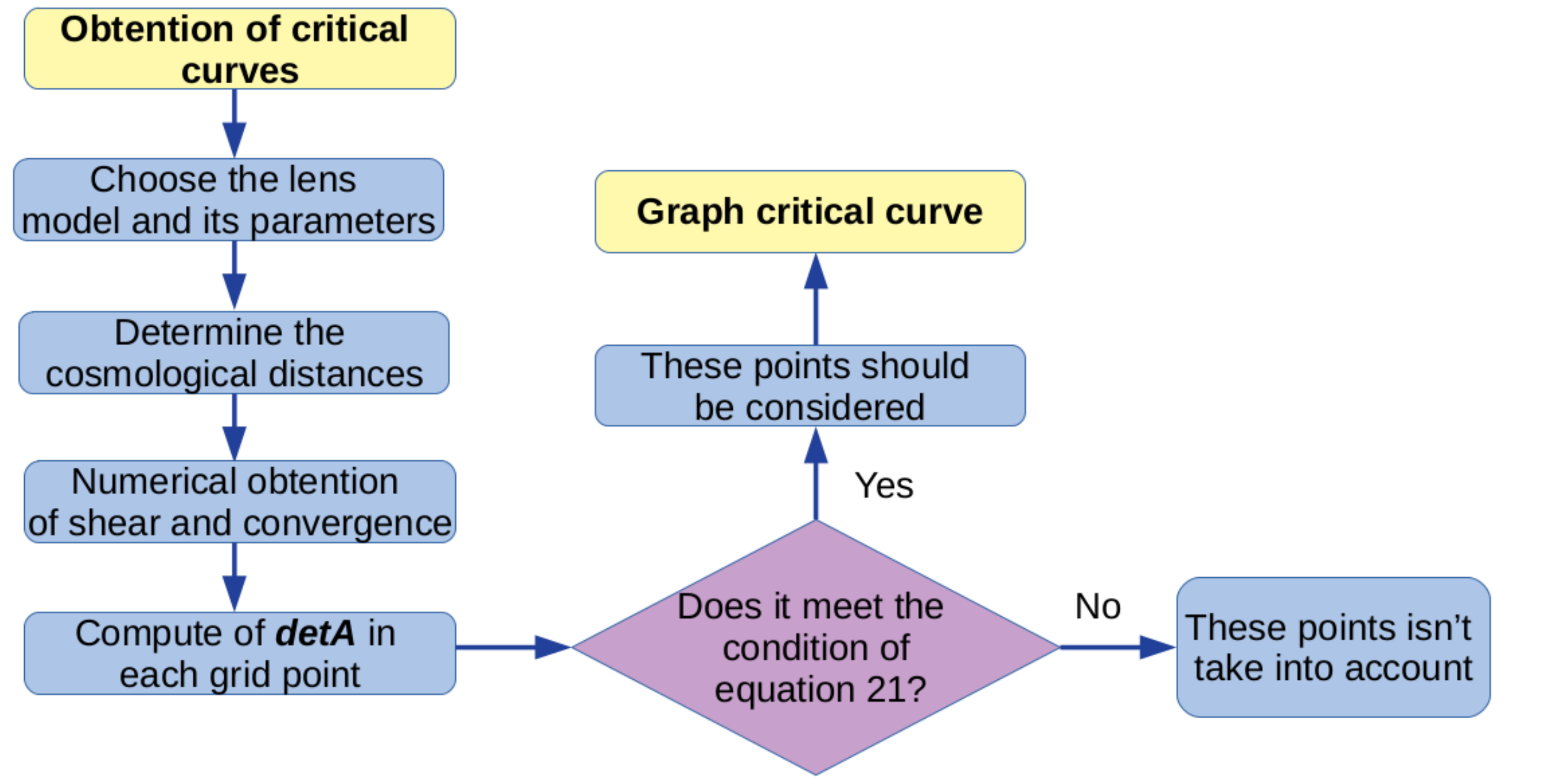}
	\caption{Process flow diagram in the computation of critical curves with \textbf{Gallenspy}.\\}
	\label{fig:flow_diagram_critical}
\end{figure}

Figure \ref{fig:flow_diagram_critical} shows the process flow diagram for the estimation of critical curves with \textbf{Gallenspy}, it is important to highlight that the error range is low due to the grid's refinement of $100X100$ points. From this perspective, the set of critical points with a percentage error of $0.1$ is estimated as those points $S_{i,j}$ in which the condition of equation \ref{restricc_curv_crit} is met.  

When the critical points are computed by \textbf{Gallenspy}, the obtention of the caustic points is an easy process because the application of the equation \ref{eq_len} is enough. 

\begin{figure}[h]
	\centering
	\includegraphics[width=2.0in]{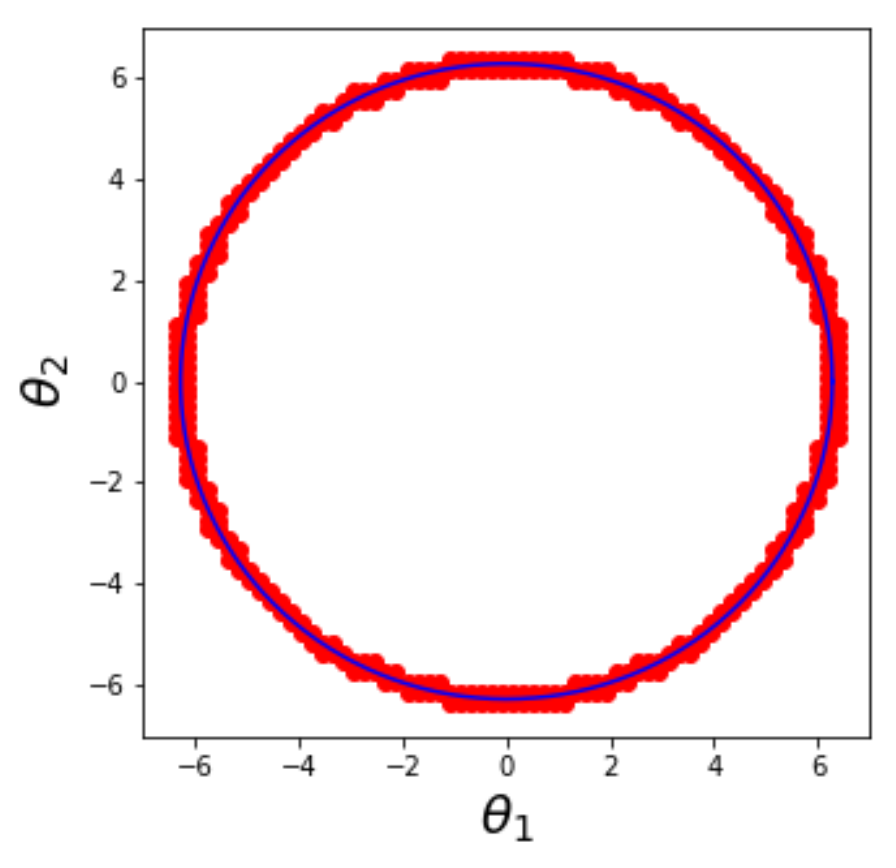}
	\caption{(Red dots) Critical curve obtained with \textbf{Gallenspy} for the case of the SIS profile developed along this work. (Blue) Plotted circle with the averaged radius  of the points obtained with \textbf{Gallenspy}.\\}
	\label{fig:crit_curv}
\end{figure}

Going back to the example of SIS profile of the previous section, the critical curve was obtained with \textbf{Gallenspy}. The result is shown in figure \ref{fig:crit_curv}.

\begin{figure}[h]
	\centering
	\includegraphics[width=3.5in]{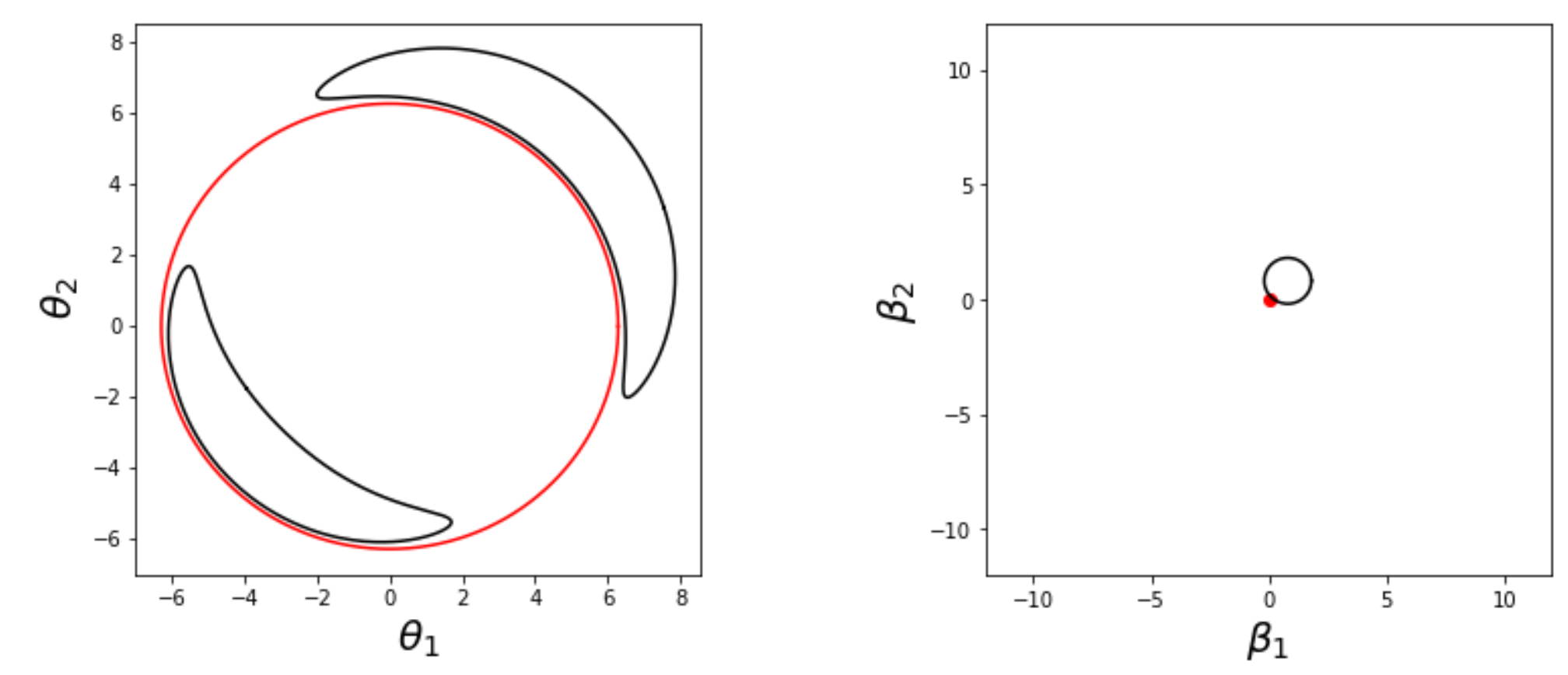}
	\caption{(Right) Caustic curve and circular source (Left) Critical curve and images formed in the SIS model.\\}
	\label{fig:critcurv+images}
\end{figure}

This graph evidences that the critical curve belongs to a circle with a radius of $6.278$ arcs. In this case, it is important to consider the analytical solution to the SIS profile for $detA=0$ \cite{Rog}, where the critical curve is a circumference of radius $r$ given by the relation 

\begin{equation}\label{radio_sis}
r = \dfrac{4\pi\sigma^{2}D_{LS}}{c^{2}D_{OS}},
\end{equation}

Based on the data provided above regarding the SIS, the radius obtained from the equation \ref{radio_sis} is of $6.283$ arcs. This is very close to the obtained result with \textbf{Gallenspy}, where it is possible to check a percentage error of $0.1$ in this numerical process. 

In this way, in the analytical solution of the SIS it it possible to conclude that the caustic curve in this mass profile belongs to a point in the origin of the plane ($\beta_{1}, \beta_{2}$)\cite{Rog}, this is evidenced in figure \ref{fig:critcurv+images} where the source is near to this caustic point and for this reason the magnification of the images is significant.

\section{GALROTPY}

\textbf{Galrotpy}\cite{galrotpy} is an interactive tool focused on the visualization and exploration of parameters through MCMC, in such a way that it is possible to make mass reconstructions from the fitting of rotational curves in disk-like galaxies.  

The main python packages used in this routine are: \textbf{matplotlib}\cite{Matplotlib} for the generation of a graphic environment, \textbf{numpy}\cite{Numpy} for data management, \textbf{astropy}\cite{Astropy}, which is useful for units assignation, \textbf{Galpy}\cite{galpy} for the construction of rotational curves with each mass profile, \textbf{emcee}\cite{emcee} used in the exploration and fitting of data with MCMC, and \textbf{corner}\cite{corner} for the reliability regions of the parametric fitting.

\begin{figure}[!tbp]
  \begin{subfigure}[b]{0.49\textwidth}
    \includegraphics[width=\textwidth, height=\textwidth]{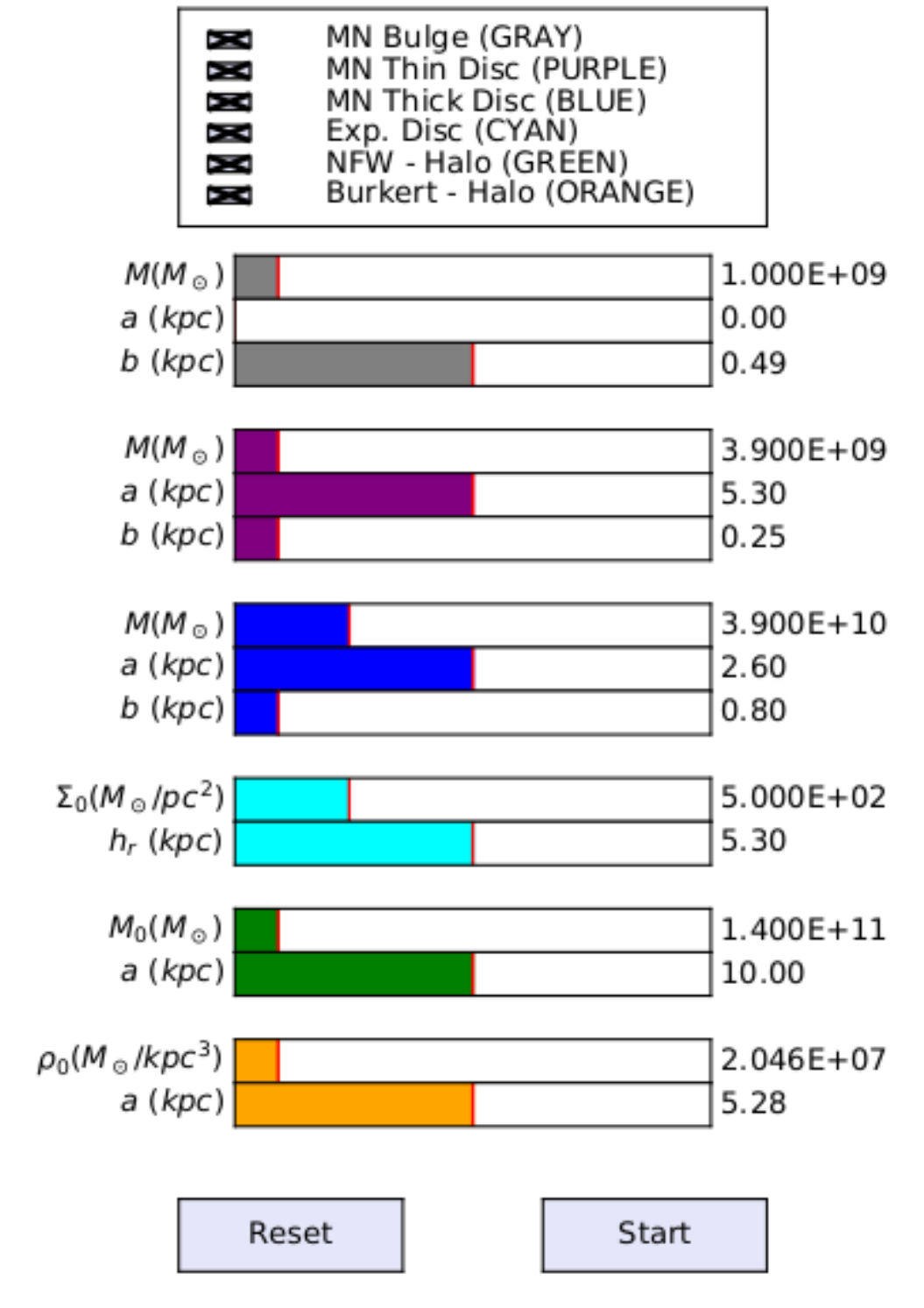}
    \caption{Panel for the selection of gravitational potentials.\\}
    \label{fig:Ej1_galrotpy}
  \end{subfigure}
  \hfill
  \begin{subfigure}[b]{0.49\textwidth}
    \includegraphics[width=\textwidth, height=\textwidth]{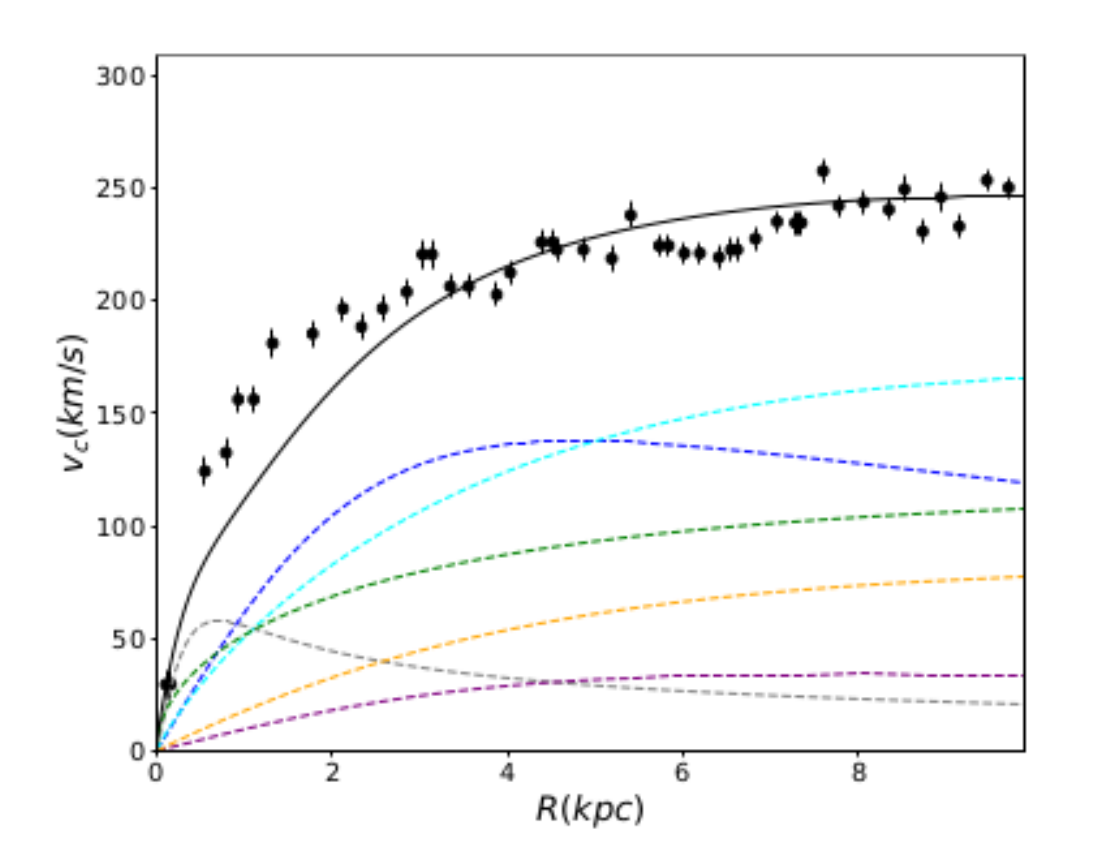}
    \caption{Rotation curve created from the superposition of distinct mass distributions.\\}
    \label{fig:Ej2_galrotpy}
  \end{subfigure}
  \caption{Graphic environment in \textbf{Galrotpy}, wherein the left side the selection of potentials for the curve fitting is possible.}
\end{figure}

The space of parametric exploration in this routine is the same as \textbf{Gallenspy} due to the reasons given in the previous section. On the other hand, the rotational velocity data from which the parametric fitting is done should be consigned in a file denominated \textbf{rot\_curve.txt}, in which the units belonging to the radial coordinates must be expressed in $Kpc$ and the velocities in $Km/s$.  
Figure \ref{fig:Ej1_galrotpy} is shows the panel for the selection of gravitational potentials in \textbf{Galrotpy}, in which the variation  parameters is done. Thus the user can do a visual fitting between the rotational curve and the observational data of rotational velocity (this is evidenced in figure \ref{fig:Ej2_galrotpy}).       

\subsection{Bayesian statistics in Galrotpy}

In the case of \textbf{Galrotpy} the MCMC has similar characteristics with \textbf{Gallenspy} even in the consideration of the \textbf{prior} and in their approach to evaluation of the \textbf{likelihood}.

However, the process of parametric exploration with \textbf{Galrotpy} presents greater facilities due to the visual fitting that is possible to do with this routine and for this reason, the parametric minimization done with \textbf{Gallenspy} is not necessary. 

In this way, with the visual parameter fitting \textbf{Galrotpy} proceeds to run the MCMC where as in \textbf{Gallenspy} the user can choose the number of walkers and steps. It is important to point out that in this routine, the \textbf{likelihood} es given by the relation

\begin{equation}
L =exp\bigg(-\dfrac{1}{2}\sum_{i=1}^{N} \Bigg[\dfrac{ v_{obs}^{i} -v^{i}_{model}}{v_{i}^{error}}\Bigg]^2\bigg),
\label{likeli_galrotpy}
\end{equation} 

with:

\begin{itemize}
    \item N the number of observationally obtained data.
    \item $v_{obs}$ the observed velocity.
    \item $v_{model}$ the rotational velocity of the mass model chosen for the fitting.
    \item $v_{error}$ the error in the velocity observational data.
\end{itemize}   

Once \textbf{Galrotpy} does the parametric exploration, the behavior of the MCMC is illustrated and the values with uncertainties of 68\% and 95\% are shown. Finally \textbf{Galrotpy} generates two types of graphics, one with the fitting rotation curve and other with the reliability regions. 

\subsection{Mass reconstruction of galaxy M33 with Galrotpy}

M33 is a spiral galaxy without bar structure \cite{galrotpy}, and its set of rotational velocity data  was obtained from Corbelli et al. \cite{Corbelli}. For the mass reconstruction with \textbf{Galrotpy}, it is important to point out that the parametric exploration was done with a number of 100 walkers and 3000 steps. 

\begin{figure}[h]
	\centering
	\includegraphics[width=2.5in]{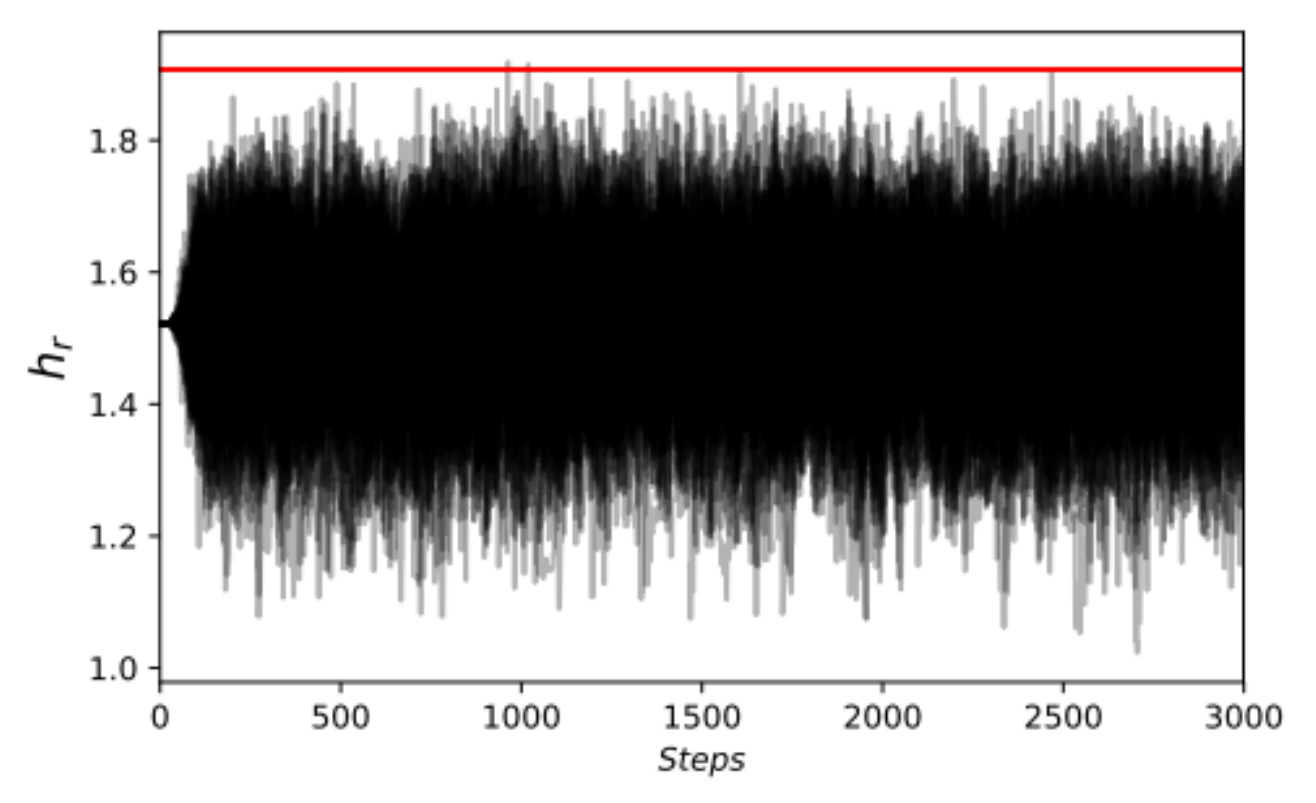}
	\caption{Behavior of the MCMC in the exploration of parameter $h_r$.\\}
	\label{fig:M33_galrotpy}
\end{figure}

In figure \ref{fig:M33_galrotpy}, is presented the way in which MCMC explores the parameter $h_{r}$, where the convergence is given in a lower number of steps less than \textbf{Gallenspy}, thanks to the visual fitting of \textbf{Galrotpy}. 

\begin{figure}[h]
	\centering
	\includegraphics[width=4.5in]{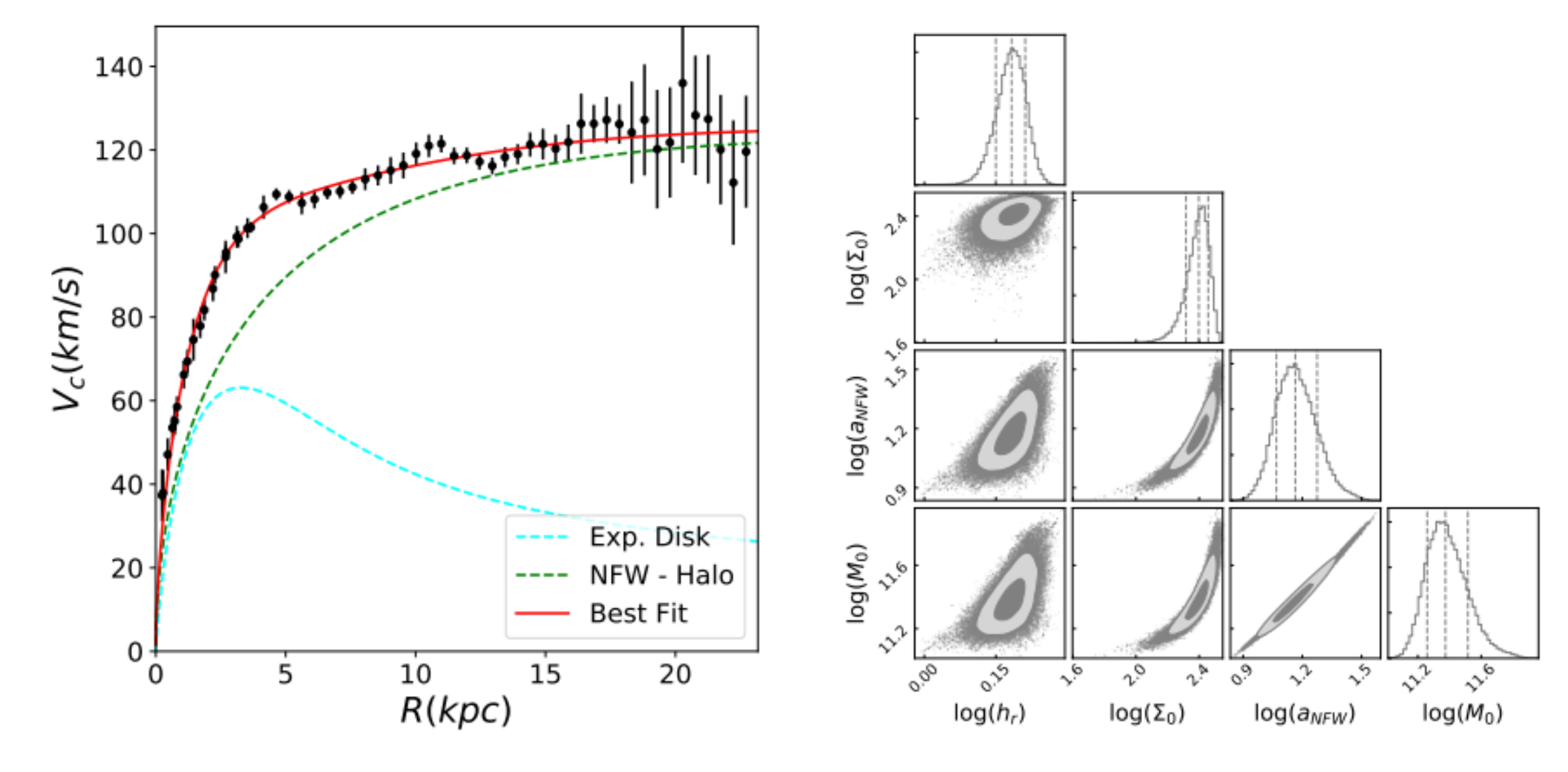}
	\caption{Rotation curves and reliability regions of the Galaxy M33 with \textbf{Galrotpy}, where the fitting was done with NFW profile for the dark matter halo and Exponential Disk in the case of baryonic matter.\\}
	\label{fig:curv_galrotpy}
\end{figure}

\begin{table}[h]
\begin{center}
\begin{tabular}{| c | c | c |}
\hline
Parameter & $68\%$ & $95\%$ \\
\hline
\multicolumn{3}{| c |}{Exponential Disk}\\
\hline

$h_{\text{r}} \, \left( \text{Kpc} \right)$ & $1.52_{-0.11}^{+0.10}$ & $1.52_{-0.23}^{+0.20}$ \\
$\Sigma_{0} \, \left(X10^{2} \text{ M}_{\odot} \text{ pc}^{-2} \right)$ & $2.50_{-0.43}^{+0.37}$ & $2.50_{-0.99}^{+0.66}$ \\
$M_{\star} \, \left( X10^{9} \text{ M}_{\odot} \right)$ & $3.61_{-0.91}^{+0.96}$ & $3.61_{-1.74}^{+1.89}$\\
\hline
\multicolumn{3}{| c |}{NFW}\\
\hline
$a \, \left( \text{X10kpc} \right)$ & $1.46_{-0.29}^{+0.42}$ & $1.46_{-0.49}^{+1.02}$ \\
$M_{\text{0}} \, \left( X10^{11} \text{ M}_{\odot} \right)$ & $2.37_{-0.55}^{+0.91}$ & $2.37_{-0.91}^{+2.45}$ \\
$\rho_{\text{\text{0}}} \, \left( X10^{6} \text{ M}_{\odot} \text{Kpc}^{-3} \right)$ & $6.05_{-2.13}^{+2.96}$ &  $6.05_{-3.55}^{+6.88}$  \\
$M_{\text{h}} \, \left(X10^{11} \text{ M}_{\odot} \right)$ & $4.16_{-0.72}^{+1.11}$ & $4.16_{-1.21}^{+2.86}$\\
\hline
\end{tabular}
\vspace{0.5cm}
\\
\caption{ \texttt{Estimated parameters with \textbf{GalRotpy} for the galaxy M33.}}
\label{tabla:Valores_m33}
\end{center}
\end{table}

Figure \ref{fig:curv_galrotpy} shows the fitting done with \textbf{Galrotpy}, where the NFW profile was used for the dark matter halo while the contribution of baryonic matter was analyzed with the Exponential Disk profile. The right side of this figure shows the reliability regions, these values are consigned in the table \ref{tabla:Valores_m33}. 

The obtained values and its uncertainties are in concordance with the results reported by L\'{o}pez Fune et.al. \cite{Corbelli2}, where  $M_{\star}\left( X10^{9} M_{\odot} \right) = 4.9 \pm1.5$ for the Exponential Disk and  $M_{h}\left( X10^{11} M_{\odot} \right) = 5.4 \pm0.6$ in the case of the dark matter halo. With this example, it was possible to show a set of results with high reliability in the use of this routine, and for this reason in the next section \textbf{Galrotpy} and \textbf{Gallenspy} are combined for the mass reconstructions of two disk-like galaxies.

\section{Mass Reconstruction of galaxies J2141 and J1331}

The galaxies  SDSSJ2141-001(J2141) and SDSSJ1331+3628(J1331) are systems that show strong lensing effect, its rotation velocities data were given by Dutton et.al. \cite{Dutt,Curv}; for these mass reconstructions the profiles of Miyamoto-Nagai, Exponential Disk, and NFW  were taken into account \cite{Binney, galrotpy}. Regarding the GLE, in the case of J2141 an extended circular source was modeled considering that the deflected image is an arc, while for J1331 it was considered a punctual source in which four images are produced in the lens plane.

As to the mass distribution of these galaxies, other authors have reported a high contribution of baryonic matter \cite{Curv,Dutt} and this coincides with the obtained results in the use of \textbf{Galrotpy} and \textbf{Gallenspy}. 

\subsection{Galaxy J2141}

J2141 is a type S0 spiral galaxy, with a dominant gravitational contribution coming from its disk\cite{Curv}. This object was initially observed in 2006 by means of the Hubble Spatial Telescope(HST)\cite{Curv}, with an ACS camera in a  F814 filter and an exposure time of 420 seconds. In 2009 the Keck telescope took again images of these object with a NIR camera and K filter.  

From these images the GLE in this galaxy was evidenced, with the formation of an arc belonging to a source with a redshift different to J2141 ($z_{L}=0.3180$,$z_{s}=0.7127$)\cite{Curv}, for this reason it was considered important to study the mass distribution of this galaxy.               

\begin{figure}[h]
	\centering
	\includegraphics[width=3.5in]{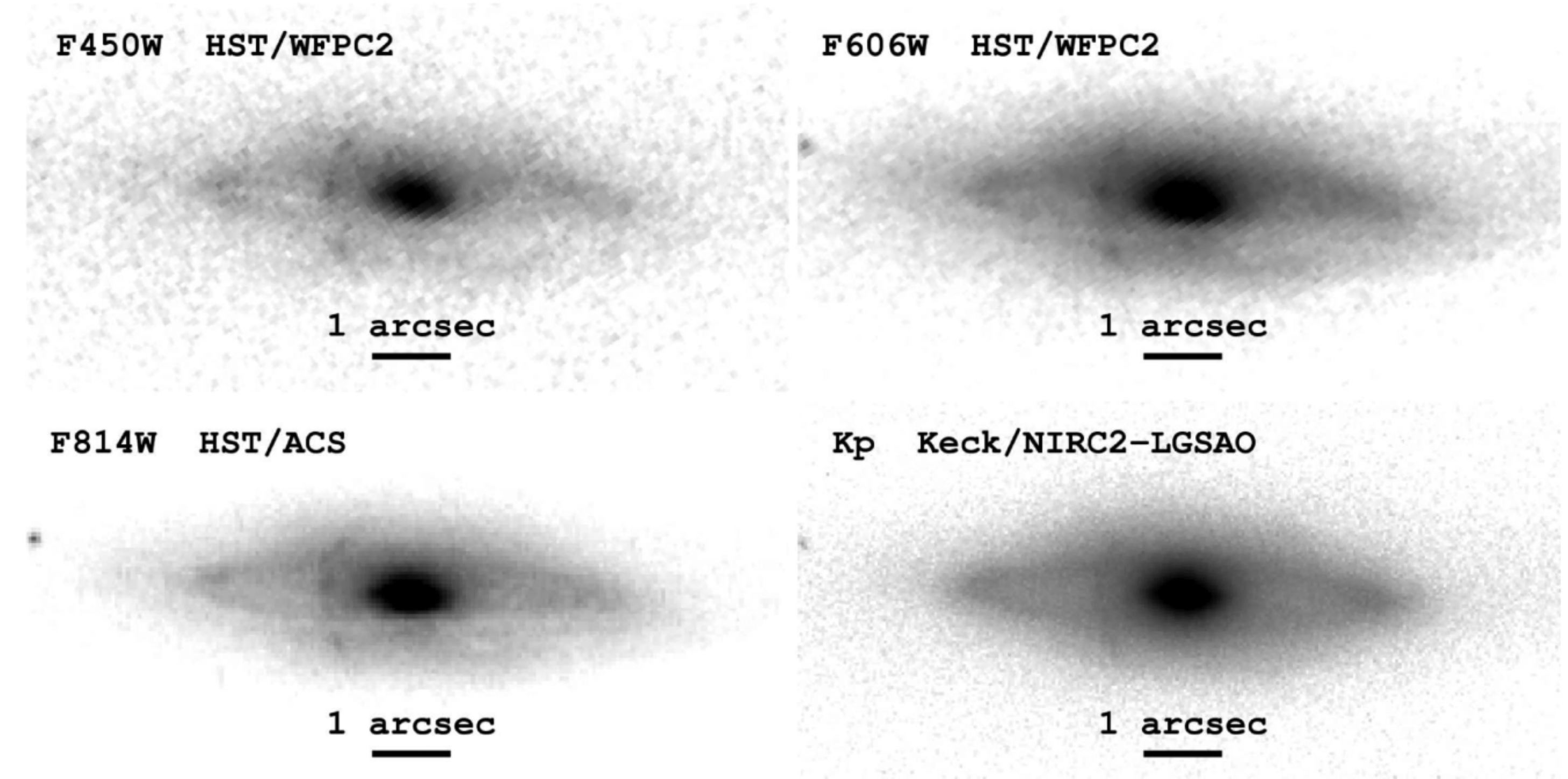}
	\caption{Images obtained of J2141 by means of HST and KeckII telescopes with distinct filters.\cite{Curv}\\}
	\label{fig:SDSSJ2141-001}
\end{figure}

\begin{figure}[!tbp]
  \begin{subfigure}[b]{0.44\textwidth}
    \includegraphics[width=\textwidth, height=\textwidth]{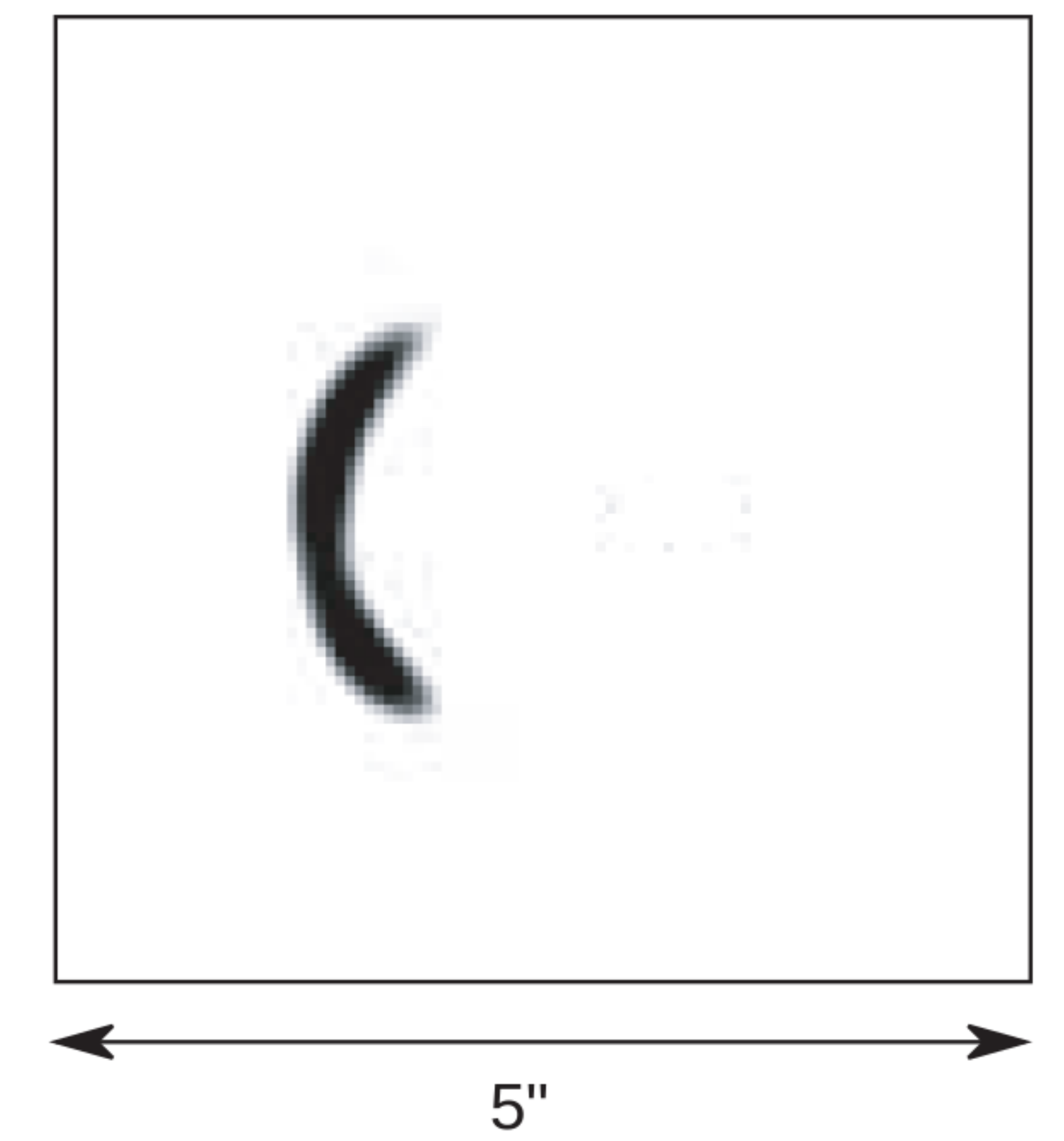}
    \caption{Adjusted image by Dutton et.al. of the arc generated by GLE in the J2141 plane. \cite{Curv}\\}
    \label{fig:arc_J2141}
  \end{subfigure}
  \hfill
  \begin{subfigure}[b]{0.47\textwidth}
    \includegraphics[width=\textwidth, height=\textwidth]{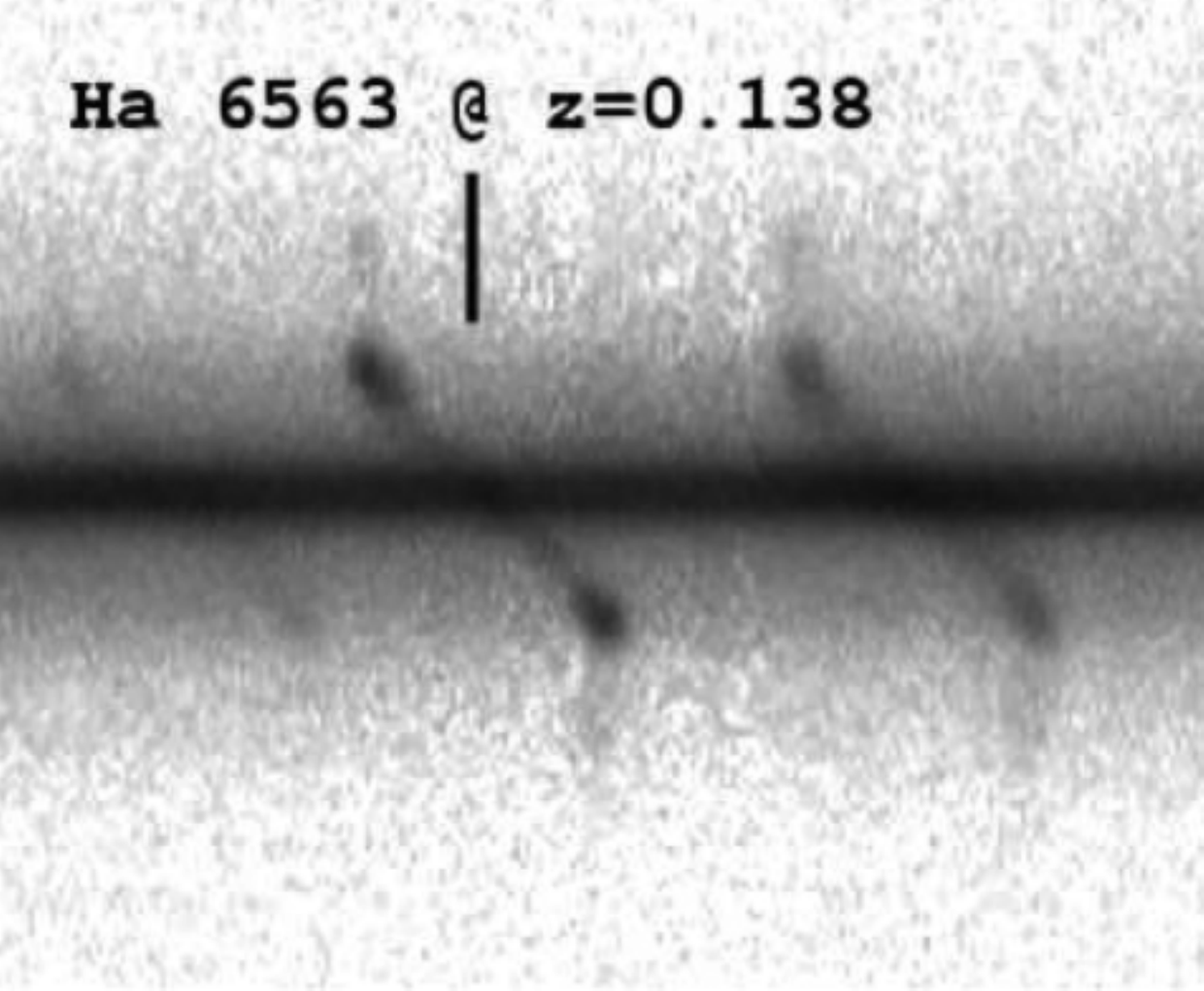}
    \caption{Spectral emission lines of $H_\alpha$ belonging to J2141. \cite{Curv}\\}
    \label{fig:H_alpha}
  \end{subfigure}
  \caption{Observational data obtained from the galaxy J2141 in relation with lensing and rotation velocities.}
\end{figure}

The rotational velocity data of this galaxy was derived from spectral lines  Mg b 5177, Fe II5270, Na D 5896, O II 3727 and $H_{\alpha}$ 6563, obtained from the Keck telescope. In figure \ref{fig:H_alpha}, the inclination of these emission lines are shown for the case of $H_{\alpha}$ 6563.

Table \ref{tabla:Velocidad_rotacional_SDSSJ2141-001} contains the rotational velocity data for different values of galactocentric radius.

\begin{table}[htb]
\begin{center}
\begin{tabular}{| c | c | c | c |}
\hline
Radius(arcsec) & Radius(kpc) & Velocity(km/s) & Error(km/s) \\
\hline
\hline
0.000 & 0.00 & 3.5 & 5.3 \\
0.593 & 1.45 & 114.1 & 5.8\\
1.185 & 2.89 & 153.8 & 2.9\\
1.778 & 4.33 & 212.7 & 2.6\\
2.370 & 5.78 & 243.8 & 2.6\\
2.963 & 7.22 & 259.8 & 2.3\\
4.148 & 10.11 & 254.9 & 7.5\\
4.740 & 11.56 & 263.4 & 2.3\\
5.333 & 13.00 & 265.9 & 3.5\\
\hline
\end{tabular}
\vspace{0.5cm}
\\
\caption{ Rotational velocity values obtained from Dutton et.al. \cite{Curv} for J2141.}
\label{tabla:Velocidad_rotacional_SDSSJ2141-001}
\end{center}
\end{table}

\subsubsection{Data concerning the GLE}

For the coordinates of the arc in the lens plane ($\theta_{1},\theta_{2}$), a restriction of its contours was made in the way of lower the computational cost in \textbf{Gallenspy}. 

\begin{figure}[h]
	\centering
	\includegraphics[height=2.0in, width=2.0in]{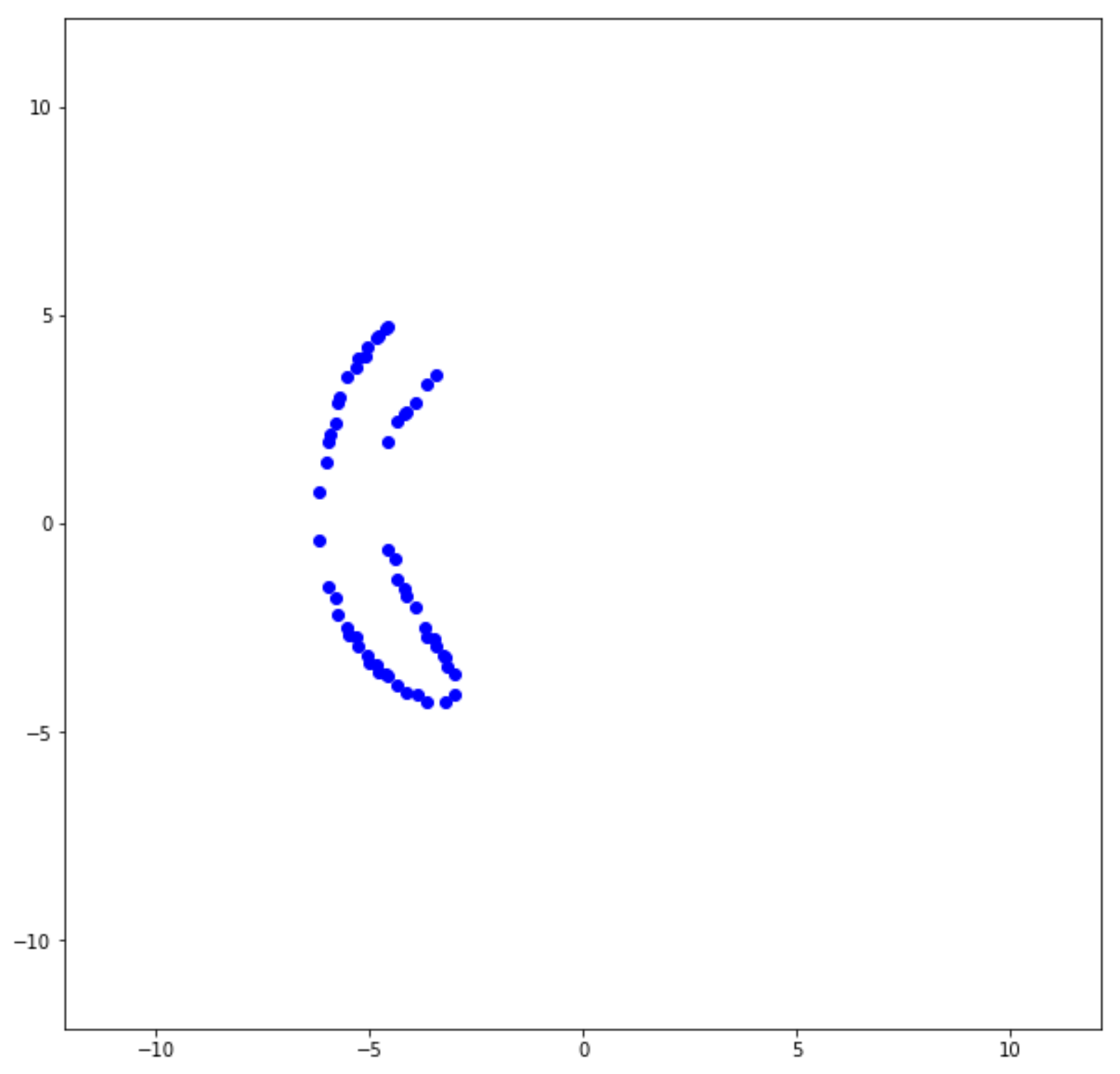}
	\caption{Arc contours generated by GLE in J2141. The scale of this plane is in $1X10^6$ radians.}
	\label{fig:contours_arc}
\end{figure}

This image treatment was made in \textbf{python}, trough a pixel to pixel discrimination based on its position and luminosity, which makes it possible to get the contour showing in figure \ref{fig:contours_arc}.  

As it is possible to observe, this obtained contour is incomplete given the noise of image \ref{fig:arc_J2141}, even so, the number of obtained coordinates was enough for an appropriate adjustment with a circular source.    

In this case the equivalence between arc seconds and pixels was made based on the observed scale of figure \ref{fig:arc_J2141},  and in this way it was possible to get each coordinate of the image in radians as observed in figure \ref{fig:contours_arc}.

For the determination of cosmological distances, the $\Lambda$CDM model was assumed due to the fact that it has been used by other authors in this galaxy \cite{Curv}; in this way the current matter density is $\Lambda_{m}=0.3$ and the Hubble parameter $H_{0}=70kms^{-1}Mpc^{-1}$.  

With these considerations, the cosmological distances given by Dutton et al. \cite{Curv} are $D_{OL}=497.6Mpc$, $D_{OS}=1510.2Mpc$ and $D_{LS}=1179.6Mpc$; which leads to $\Sigma_{crit}=4285.3M_{\odot}pc^{-2}$.

In relation to the source, its ranges of possible values for the position and the radius are displayed in table \ref{tabla:Valores_fuente}.

\begin{table}[h]
\begin{center}
\begin{tabular}{| c | c | c |}
\hline
\hline
Parameter & Range & Units  \\
\hline
\hline
Radius & 0.05<r<1.5 & arcs \\
x-center & -2.5<h<2.5 & arcs\\
y-center & -2.5<k<2.5 & arcs\\
\hline
\end{tabular}
\vspace{0.5cm}
\\
\caption{Range of values to explore for the source.}
\label{tabla:Valores_fuente}
\end{center}
\end{table}

\subsubsection{Mass reconstruction of J2141.}

\begin{figure}[h]
	\centering
	\includegraphics[height=2.0in, width=3.0in]{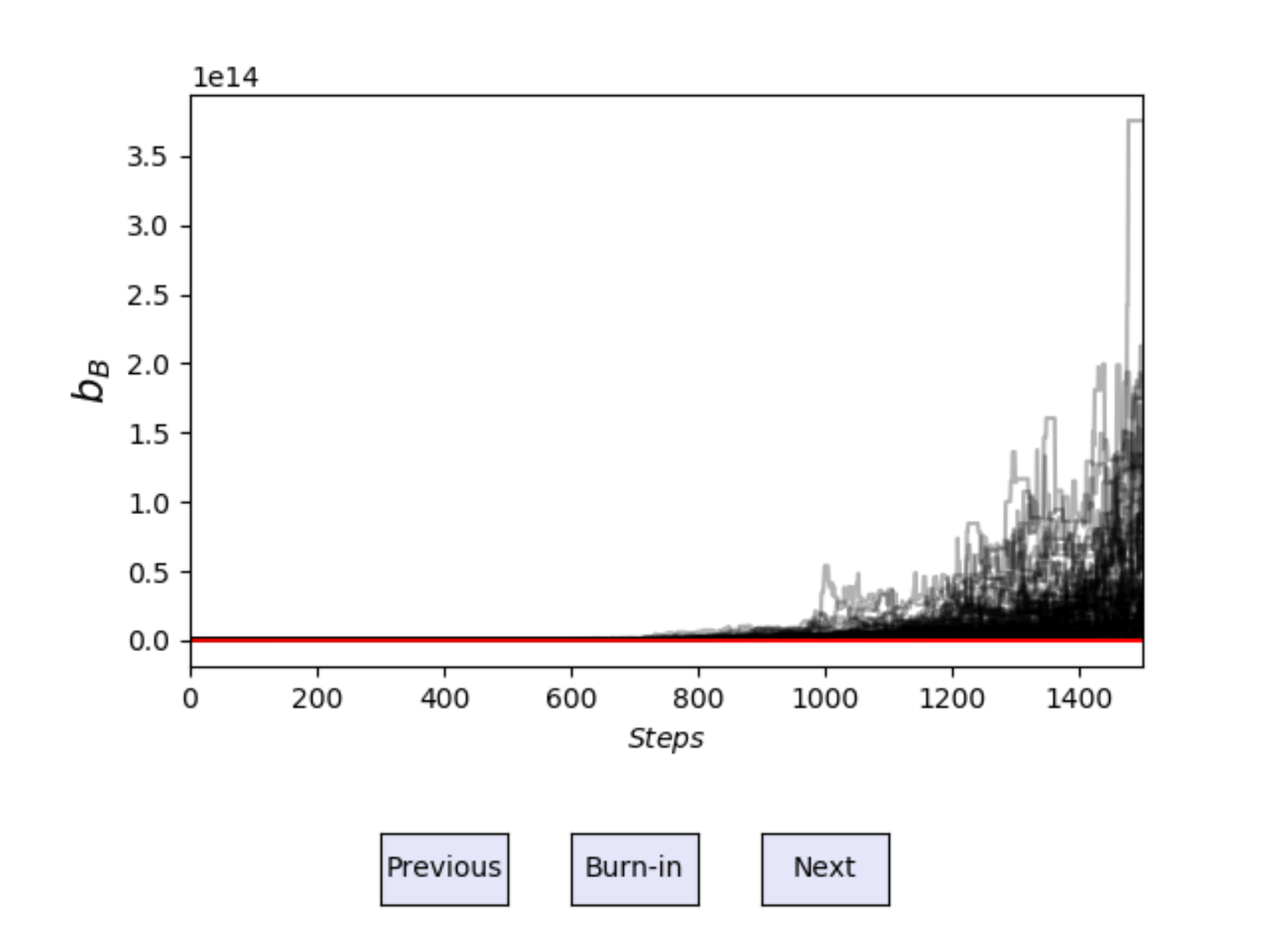}
	\includegraphics[height=2.0in, width=3.0in]{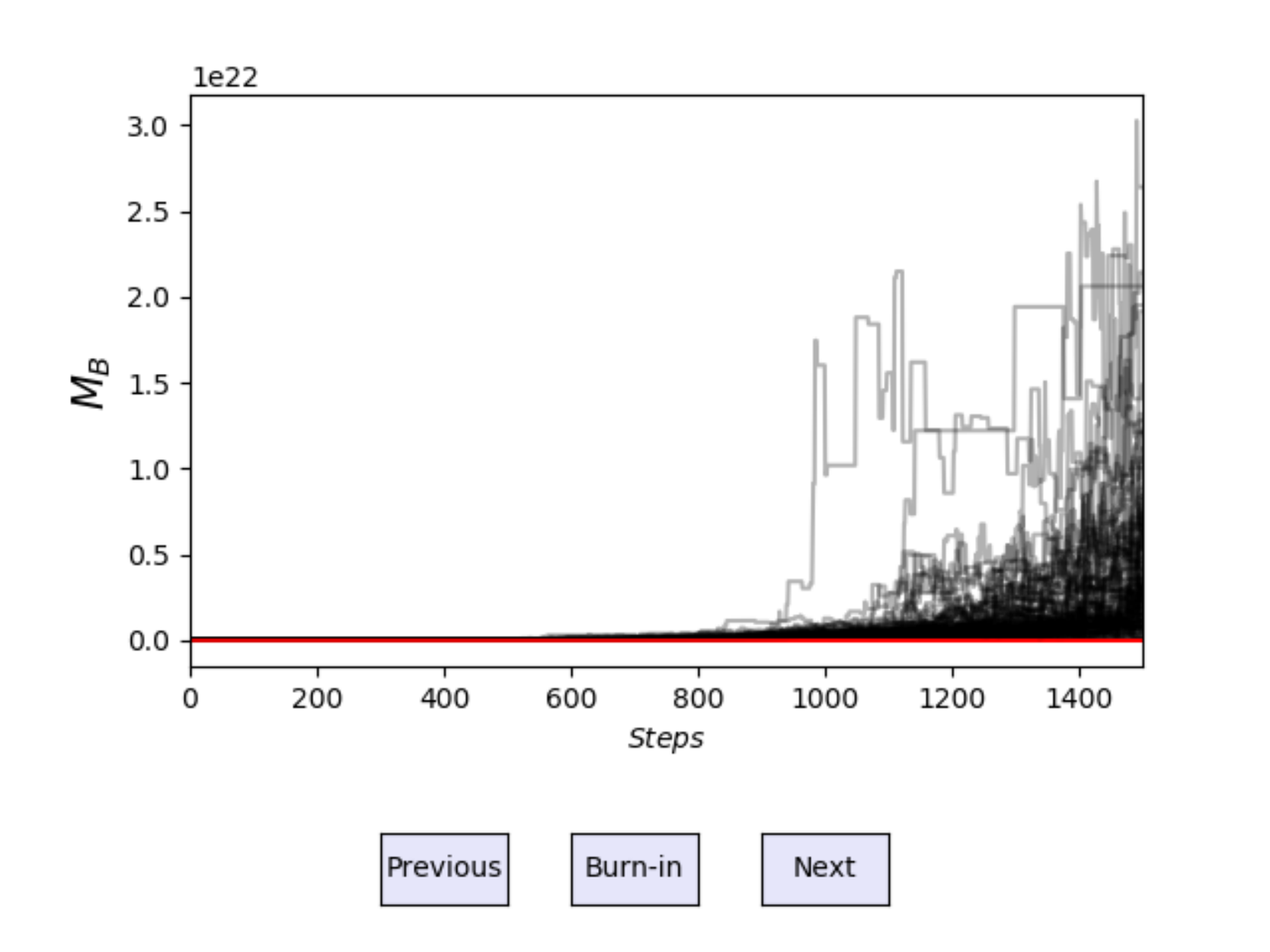}
	\caption{Exploration roads for the parameters of the Miyamoto-Nagai profile.}
	\label{fig:MN_road_exploration}
\end{figure}

\begin{figure}[h]
	\centering
	\includegraphics[height=2.0in, width=3.0in]{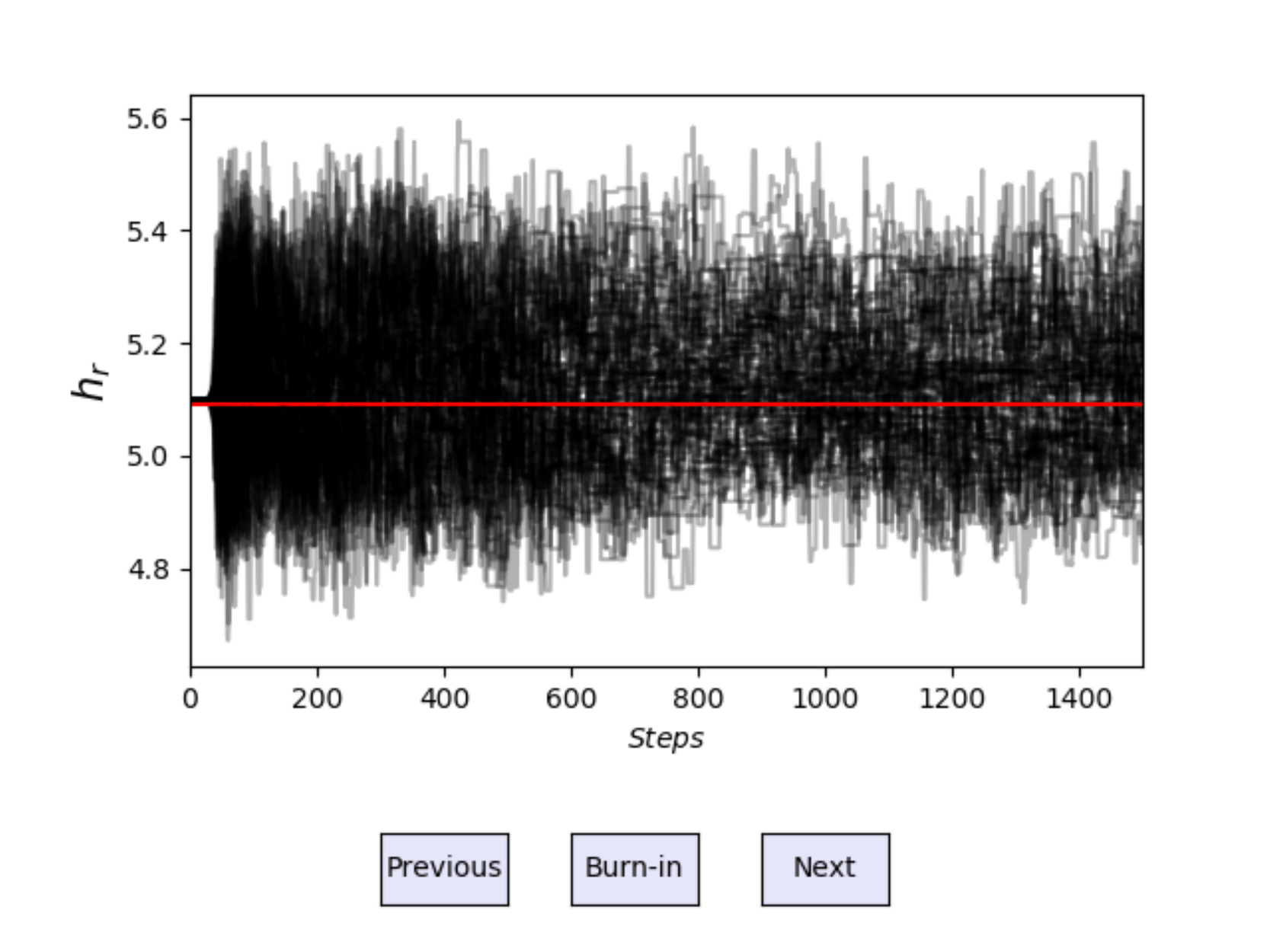}
	\includegraphics[height=2.0in, width=3.0in]{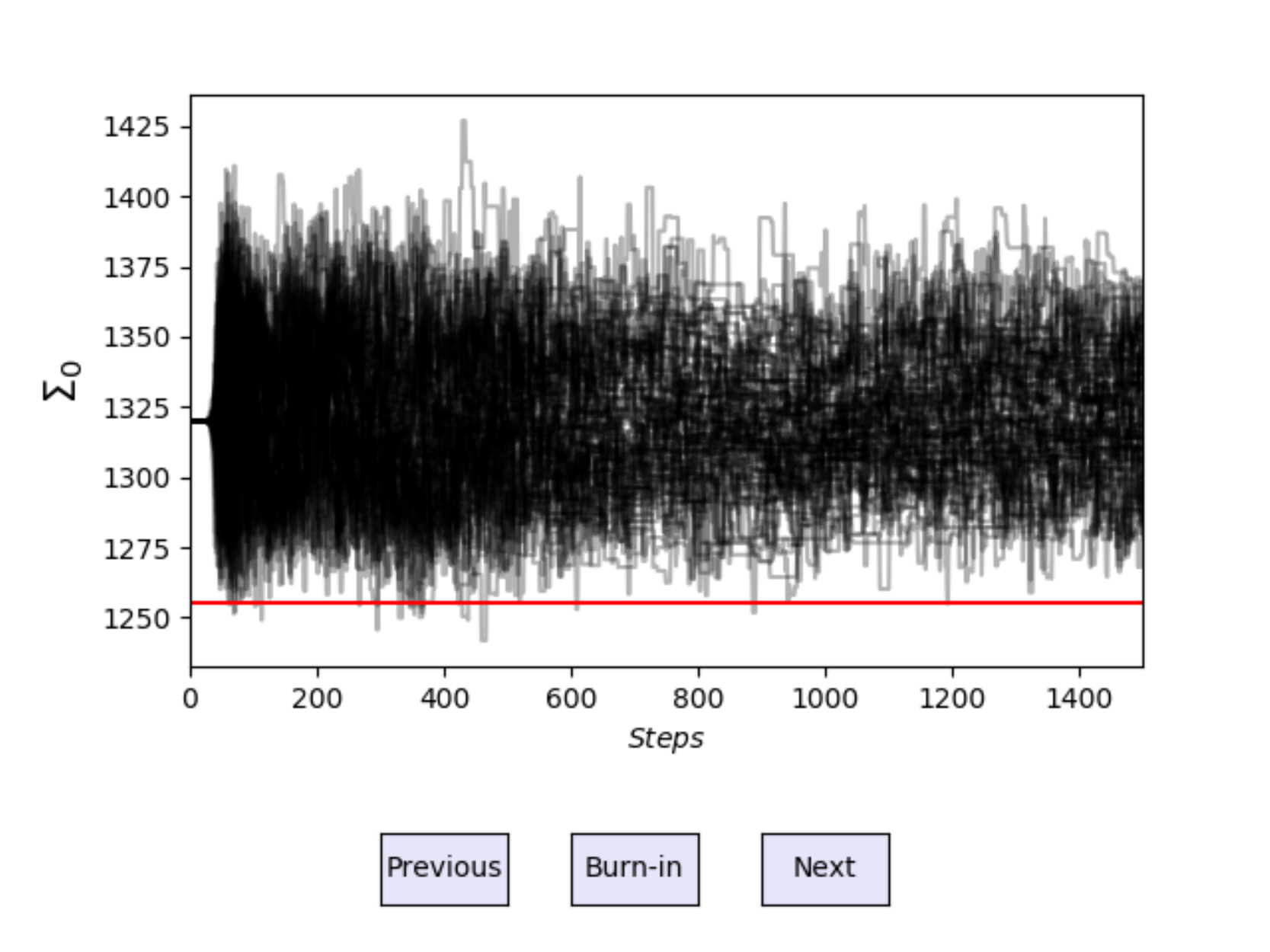}
	\caption{Exploration roads for the parameters of the Exponential Disk profile.}
	\label{fig:ED_road_exploration}
\end{figure}

\begin{figure}[h]
	\centering
	\includegraphics[height=2.0in, width=3.0in]{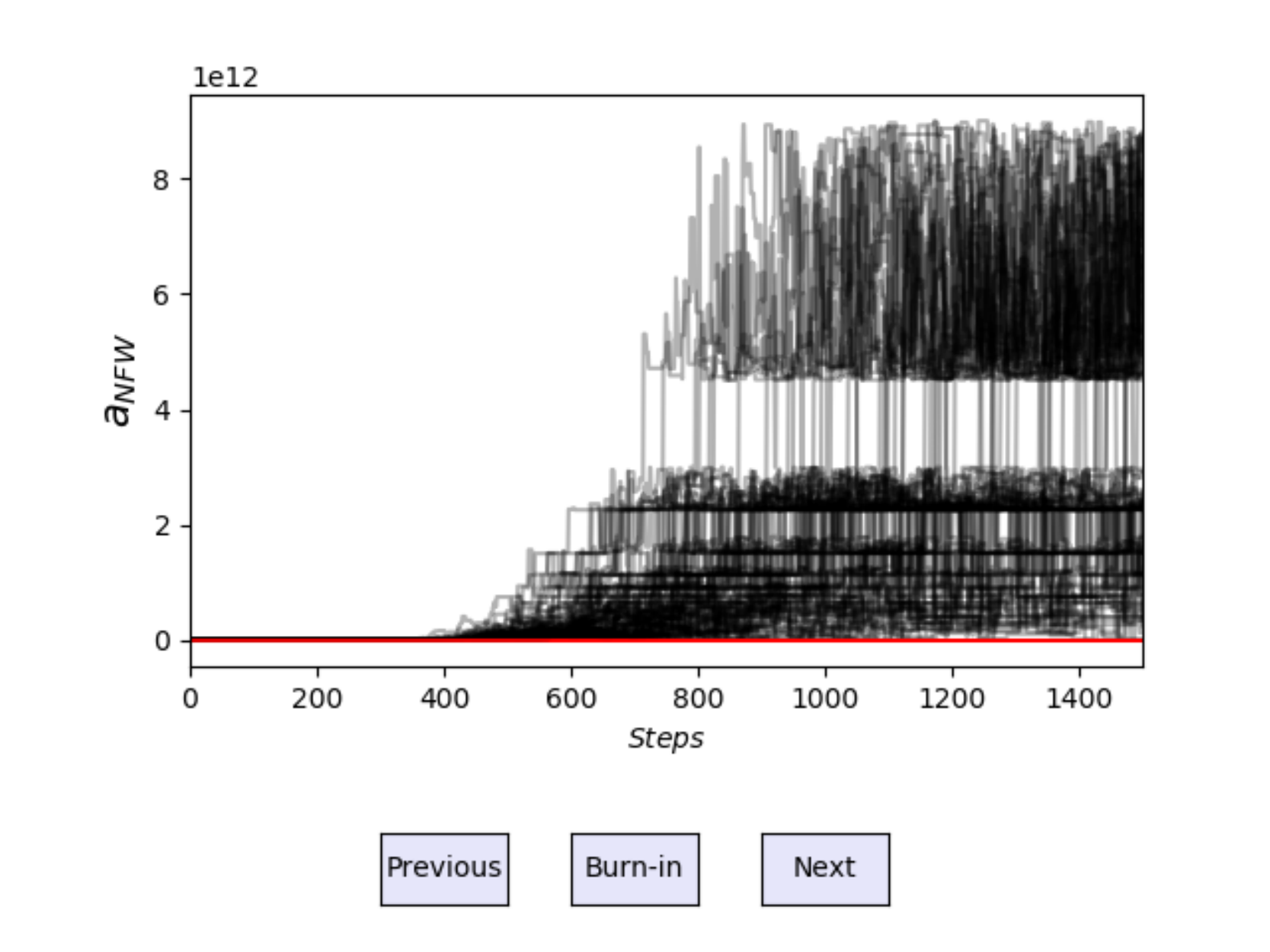}
	\includegraphics[height=2.0in, width=3.0in]{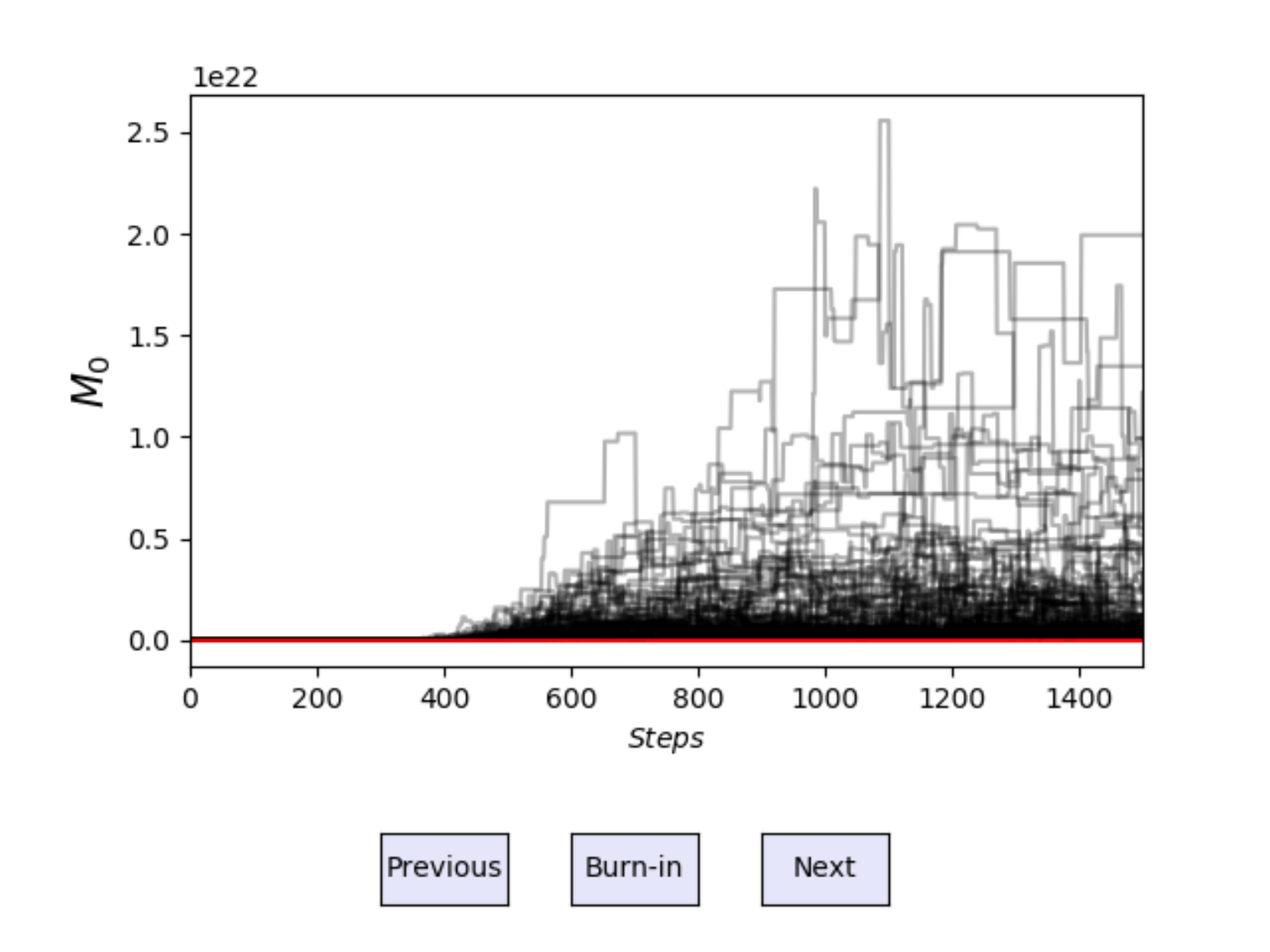}
	\caption{Exploration roads for the parameters of the NFW profile.}
	\label{fig:NFW_road_exploration}
\end{figure}

In this mass reconstruction, the profiles selected in table \ref{tabla:Param_Gallenspy} were Bulge I, Exponential Disk, and NFW.  The first parametric exploration was done with \textbf{Galrotpy}, where the more reliable result for the MCMC was obtained for a number of 100 walkers and 1500 steps. The parametric exploration roads are shown in figures \ref{fig:MN_road_exploration}, \ref{fig:ED_road_exploration} and \ref{fig:NFW_road_exploration}.   

In these graphics, the initial values of the MCMC are in red lines, whereas for the dark matter halo and bulge is not evidenced a convergence of the values; this is pointed out by Dutton as the problem of degeneracy among the disk and the different mass components of the galaxy.\cite{Curv} 

These authors \cite{Curv} affirm that the main reason of this degeneracy is the gravitational dominance of the disc in this system, to adress this problem, it is possible to adjust the rotation curve only with the profile of this mass component. Therefore it is very difficult to know with clarity the circular velocity contribution of other components.    

Because of this situation, this mass reconstruction was done through the integration of galactic dynamics and GLE, where the restrictions of each method occur in different geometries. Therefore this combination is a powerful tool to break this degeneracy.  

\begin{figure}[!tbp]
  \begin{subfigure}[b]{0.55\textwidth}
    \includegraphics[width=\textwidth, height=\textwidth]{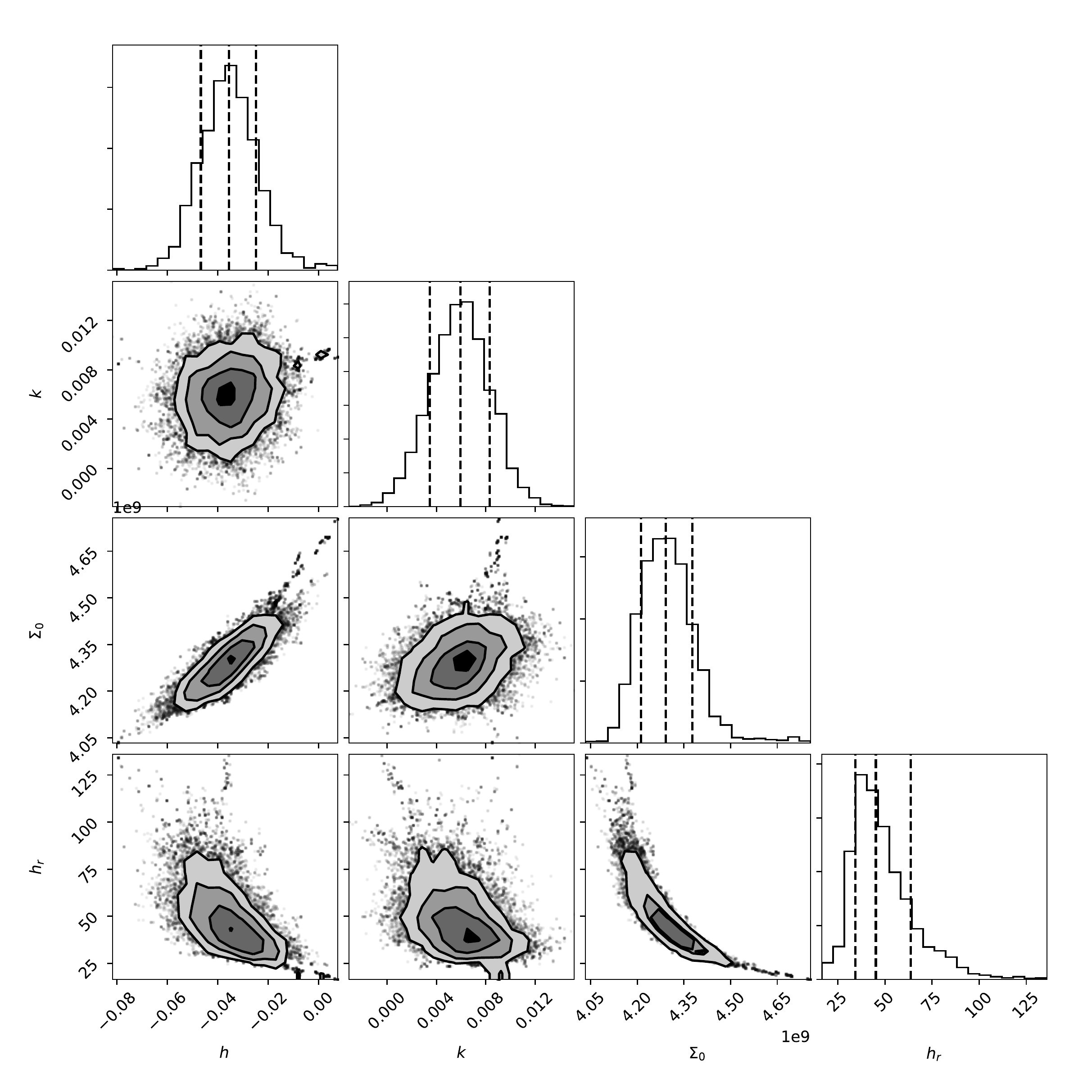}
    \caption{Estimation of circular source with Gallenspy for the galaxy J2141.\\}
    \label{fig:circular_source_J2141}
  \end{subfigure}
  \hfill
  \begin{subfigure}[b]{0.50\textwidth}
    \includegraphics[width=\textwidth, height=\textwidth]{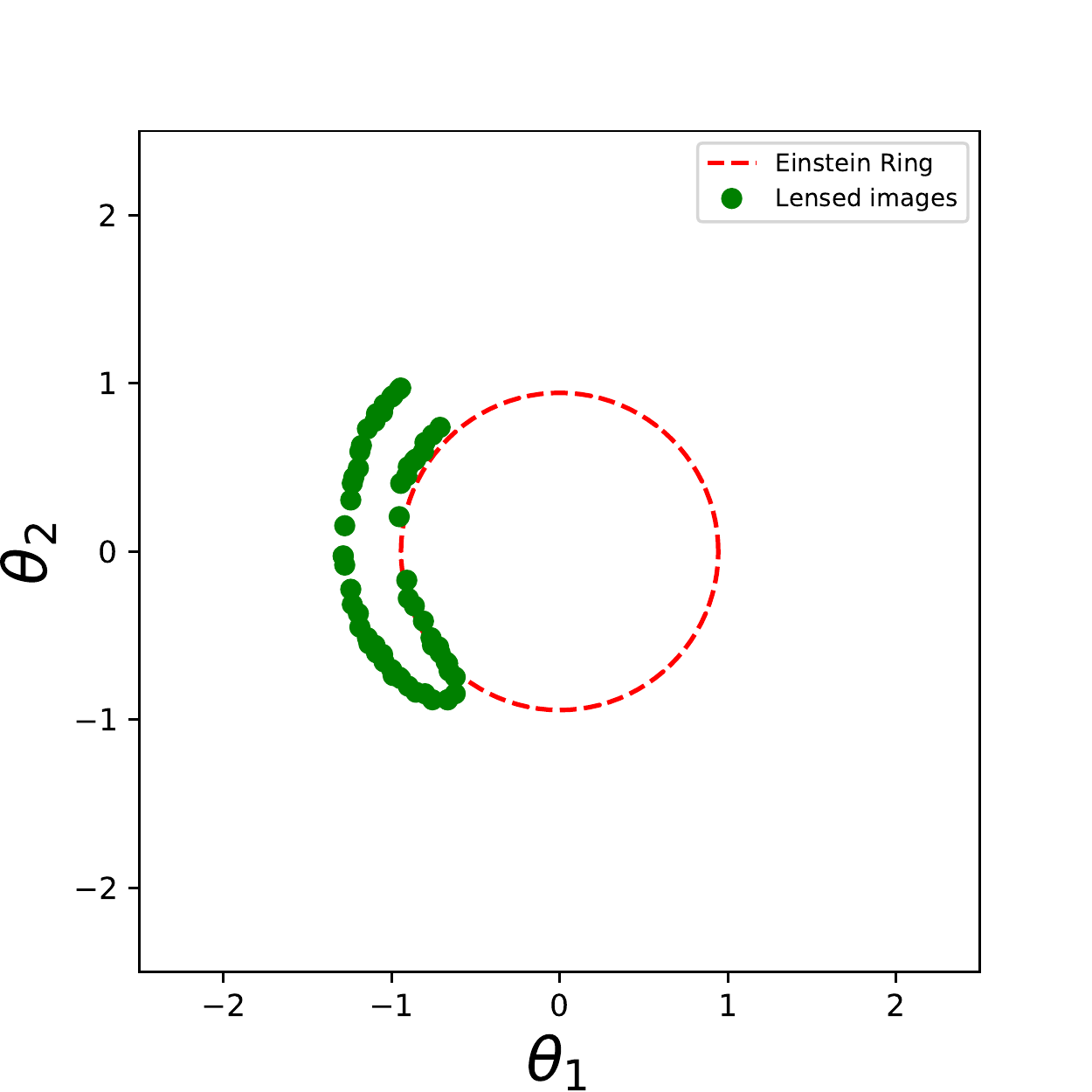}
    \caption{Deflected image and Einstein ring (scale in arcs).\\}
    \label{fig:curvs_crit_gal1}
  \end{subfigure}
  \caption{Compute of \textbf{Gallenspy} for source parameters (left side) and Einstein radius (right side).}
\end{figure}

In this way and with the arc coordinates, the parameters of the source were estimated as shown in figure \ref{fig:circular_source_J2141} under a lens model of Exponential Disk, where these values are presented in table \ref{tabla:fuente_gal1} with a source radius of 0.03$arcs$. 

\begin{table}[h]
\begin{center}
\begin{tabular}{| c | c | c |}
\hline
Parameter & $68\%$ & $95\%$ \\
\hline
$h \, \left( X10^{2}\text{arcs} \right)$ & $3.56_{-1.108}^{+1.082}$ &  $3.56_{-2.106}^{+2.395}$  \\
$k \, \left( X10^{3}\text{arcs} \right)$ & $5.938_{-2.470}^{+2.372}$ & $5.938_{-4.849}^{+4.546}$ \\
\hline
\end{tabular}
\vspace{0.5cm}
\\
\caption{ Source values estimated by \textbf{Gallenspy} in the case of J2141.}
\label{tabla:fuente_gal1}
\end{center}
\end{table}

With the position and size of the source, the mass reconstruction of J2141 based on the GLE was done with the restrictions obtained of the rotation curves in relation with the parametric space. For this specific case, the amplitude of the parametric exploration established was three times of the table \ref{tabla:Param_Gallenspy}.  

\begin{figure*}[]
	\centering
	\includegraphics[height=8.5in, width=6.5in]{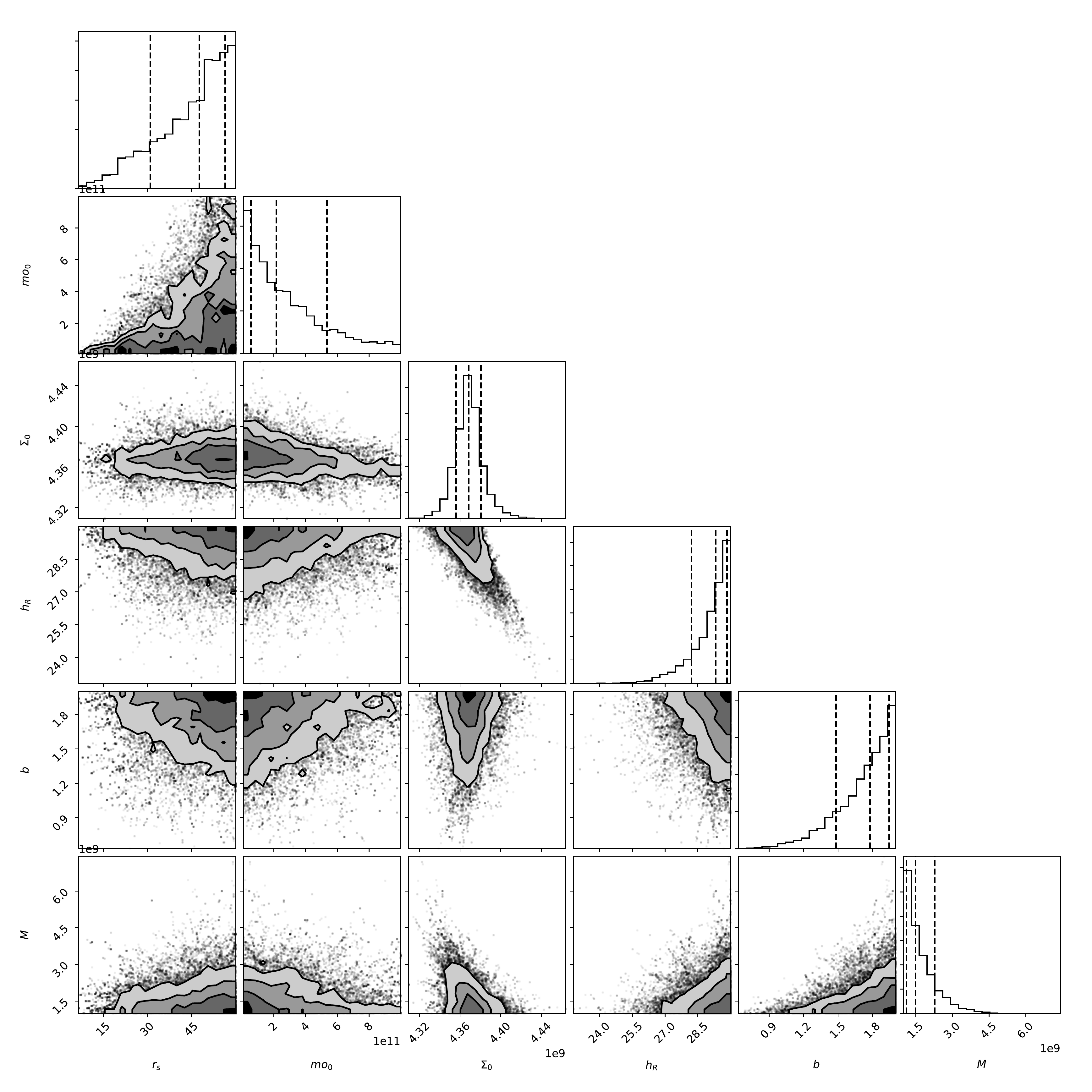}
	\caption{Contours obtained through the combination of restrictions between \textbf{Galrotpy} and \textbf{Gallenspy} for the galaxy J2141.}
	\label{fig:contours_gallenspy1}
\end{figure*}

The combination of lensing and Galactic Dynamics was a very efficient process in breaking of the degeneracy between the mass components of J2141, this allowed a lower mass density value for the disc as it is shown in figure \ref{fig:contours_gallenspy1}. Based on the result, it is important to point out how the combination of \textbf{Galrotpy} and \textbf{Gallenspy} is a great alternative for galaxies where the gravitational contribution of each mass component is not easy to distinguish.

In table \ref{tabla:Gallenspy_gal1} the parameters and its uncertainties are consigned, where the parameters with major dispersion belong to NFW profile, which is related to the degeneracy of this system.

\begin{table}[h]
\begin{center}
\begin{tabular}{| c | c | c |}
\hline
Parameter & $68\%$ & $95\%$ \\
\hline
\multicolumn{3}{| c |}{NFW}\\
\hline
$a \, \left( \text{kpc} \right)$ & $47.653_{-16.696}^{+8.786}$ & $47.653_{-0.036}^{+0.020}$ \\
$m_{\text{\text{0}}} \, \left( X10^{11} \text{ M}_{\odot} \right)$ & $2.171_{-1.610}^{+3.18}$ &  $2.171_{-2.009}^{+6.815}$  \\
\hline
\multicolumn{3}{| c |}{Exponential Disc}\\
\hline
$h_{\text{r}} \, \left( \text{Kpc} \right)$ & $29.312_{-1.111}^{+0.520}$ & $29.312_{-2.620}^{+0.663}$ \\
$\Sigma_{0} \, \left(X10^{9} \text{ M}_{\odot} \text{ pc}^{-2} \right)$ & $4.368_{-0.012}^{+0.012}$ & $4.368_{-0.026}^{+0.031}$ \\
\hline
\multicolumn{3}{| c |}{Miyamoto-Nagai}\\
\hline
$b \, \left( \text{Kpc} \right)$ & $1.778_{-0.296}^{+0.166}$ & $1.778_{-0.643}^{+0.212}$ \\
$M \, \left(X10^{9} \text{ M}_{\odot} \right)$ & $1.492_{-0.367}^{+0.781}$ & $1.492_{-0.043}^{+0.474}$ \\
\hline
\end{tabular}
\vspace{0.5cm}
\\
\caption{ Values of parameters obtained in the mass reconstruction for J2141.}
\label{tabla:Gallenspy_gal1}
\end{center}
\end{table}

Based on the values of these parameters, the Einstein ring and Einstein radius $(\theta_{Eins})$ were computed. In figure \ref{fig:curvs_crit_gal1}, this curve obtained by \textbf{Gallenspy} is evidenced, where the Einstein radius $\theta_{Eins}$ presented a value of $0.943_{-0.144}^{+0.128}$. 

The compute of enclosed mass for J2141, was done taking into account the relation:

\begin{equation}\label{calc_masa}
M=\dfrac{\Sigma_{cr}}{2}\int_{S} \nabla^{2} \psi(\theta_{1},\theta_{2}) d^{2}\theta,
\end{equation}

Table \ref{tabla:Gallens_gal1} shows the estimated values.

\begin{table}[h]
\begin{center}
\begin{tabular}{| c | c |}
\hline
Parameter & $Log_{10} \, \left( \dfrac{\text{M}}{\text{M}_{\odot}} \right)$ \\
\hline
$\text{M}_{Eins}$& $10.906_{-0.160}^{+0.030}$\\
\hline
$\text{Mbar}_{Eins}$& $10.905_{-0.027}^{+0.023}$\\
\hline
\end{tabular}
\vspace{0.5cm}
\\
\caption{ Enclosed mass with Einstein radius, M being the total mass and Mbar the baryonic matter.}
\label{tabla:Gallens_gal1}
\end{center}
\end{table}

These results are in concordance with the range of  values given by Dutton et al. \cite{Curv}, where they reported for J2141 a value of $Log_{10} \, \left(\dfrac{\text{Mbar}}{\text{M}_{\odot}}\right)=  10.99_{-0.25}^{+0.11}$ within the Einstein radius.

\begin{figure}[!tbp]
  \begin{subfigure}[b]{0.49\textwidth}
    \includegraphics[width=\textwidth, height=\textwidth]{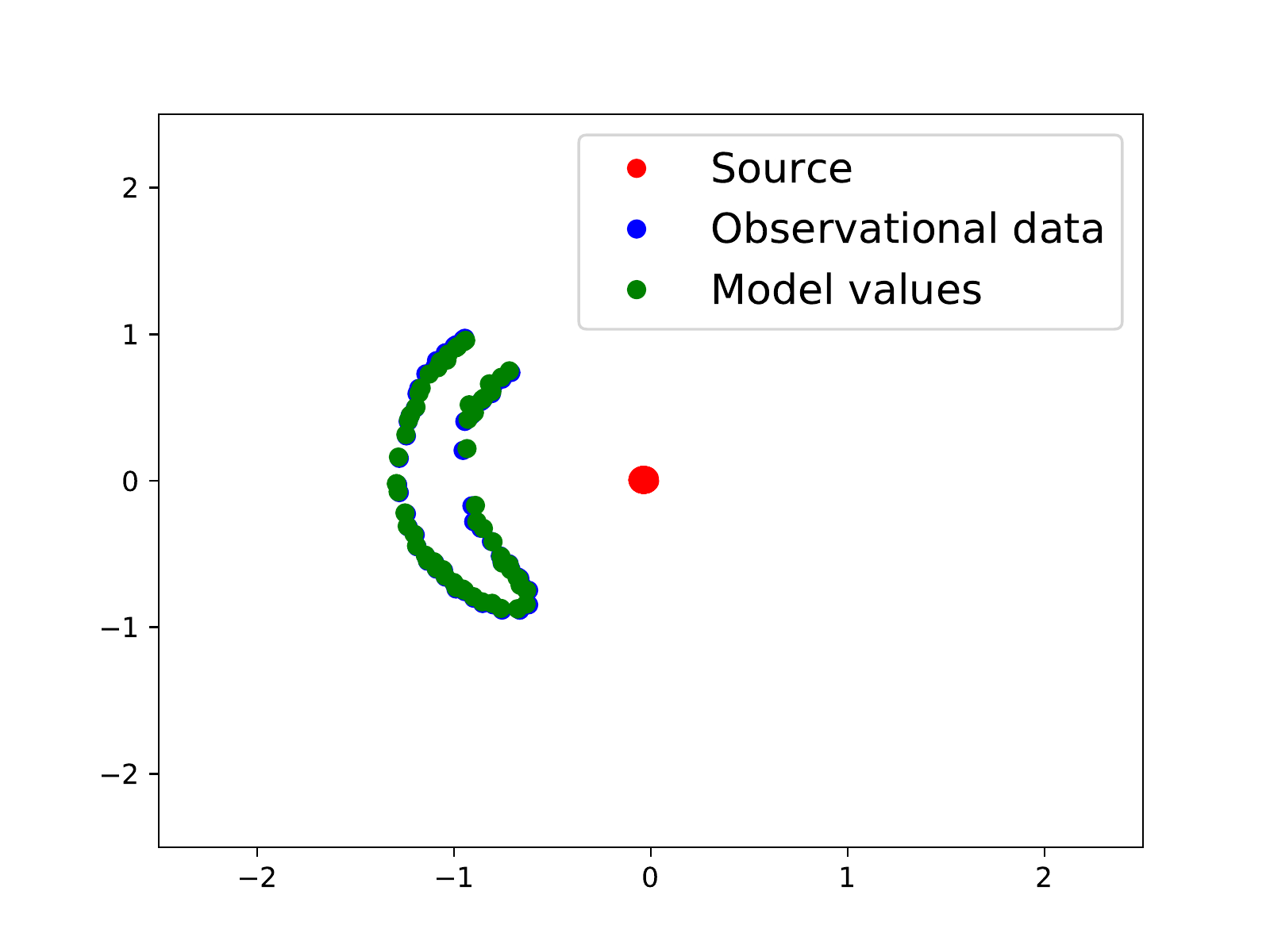}
    \caption{Comparison of the observational data and lens model data for a circular source in galaxy J2141 with \textbf{Gallenspy}, in this case the lens model choose let us obtain a set of images which overlap the observational images.\\}
    \label{fig:ajust_lens_gal1}
  \end{subfigure}
  \hfill
  \begin{subfigure}[b]{0.49\textwidth}
    \includegraphics[width=\textwidth, height=\textwidth]{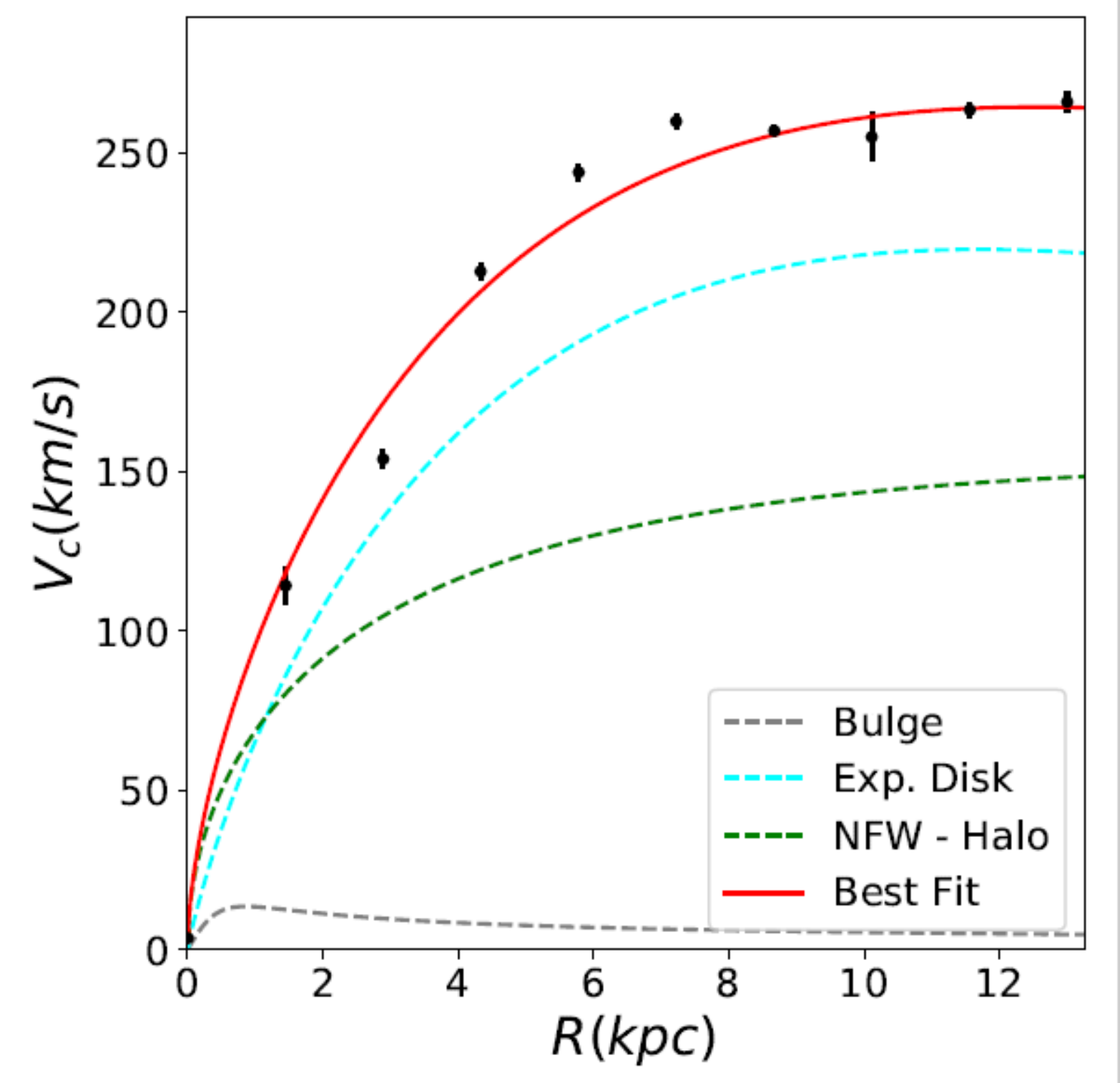}
    \caption{Fitting of rotation curve with the model data.\\}
    \label{fig:ajust_curv_gal1}
  \end{subfigure}
  \caption{Fit obtained for lensing and rotation curves through the combination of restrictions between \textbf{Gallenspy} and \textbf{Galrotpy}.}
\end{figure}

The results obtained in this work show separately mass estimation of the bulge and disc, different from results of Dutton et al., where they obtain the stellar mass without discriminating each contribution of these baryonic matter components. These mass values are shown in Table \ref{tabla:Masa_comp_gal1}, and the fitting made to the rotation curve and arc generated in the GLE with these results is illustrated in images \ref{fig:ajust_lens_gal1} and \ref{fig:ajust_curv_gal1}.

\begin{table}[h]
\begin{center}
\begin{tabular}{| c | c |}
\hline
Component of the galaxy & $log_{10}\bigg(\dfrac{M}{M_{\odot}}\bigg)$ \\
\hline
Bulge & $8.004_{-0.0009}^{+0.010}$\\
Disk & $10.905_{-0.028}^{+0.034}$\\
Dark Matter Halo & $7.740_{-0.0027}^{+0.0094}$\\
\hline
\end{tabular}
\vspace{0.5cm}
\\
\caption{Mass values for each component of J2141.}
\label{tabla:Masa_comp_gal1}
\end{center}
\end{table}

In the rotational curve's fitting, it is possible to evidence how the dark matter halo is gravitationally dominant in a radius less than $1.5\text{Kpc}$, this observation could be done due to the combination of GLE and galactic dynamics, since the generated arc is in a near radius to this galaxy zone and therefore the combination of restrictions between \textbf{Gallenspy} and \textbf{Galrotpy} is a great option.

\subsection{Galaxy J1331}

SDSSJ1331+3628 (J1331) is a spiral galaxy  with a counter-rotating massive core \cite{Dutt}, where just like J2141, the inclination of this galaxy allows to get its rotational velocity values in function of the galactocentric radius.

J1331 is localet in RA = 202.9188$^{\circ}$ and DEC = 36.469990$^{\circ}$, and in the observation of this system Treut et.al (2011) \cite{Treu} observed 2 distinct redshifts within a radius of 1 arcs ($z_{L}=0.113, z_{s}=0.254$), given the GLE in this galaxy.

\begin{figure}[h]
	\centering
	\includegraphics[height=2.5in, width=3.5in]{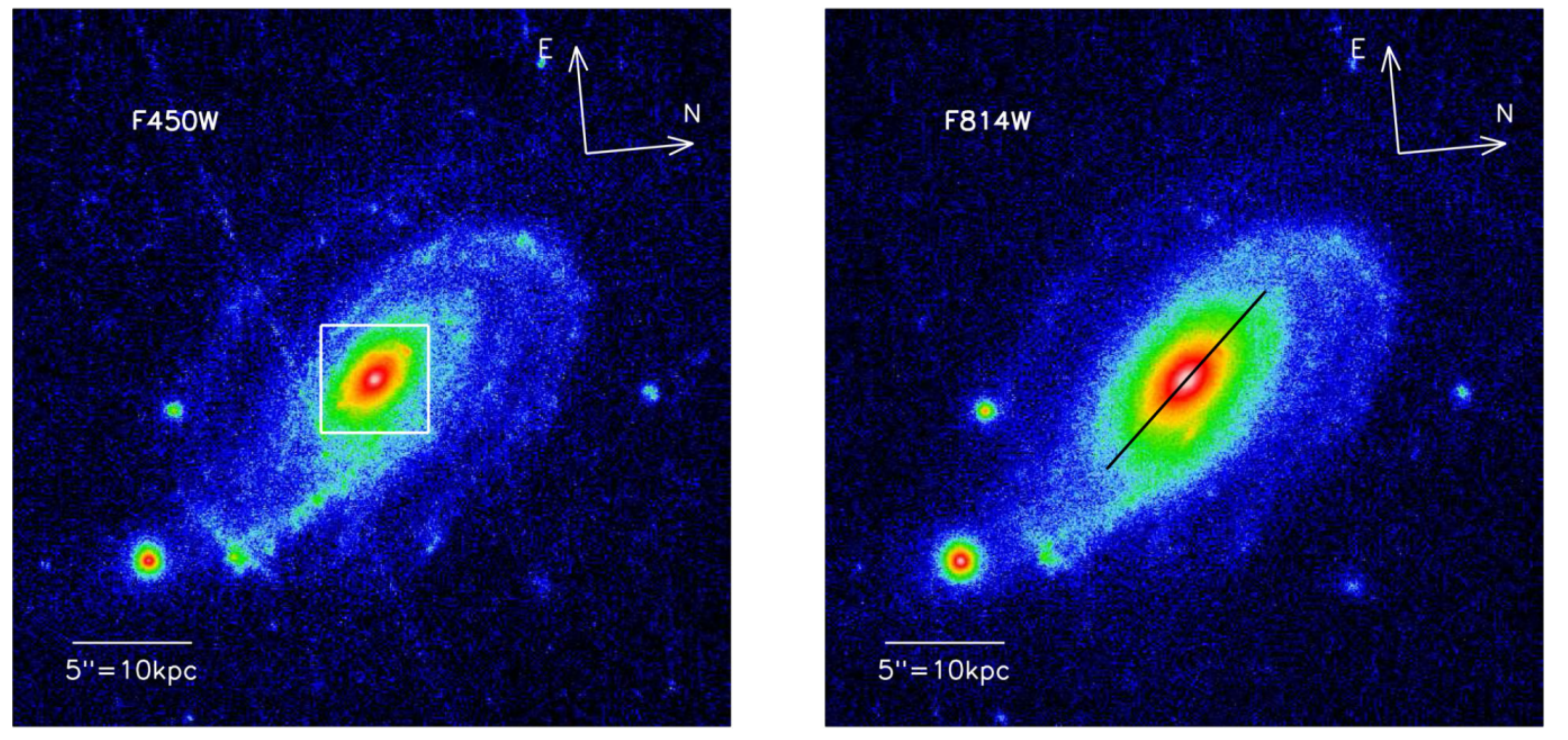}
	\caption{Images of J1331 obtained with the HST telescope in F450W and F814W filters.(Image taken from Trick et.al \cite{Dutt})}
	\label{fig:image_gal2}
\end{figure}

The images of this galaxy, were obtained by means of the SWELLS(Sloan WFC Edge-on late-type lens survey)(WFC-Wide Field Camera)\cite{SWELLS}; In figure \ref{fig:image_gal2} these images are illustrated with the HST telescope in the F450W and F814W filters, where the high size of its core  is evidenced, for this reason in the right side it is evidenced a slit done by Trick et. al \cite{Dutt}, where they reconstructed the brightness surfaces. 

\begin{figure}[!tbp]
  \begin{subfigure}[b]{0.5\textwidth}
    \includegraphics[width=\textwidth, height=\textwidth]{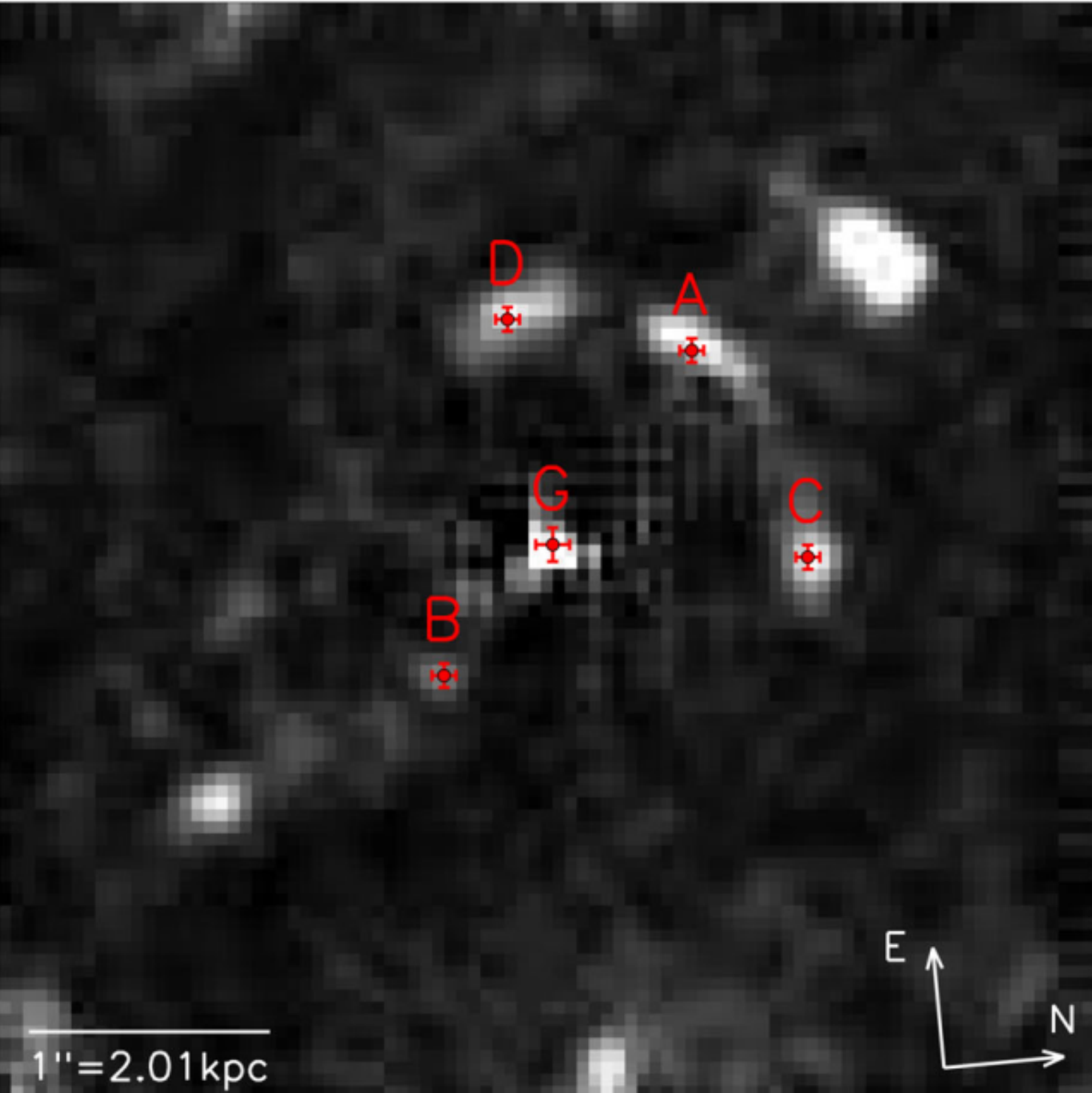}
    \caption{Quadruplet of images formed through the ELG for a point source, in this case G is the galactic center. (Image taken from Trick et.al \cite{Dutt}).\\}
    \label{fig:lens_gal2}
  \end{subfigure}
  \hfill
  \begin{subfigure}[b]{0.5\textwidth}
    \includegraphics[width=\textwidth, height=\textwidth]{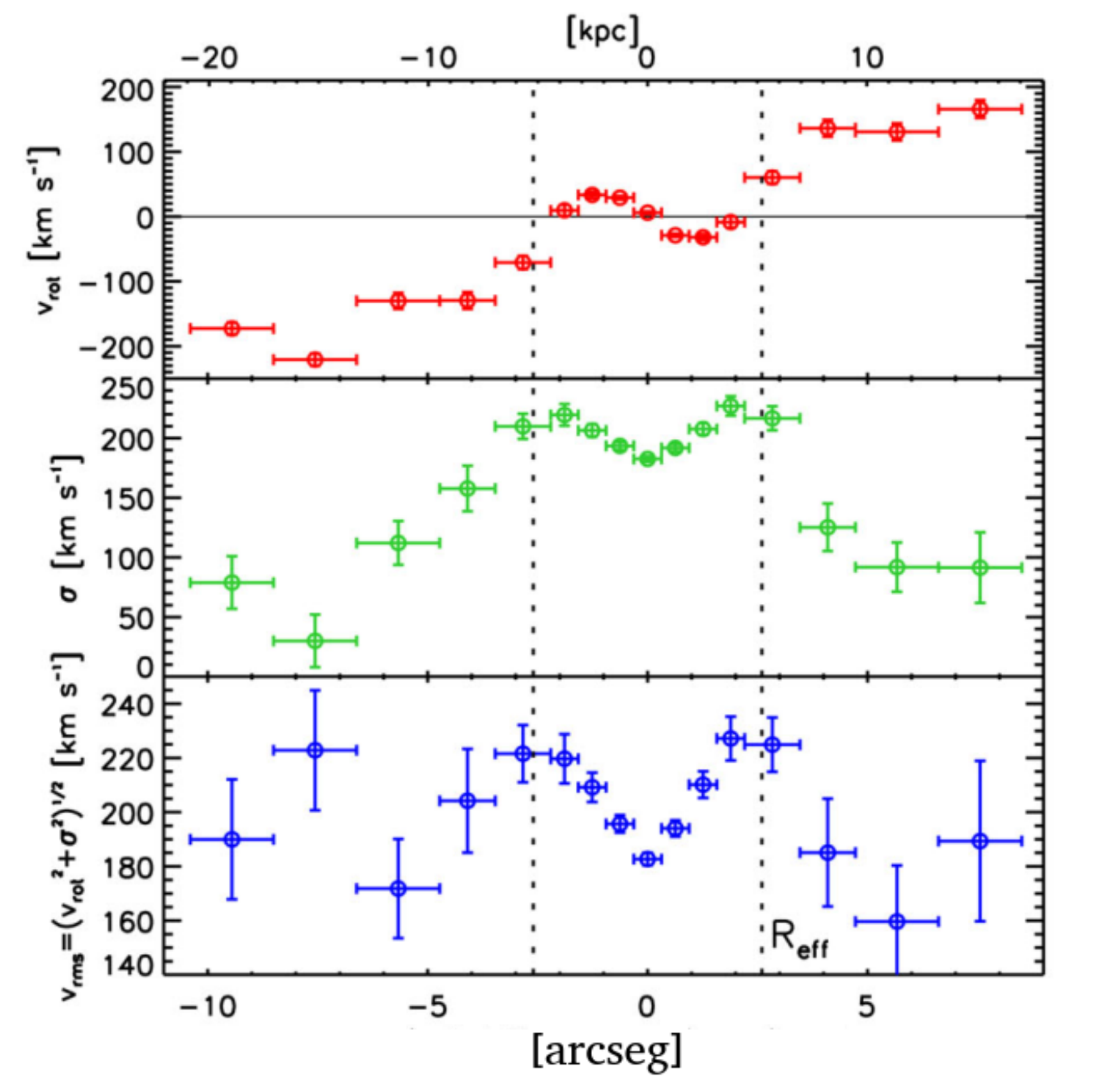}
    \caption{Rotational velocity values of J1331, where the effective radius is distinguished by  Trick et.al with 2.6 arcs, which contains is a supermassive and counter-rotating core \cite{Dutt}.\\}
    \label{fig:curvrot_gal2}
  \end{subfigure}
  \caption{Observational data of lensing and rotational velocities for the galaxy J1331.}
\end{figure}

The images produced for the GLE are shown in figure \ref{fig:lens_gal2}, which are labeled with letters A,B,C y D, also it is important to clarify that the other 3 unlabeled images do not belong to this group, since according to what Trick et.al indicate\cite{Dutt} these are part of a stellar formation region.

Regarding dynamical aspect of J1331, Dutton et al.(2013) obtained its rotational velocity values with the use of Keck I telescope by means of a LRIS spectrograph (Low Resolution Imaging Spectrograph)\cite{Dutt}, where these data were obtained of the absortion lines $M_{gb}(5177$ $\textup{\r{A}}$), $Fe_{II}(5270.5335$ $\textup{\r{A}}$) and $Fe_{II}(5406$ $\textup{\r{A}}$), while the gas velocity was estimated with emision lines $H_{\alpha}(6563$ $\textup{\r{A}}$) and $N_{II}(6583$ $\textup{\r{A}}$).

An important aspect of J1331 is its supermassive core, according to what is exposed by other authors \cite{Dutt} about half the brightness is enclosed in the effective radius illustrated in figure \ref{fig:curvrot_gal2}. For this reason, this galaxy has been and object of interest in different works \cite{Treu, Brewer}, where a possible merger event in the past of this system which changes its structure and kinematics is speculated.  

\subsubsection{Mass reconstruction based on the GLE}

Unlike J2141, the mass profiles for J1331 within its core can not be used by \textbf{Galrotpy}, due to the rotational negative velocity values present in this part of the galaxy. 

Therefore, the galaxy region enclosed in the effective radius was analyzed totally with lensing, such that the only restrictions applied in \textbf{Gallenspy} are in the parametric ranges used with \textbf{Galrotpy} for the fitting of the rotation curves in close radii to the galaxy periphery.    

In this way, other authors note how the breaking of the degeneracy is not an easy task \cite{Dutt,Dutt2}, and even if different advances have been obtained this objective has not been achieved yet.

The observational data of the images (A-D) are consigned in table \ref{tabla:Datos_Lensing}, where it was necessary to express these coordinates in arcs for a scale $1$ pixel $=0.05$ arcs.

\begin{table}[h]
\begin{center}
\begin{tabular}{| c | c | c | c | c | c |}
\hline
Coordinates & A & B & C & D & G \\
\hline
$\theta_{1}$ & 12.1 & -8.5 & 21.7 & -3.3 & 0.5\\
$\theta_{2}$ & 16.6 & -10.4 & -0.5 & 19.2 & 0.5\\
\hline
\end{tabular}
\vspace{0.5cm}
\\
\caption{(In pixels) Positions of the images (A-D) and galactocentric center (G) given by Trick et al.\cite{Dutt}. The error in each image is of 0.05, while of the G is 0.07.}
\label{tabla:Datos_Lensing}
\end{center}
\end{table}

For the determination of the cosmological distances, the redshifts were taken into account in the numerical solution to the Dyer-Roeder equation \cite{Rog}, which \textbf{Gallenspy} carries out based on the Jimenez code \cite{Julian}. For this case, the cosmological model $\Lambda$CDM was used. The obtained results were $D_{LS} = 442.7X10^3$Kpc, $D_{OL} = 422X10^3$Kpc and $D_{OS} = 817.9X10^{3}$Kpc.

The next step in \textbf{Gallenspy} was the estimation of the source position, for which the obtained values are in table \ref{tabla:fuente_gal2}. 

\begin{table}[h]
\begin{center}
\begin{tabular}{| c | c | c |}
\hline
Parameter & $68\%$ & $95\%$ \\
\hline
$\beta_1 \, \left( X10^{-3}\text{arcseg} \right)$ & $6.496_{-0.078}^{+0.085}$ & $6.496_{0.0787}^{+0.0853}$ \\
$\beta_2 \, \left( X10^{11}\text{arcseg} \right)$ & $2.104_{-0.037}^{+0.038}$ &  $2.104_{-0.074}^{+0.079}$  \\
\hline
\end{tabular}
\vspace{0.5cm}
\\
\caption{Source position obtained with \textbf{Gallenspy} for the case of the GLE in J1331.}
\label{tabla:fuente_gal2}
\end{center}
\end{table}

\begin{figure}[h]
	\centering
	\includegraphics[width=2.5in]{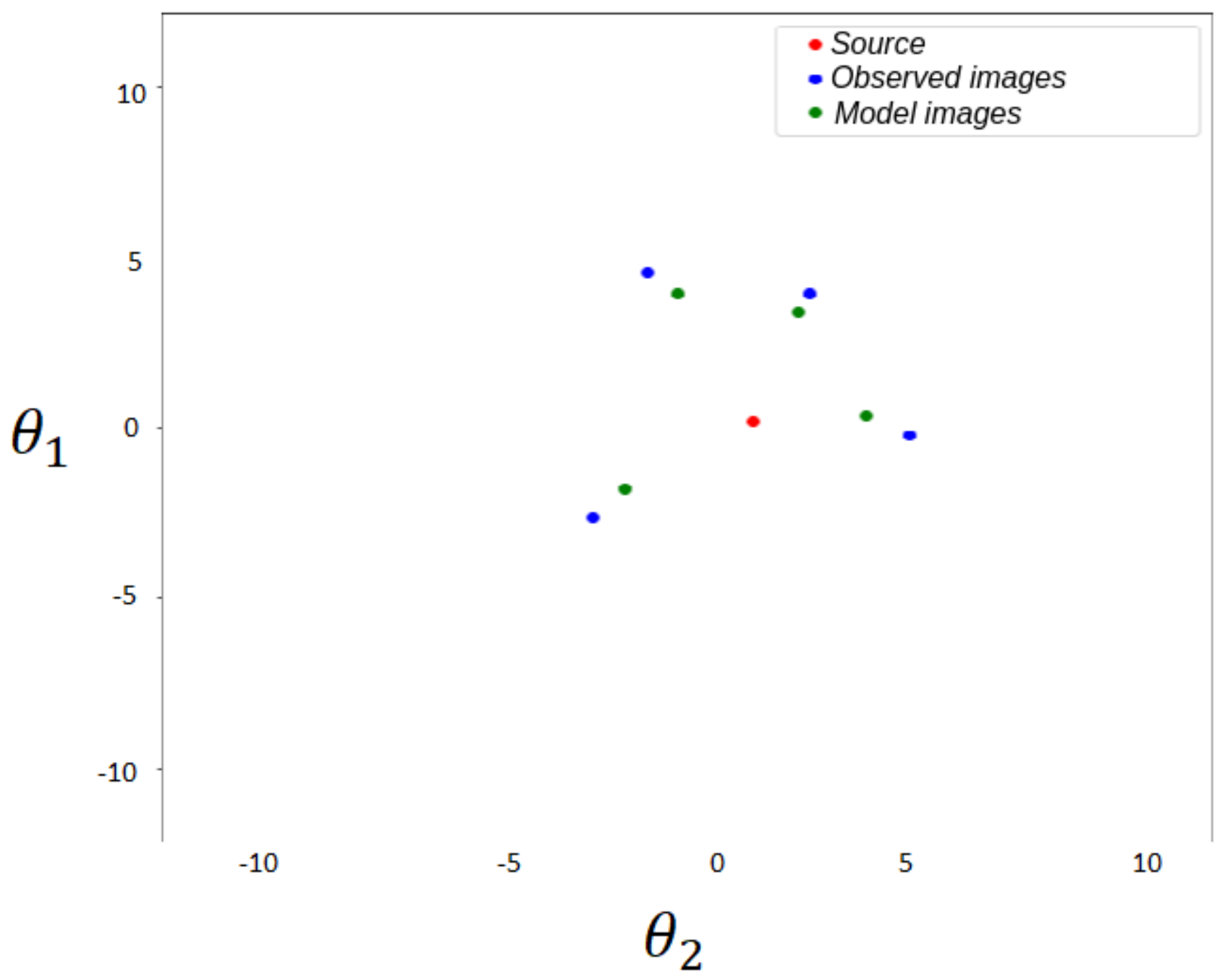}
	\caption{Adjustment obtained by \textbf{Gallenspy} with the parametric space of the Bulge II belonging to the table \ref{tabla:Param_Gallenspy}.}
	\label{fig:ajuste_antiguo}
\end{figure}

In the parameters exploration for the mass reconstruction, the established ranges of the bulges I and II belonging to the table \ref{tabla:Param_Gallenspy} were not enough for the fitting of the observational images, and this is shown in figure \ref{fig:ajuste_antiguo}. For this reason a very massive bulge was considered, where the selected most appropriate profile  is of the Miyamoto-Nagai with parametric exploration ranges of thick disc  evidenced in table  \ref{tabla:Param_Gallenspy}.  

\begin{figure*}[]
	\includegraphics[height=4.0in, width=5.0in]{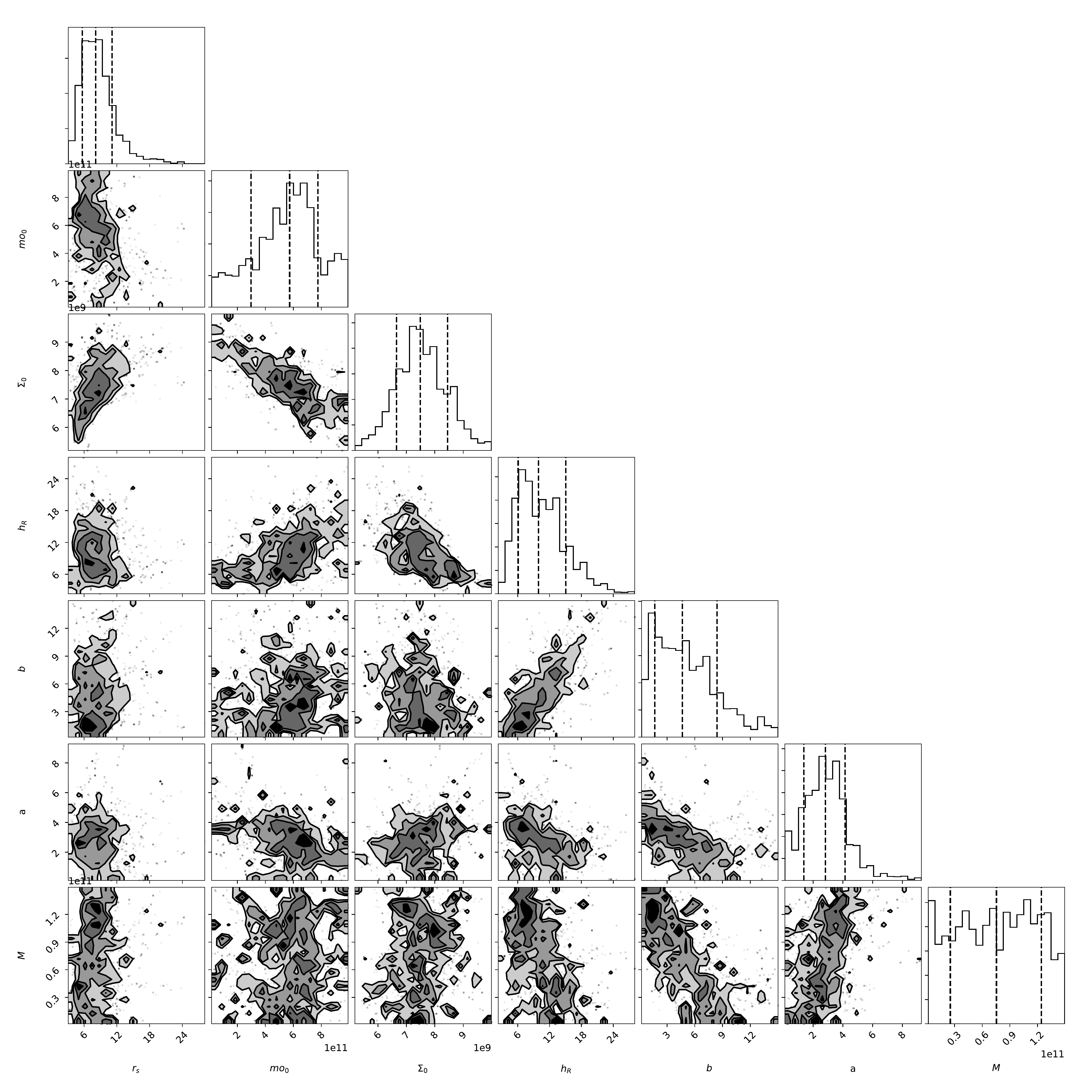}
	\caption{Reliability regions of obtained parameters in the case of J1331.}
	\label{fig:triangle}
\end{figure*}

This process was done in \textbf{Gallenspy} with 100 walkers and 100 steps in the MCMC, and in figure \ref{fig:triangle} these contours are shown.

\begin{figure}[!tbp]
  \begin{subfigure}[b]{0.49\textwidth}
    \includegraphics[width=\textwidth, height=\textwidth]{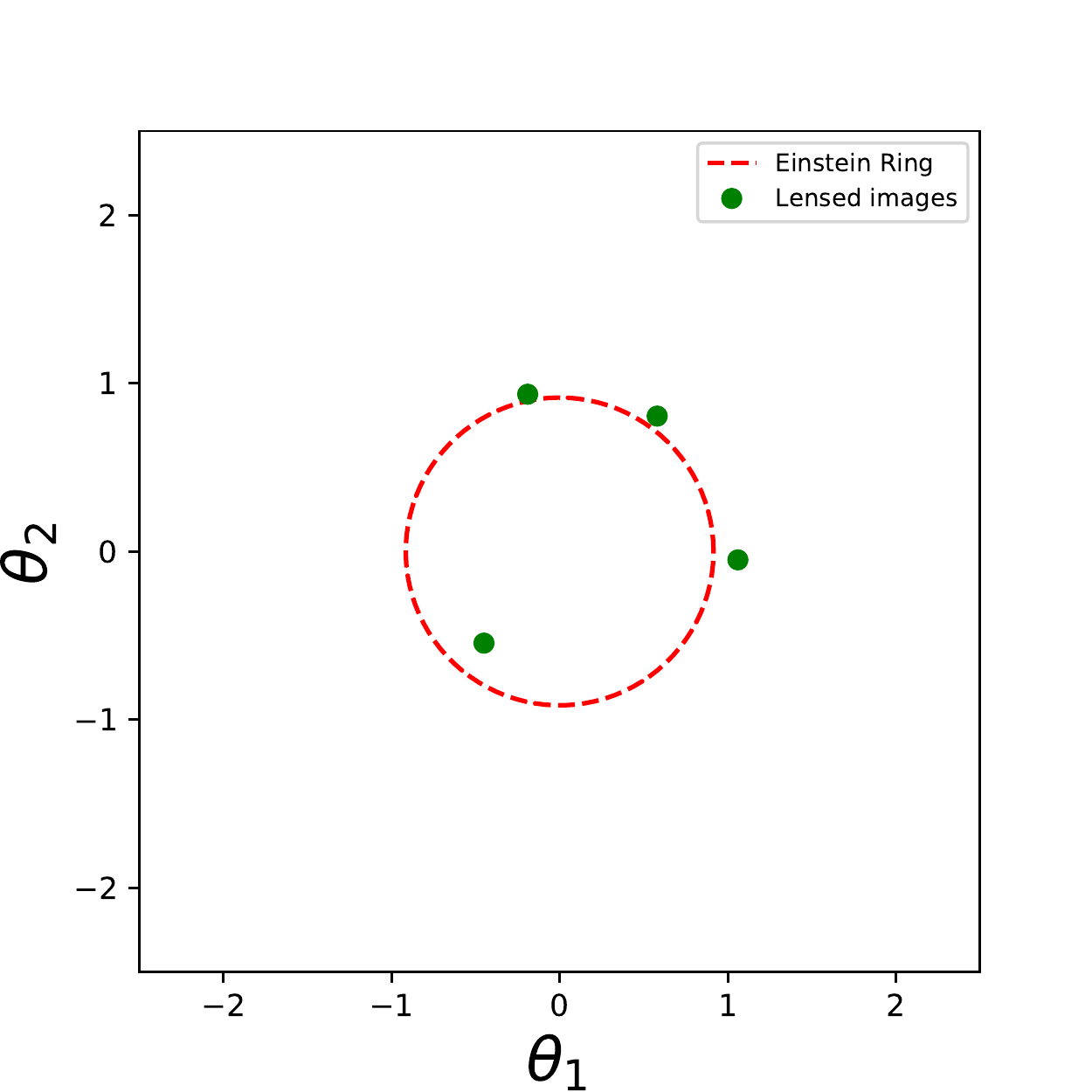}
    \caption{Einstein ring obtained with the mass distribution of J1331, the lens plane is in an $arcs$ scale.\\}
    \label{fig:einsRadius_gal2}
  \end{subfigure}
  \hfill
  \begin{subfigure}[b]{0.4\textwidth}
    \includegraphics[width=\textwidth, height=\textwidth]{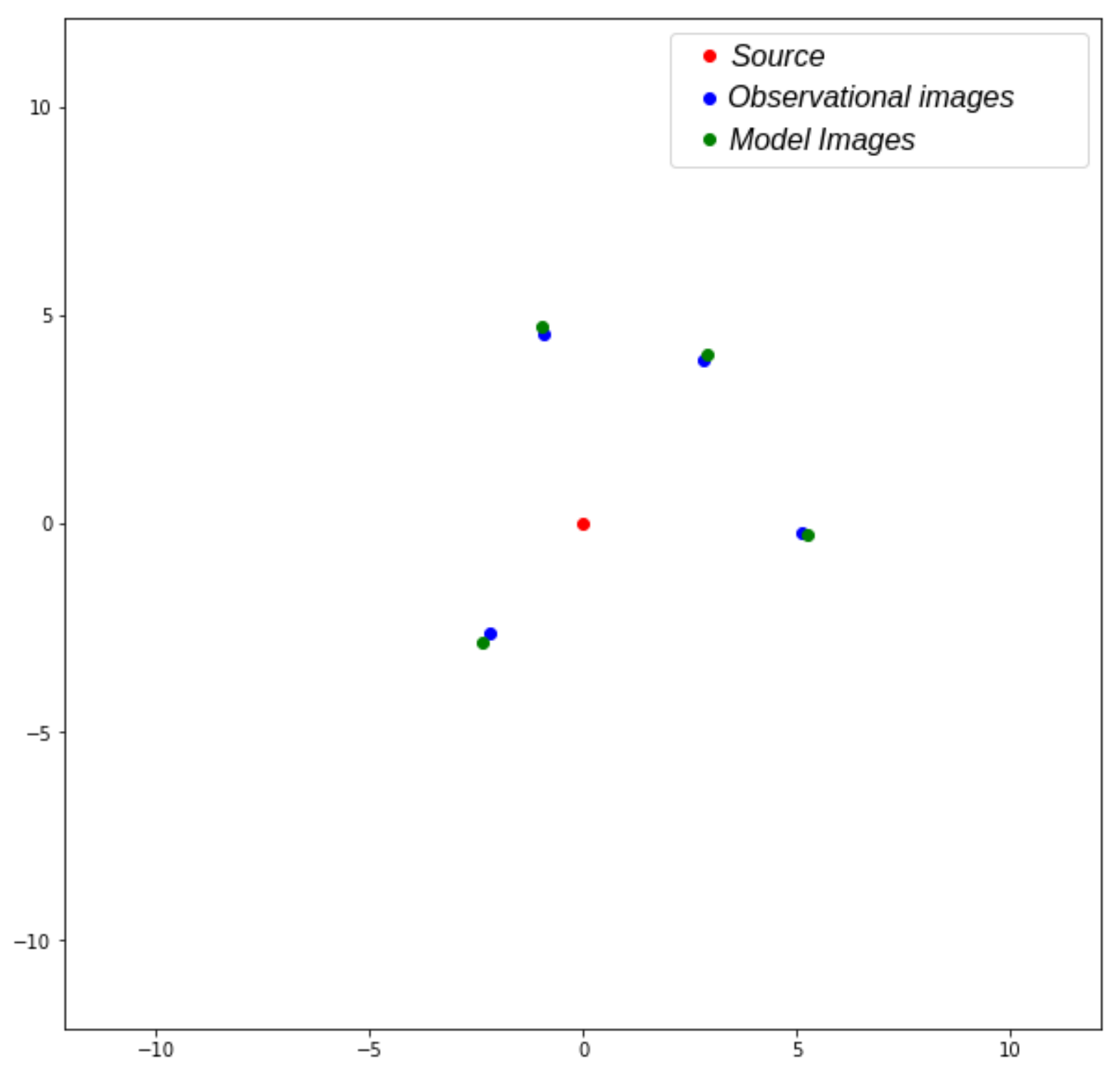}
    \caption{Adjustment obtained with \textbf{Gallenspy} for J1331, where it is important to highlight how the model images are very close to the observational images for the lens model choose.\\}
    \label{fig:Ajuste_Gal2}
  \end{subfigure}
  \caption{Comparison of observational images with Einstein ring and lens model images computed in \textbf{Gallenspy}.}
\end{figure}

Under this estimation, the most appropriate set of parameters for the mass distribution of J1331 obtained with \textbf{Gallenspy} is in table \ref{tabla:Gallenspy_J1331}, these values allowed to get the fit shown in figure \ref{fig:Ajuste_Gal2}.

\begin{table}[h]
\begin{center}
\begin{tabular}{| c | c | c |}
\hline
Parameter & $95\%$ & $68\%$ \\
\hline
\multicolumn{3}{| c |}{NFW}\\
\hline
$a \, \left( \text{Kpc} \right)$ & $8.131_{-2.451}^{+2.982}$ & $8.131_{-3.972}^{+8.973}$\\
$m_{0} \, \left(X10^{11} \text{ M}_{\odot} \right)$ & $5.734_{-2.095}^{+2.752}$ & $5.734_{-5.066}^{+3.820}$\\
\hline
\multicolumn{3}{| c |}{Exponential Disk}\\
\hline
$h_{\text{r}} \, \left(\text{Kpc} \right)$ & $9.902_{-3.824}^{+5.140}$ & $9.902_{-5.851}^{+10.923}$ \\
$\Sigma_{\text{0}} \, \left(10^{9} \text{ M}_{\odot} \text{Kpc}^{-2} \right)$ & $7.487_{-0.838}^{+0.958}$ & $7.487_{-1.645}^{+1.826}$  \\
\hline
\multicolumn{3}{| c |}{Miyamoto-Nagai}\\
\hline
$b \, \left( \text{kpc} \right)$ & $4.647_{-2.968}^{+3.759}$ & $4.647_{-3.882}^{+8.607}$ \\
$a \, \left( \text{kpc} \right)$ & $2.826_{-1.442}^{+1.324}$ & $2.826_{-2.551}^{+3.756}$ \\
$M \, \left( 10^{10} \text{ M}_{\odot} \right)$ & $7.542_{-5.010}^{+4.871}$ & $7.542_{-7.221}^{+7.055}$  \\
\hline
\end{tabular}
\vspace{0.5cm}
\\
\caption{Set of parameters obtained with \textbf{Gallenspy} for  J1331.}
\label{tabla:Gallenspy_J1331}
\end{center}
\end{table}

\begin{figure}[!tbp]
  \begin{subfigure}[b]{0.4\textwidth}
    \includegraphics[width=\textwidth, height=\textwidth]{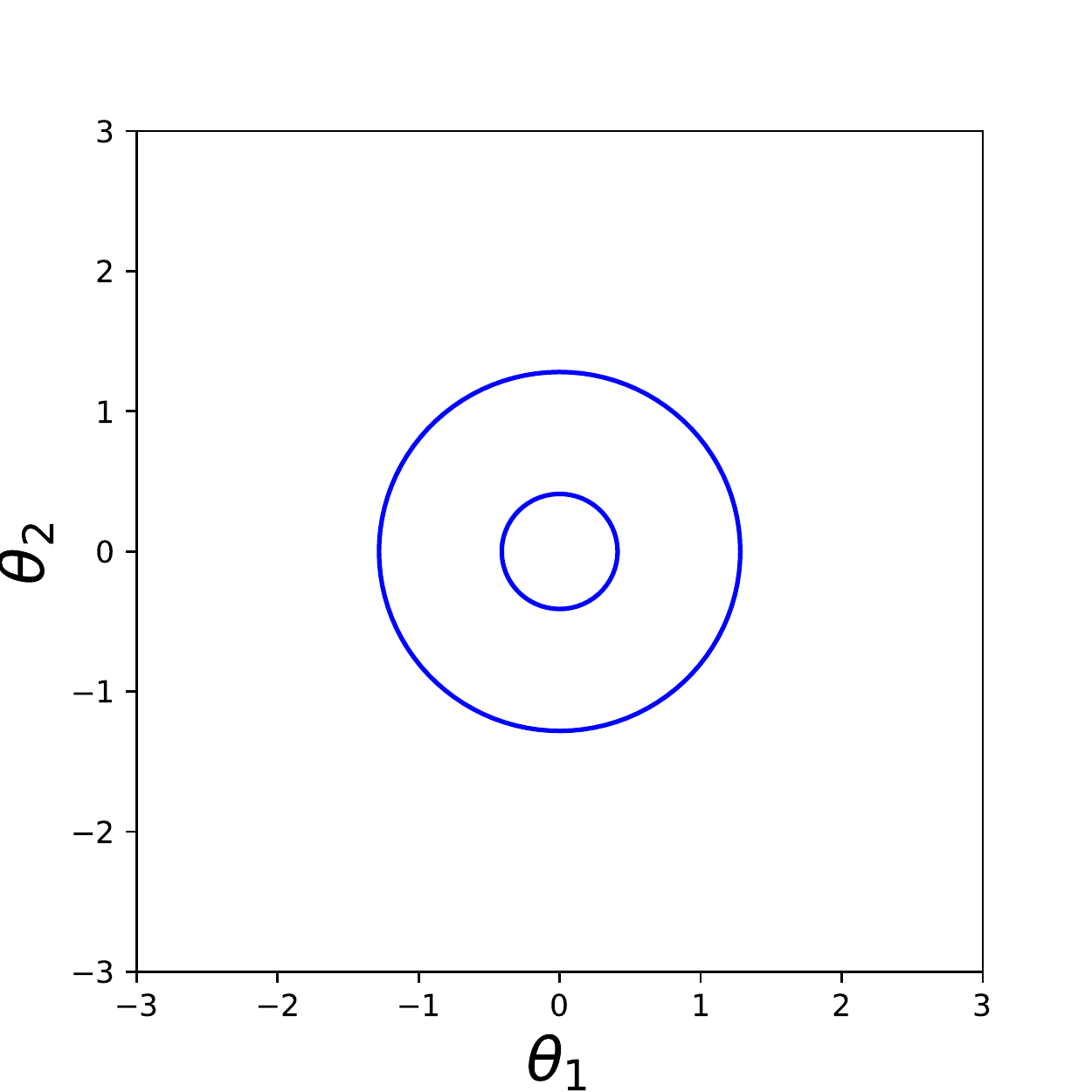}
    \caption{Critical curves obtained with the mass distribution of J1331, the lens plane is in an $arcs$ scale.\\}
    \label{fig:critica_gal2}
  \end{subfigure}
  \hfill
  \begin{subfigure}[b]{0.38\textwidth}
    \includegraphics[width=\textwidth, height=\textwidth]{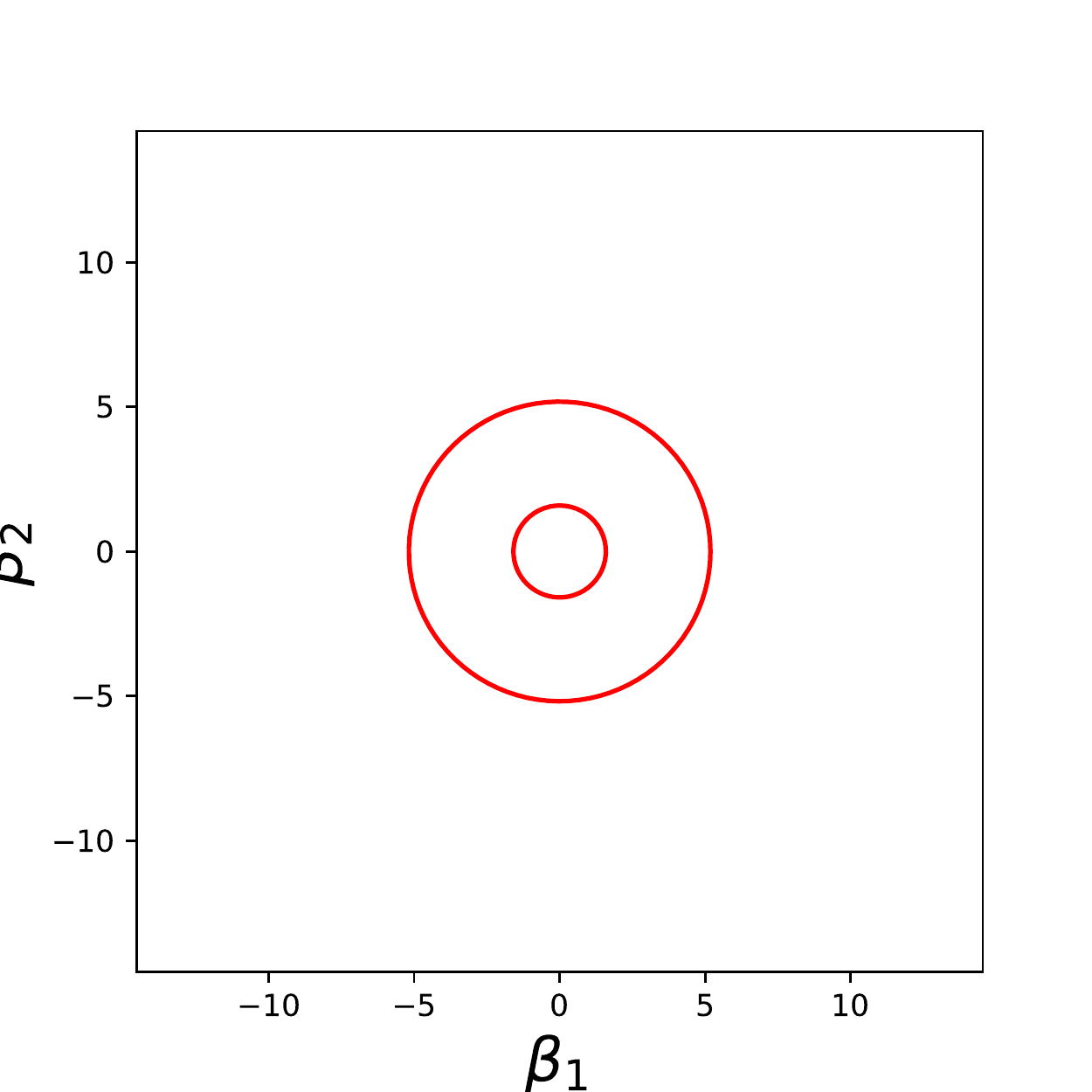}
    \caption{Caustic curves obtained with the mass distribution of J1331, the source plane is in a scale of $1X10^6$ radians.\\}
    \label{fig:caustica_gal2}
  \end{subfigure}
  \caption{Critical and caustic curves obtained with \textbf{Gallenspy} for the selected lens model.}
\end{figure}

With this mass distribution, the critical and caustic curves and the Einstein ring are presented in figure \ref{fig:critica_gal2},\ref{fig:caustica_gal2} and \ref{fig:einsRadius_gal2} where the Einstein radius and critical radii have values of $0.915\text{arcseg}_{-0.020}^{+0.013}$, $1.280\text{arcseg}_{-0.001}^{+0.001}$ and $0.410\text{arcseg}_{-0.001}^{+0.001}$ respectively. It is important to point that with this lens model was possible to estimate the mass within the effective radius under which Trick et al. \cite{Dutt} got restrictions for the mass estimation through the luminosity of J1331.

In table \ref{tabla:Masa_eins_gal2} the mass values restricted by the Einsten radius are consigned; the results reported by Trick et al. under the lens model that they assumed \cite{Dutt}, the Einstein radius estimated is  $0.91\pm0.02\text{arcseg}$ with an enclosed mass of $7.8\pm0.3\text{X}10^{10}\text{M}_\odot$. This is in concordance with the results obtained in this work with \textbf{Gallenspy}.

\begin{table}[h]
\begin{center}
\begin{tabular}{| c | c |}
\hline
Mass components & Mass value $\bigg(1\text{X}10^{10}M_{\odot}\bigg)$ \\
\hline
Bulge & $0.052_{-0.035}^{+0.045}$\\
Disk & $7.830_{-0.758}^{+1.210}$\\
Dark matter halo & $0.298_{-0.198}^{+0.161}$\\
\textbf{Einstein Mass} & $\mathbf{8.181_{-0.959}^{+1.417}}$\\
\hline
\end{tabular}
\vspace{0.5cm}
\\
\caption{Mass Values within Einstein radius.}
\label{tabla:Masa_eins_gal2}
\end{center}
\end{table}

Regarding the restriction within the critical radius, the values obtained of baryonic and dark matter are $2.190_{-0.403}^{+0.257}\text{X}10^{11}\text{M}_\odot$ and $2.179_{-0.399}^{+0.289}\text{X}10^{11}\text{M}_\odot$ respectively, these results are also consistent with the reported by Trick et al. \cite{Curv}, where for the effective radius these values are $2.352\pm0.2\text{X}10^{11}\text{M}_\odot$ for the total mass and $1.970\pm0.39\text{X}10^{11}\text{M}_\odot$ of baryonic matter, also it should be clarified that these authors obtain their results with alternative methods to the GLE of J1331\cite{Dutt}.

The results obtained with \textbf{Gallenspy} show that this is a very effective tool for the mass reconstructions within the critical curve and Einstein radius. However for J1331, the estimation is not enough with radii greater than $2.6$ arcs, and therefore in this case \textbf{Galrotpy} was used.

In other works \cite{Dutt, Dutt2} it is evidenced how the mass reconstructions for J1331 from a dynamics analysis have many complications due to the complexity of its rotation curve. This is the reason why Trick et al. \cite{Dutt} restricted this mass reconstruction to the effective radius, while Dutton et al. in 2013 \cite{Dutt2} dedicated efforts in studying the periphery of the galaxy. Based on what was previously shown, each routine in this work was applied separately in different galaxy regions.

\subsubsection{Mass reconstruction of J1331 with Galrotpy}

\begin{figure*}[]
	\centering
	\includegraphics[height=4.5in, width=6.0in]{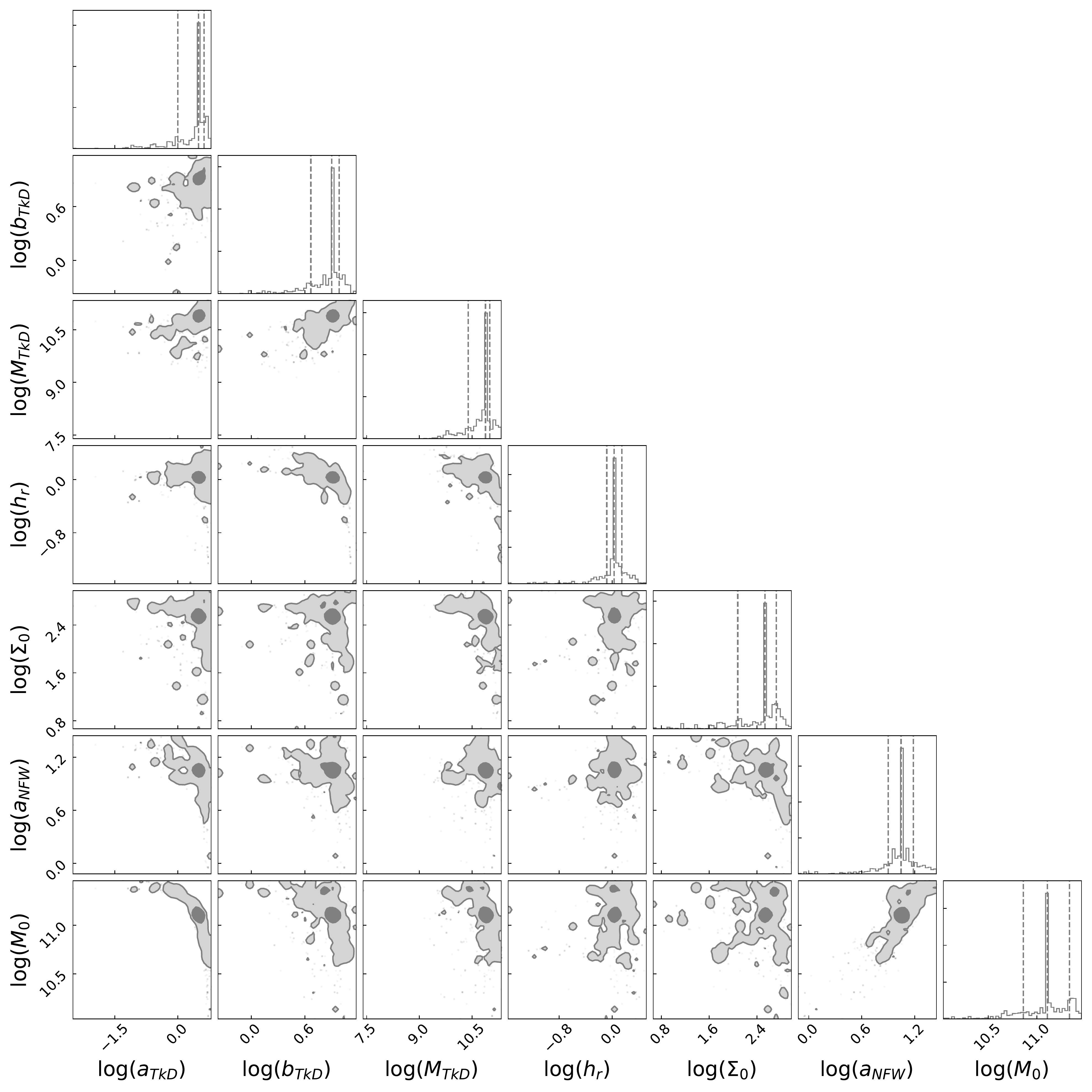}
	\caption{Credibility regions of obtained parameters with \textbf{Galrotpy} for J1331.}
	\label{fig:contornos_galrotpy_gal2}
\end{figure*}

\begin{figure}[h]
	\centering
	\includegraphics[width=2.5in]{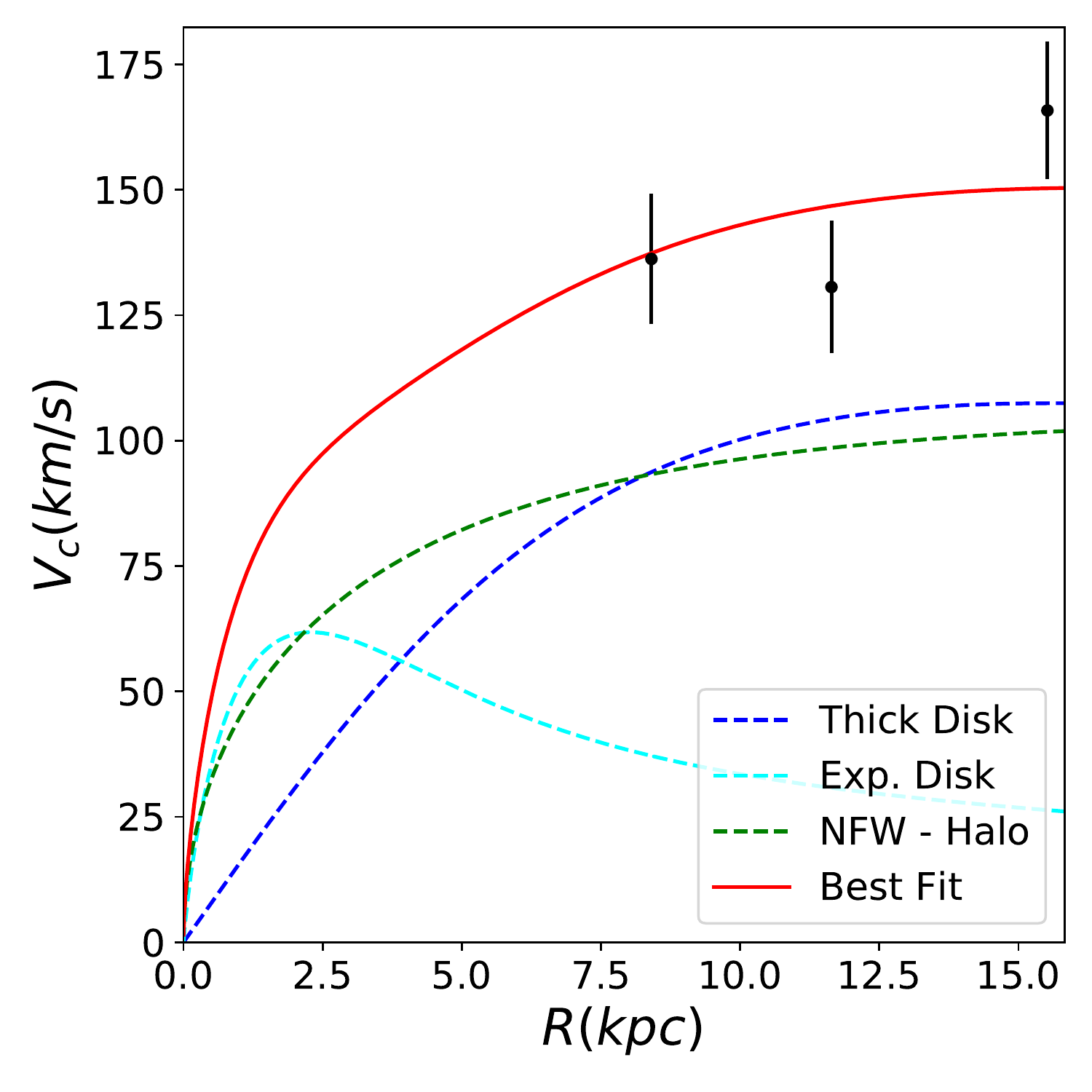}
	\caption{Rotation curve of J1331 obtained with \textbf{Galrotpy}, in this case it is possible to observe how the bulge is dominant gravitationally which is in concordance with the analysis done with \textbf{Gallenspy} from lensing.  }
	\label{fig:Ajustecurva_Gal2}
\end{figure}

The best result in the fitting of the rotation curve with \textbf{Galrotpy} was made with 20 walkers and 100 steps, wherein the image \ref{fig:contornos_galrotpy_gal2} the contours of each parameter obtained are presented in figure \ref{fig:Ajustecurva_Gal2}.

In table \ref{tabla:Galrotpy_J1331} are presented these values and its uncertainties for each parameter. The estimated mass distribution with these indicated parameters was restricted to 7.56 arcs (in this radius are all data of rotational velocity), and therefore the enclosed mass in this amplitude was estimated in $log_{10}\bigg(\dfrac{M}{M_{\odot}}\bigg)=11.448_{-0.131}^{+0.224}$ where the baryonic matter has a value of  $log_{10}\bigg(\dfrac{M}{M_{\odot}}\bigg)=10.898_{-0.164}^{+0.303}$.

\begin{table}[h]
\begin{center}
\begin{tabular}{| c | c | c |}
\hline
Parameter & $95\%$ & $68\%$ \\
\hline
\multicolumn{3}{| c |}{NFW}\\
\hline
$a \, \left( \text{Kpc} \right)$ & $11.080_{-0.801}^{+0.129}$ & $11.080_{-0.314}^{+0.417}$\\
$m_{0} \, \left(X10^{11} \text{ M}_{\odot} \right)$ & $1.277_{-0.098}^{+0.136}$ & $1.277_{-0.055}^{+0.088}$\\
\hline
\multicolumn{3}{| c |}{Exponential Disk}\\
\hline
$h_{\text{r}} \, \left(\text{Kpc} \right)$ & $1.066_{-0.087}^{+0.114}$ & $1.066_{-0.023}^{+0.033}$ \\
$\Sigma_{\text{0}} \, \left(10^{2} \text{ M}_{\odot} \text{pc}^{-2} \right)$ & $3.423_{-0.327}^{+0.393}$ & $3.423_{-0.222}^{+0.185}$ \\
\hline
\multicolumn{3}{| c |}{Miyamoto Nagai}\\
\hline
$b \, \left( \text{kpc} \right)$ & $7.892_{-0.648}^{+0.451}$ & $7.892_{-0.325}^{+0.167}$ \\
$a \, \left( \text{kpc} \right)$ & $3.114_{-0.299}^{+0.228}$ & $3.114_{-0.211}^{+0.111}$ \\
$M \, \left( 10^{10} \text{ M}_{\odot} \right)$ & $7.677_{-0.715}^{+0.112}$ & $7.677_{-0.520}^{+0.249}$  \\
\hline
\end{tabular}
\vspace{0.5cm}
\\
\caption{Estimated parameters with \textbf{Galrotpy} for J1331 galaxy.}
\label{tabla:Galrotpy_J1331}
\end{center}
\end{table}

The results given by Dutton et al. in 2013 \cite{Dutt2} indicate that the baryonic matter in this radius is of $log_{10}\bigg(\dfrac{M}{M_{\odot}}\bigg)=11.03\pm0.07$ which is in concordance with the result obtained through of \textbf{Galrotpy}. Also, it is important to note, the great relation in the estimation of the bulge mass, where they report a value of $log_{10}\bigg(\dfrac{M}{M_{\odot}}\bigg)=10.89\pm0.10$ and in this work the obtained value is $log_{10}\bigg(\dfrac{M}{M_{\odot}}\bigg)=10.885_{-0.520}
^{+0.249}$. 

\subsubsection{Analysis of the mass reconstruction for J1331}

From the mass reconstructions performed for this galaxy with \textbf{Galrotpy} and \textbf{Gallenspy}, it follows that about of the 78\% of the mass of J1331 is enclosed in the effective radius, this confirmed the presence of a supermassive core, which due to presenting a negative direction in its rotation opens the possibility to think that this galaxy is the result of a merger process between two stellar systems, with angular momentum oriented in distinct orientations.\cite{Dutt}

Also it is important to mention the high effectiveness of \textbf{Galrotpy} in this process, where the obtained results for radii close to the galaxy periphery were very successful in comparison with the results of Dutton et al. in 2013 \cite{Dutt2}, all this taking into account that these estimations were done with the fitting from just 3 values of rotational velocity. 

Besides, it is important to remember that the degeneracy between the disc and halo is still a research topic \cite{Dutt,Dutt2}, and for this reason, the possibility of adjusting \textbf{Galrotpy} for negative values of the rotational velocity is open, since this would allow to combine lensing and galactic dynamics for similar galaxies to J1331.

\section{HE 0435-1223 test case}

An additional case of tested for \textbf{Gallenspy} was  the Quasar HE 0435-1223, in which the GLE is evidenced through a quadruply imaged belonging to a background source \cite{Courbin_circular}. HE 0435-1223 was discovered by Wisotzki et al. (2000) and since then it has been a research object in distinct works\cite{Courbin_circular}. 

\begin{figure}[h]
	\centering
	\includegraphics[width=2.0in]{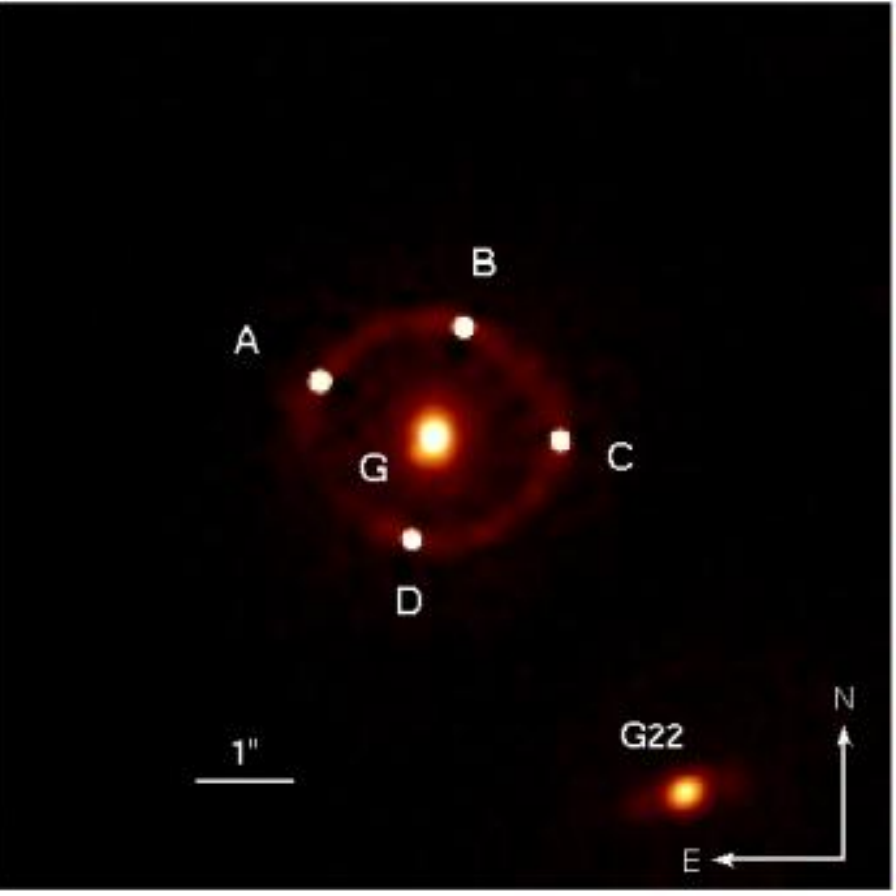}
	\caption{Quadruply imaged formed through the GLE in the case of quasar 0435-1223. (Image taken from Courbin et.al.(2011)\cite{Courbin_circular}.}
	\label{fig:Courbin_circular}
\end{figure}

Figure \ref{fig:Courbin_circular} shows these formed images by means of GLE, where the redshifts of the lens and source are $z_{s}=1.689$ and $z_{L}=1.4546$ respectively, with a value of $6.666Kpc$ for the Einstein radius.  

Based on these redshifts value, the cosmological distances estimated by \textbf{Gallenspy} are $D_{ds}=1070.3Mpc$, $D_d=1163.3Mpc$ and $D_{s}=1700.4Mpc$. Also it is important to point out that the positions of the images formed in the GLE were obtained from Courbin et al.(2011) \cite{Courbin_circular}, and in this way it was possible to perform the mass reconstruction for this quasar with the Exponential Disk and NFW profiles, where the estimated parameters are presented in table \ref{tabla:Quasar}.

\begin{table}[h]
\begin{center}
\begin{tabular}{| c | c | c |}
\hline
Parameter & $95\%$ & $68\%$ \\
\hline
\multicolumn{3}{| c |}{NFW}\\
\hline
$a \, \left( \text{Kpc} \right)$ & $52.792_{-27.014}^{+0.133}$ & $52.792_{-0.122}^{+0.073}$\\
$m_{0} \, \left(X10^{11} \text{ M}_{\odot} \right)$ & $19.961_{-13.156}^{+0.037}$ & $19.961_{-0.054}^{+0.030}$\\
\hline
\multicolumn{3}{| c |}{Exponential Disk}\\
\hline
$h_{\text{r}} \, \left(\text{Kpc} \right)$ & $11.999_{-0.007}^{+0.0005}$ & $11.999_{-0.001}^{+0.0003}$ \\
$\Sigma_{\text{0}} \, \left(10^{2} \text{ M}_{\odot} \text{pc}^{-2} \right)$ & $2.999_{-0.0025}^{+0.00004}$ & $2.999_{-0.0007}^{+0.0003}$ \\
\hline
\end{tabular}
\vspace{0.5cm}
\\
\caption{Values of the obtained parameters with \textbf{Gallenspy}}
\label{tabla:Quasar}
\end{center}
\end{table}

\begin{figure}[h]
	\centering
	\includegraphics[height=2.0in, width=2.5in]{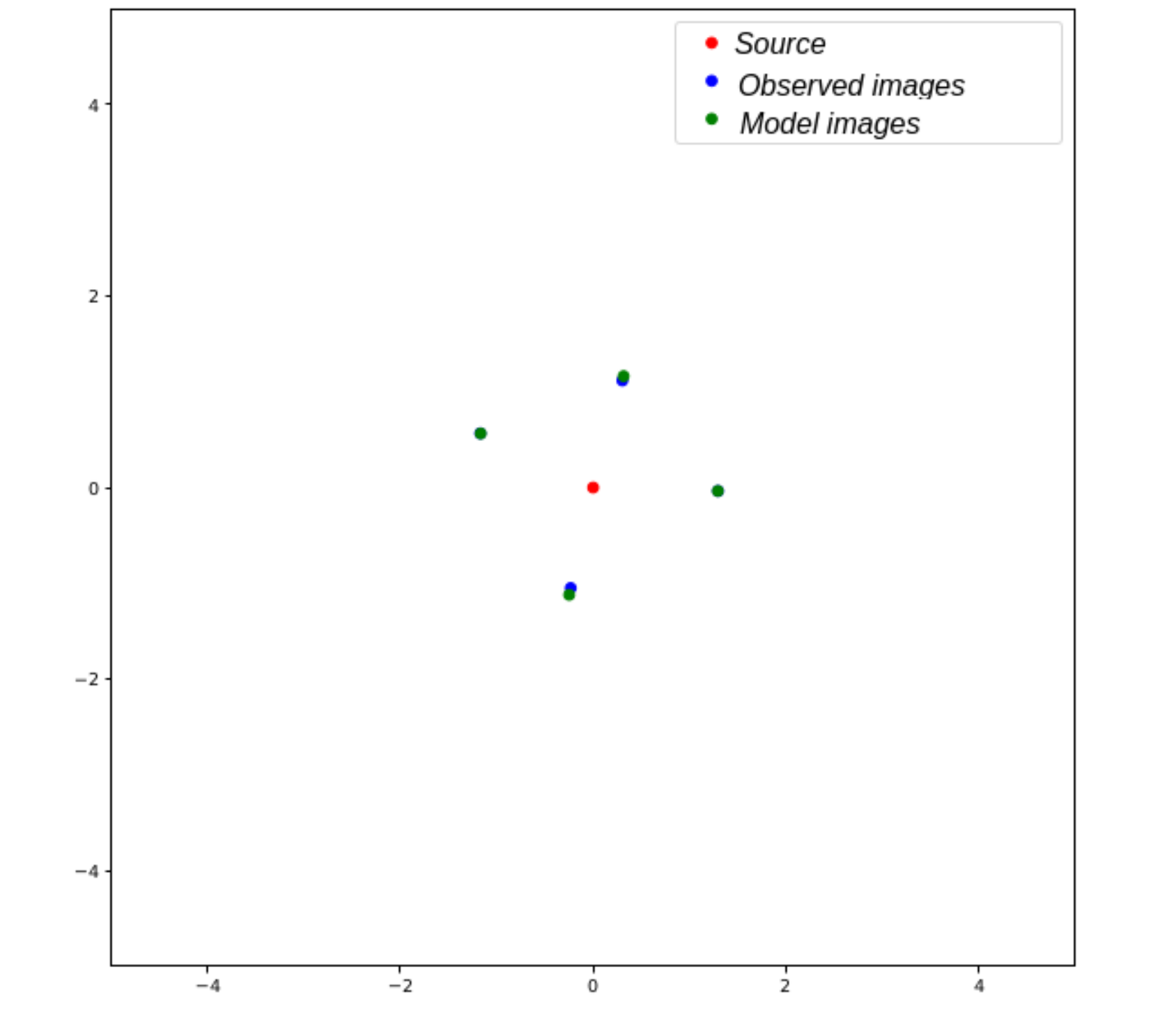}
	\caption{Comparison between model and observational images formed through the GLE.}
	\label{fig:ajuste_cuasar}
\end{figure}

With these parameters, the obtained images are illustrated in figure \ref{fig:ajuste_cuasar}, where the values of baryonic and dark matter are consigned in table \ref{tabla:masa_cuasar}. The results given by Courbin et al.(2011) for this system, reveal that the total mass  of this quasar is of $3.16\pm0.31X10^{11}M_{\odot}$ which is very close to the obtained value in this work, other important aspect is the baryonic matter fraction which in this work is of $0.764\pm0.15$ while Courbin et al.(2011) reported $0.65^{+0.13}_{-0.10}$ with the Sapelter IMF; in this way it is possible to confirm that \textbf{Gallenspy} is an efficient tool.   

\begin{table}[h]
\begin{center}
\begin{tabular}{| c | c |}
\hline
Mass & Value $\bigg(1\text{X}10^{11}M_{\odot}\bigg)$ \\
\hline
Baryonic & $2.395_{-0.0031}^{+0.0003}$\\
Dark & $0.618_{-0.066}^{+0.001}$\\
\textbf{Einstein Mass} & $\mathbf{3.014_{-0.069}^{+0.006}}$\\
\hline
\end{tabular}
\vspace{0.5cm}
\\
\caption{Mass Values within Einstein radius.}
\label{tabla:masa_cuasar}
\end{center}
\end{table}

\section{Conclusions}

In this work  the \textbf{Gallenspy} code was presented, which is very useful in mass reconstructions based on GLE; in this way it is important to highlight, the way in which this routine allows to obtain the mass distribution of bulge and disc separately unlike to the methods of reconstruction applied by other authors \cite{Curv, Dutt, Dutt2}. 

Also, the advantages of combining Lensing and Galactic Dynamics were illustrated with the use of \textbf{Galrotpy} and \textbf{Gallenspy}, in this case, with the restrictions given by each routine, it was possible to have significant progress in breaking the degeneracy in J1331 and J2141. Additionally,  \textbf{Gallenspy} was used for J1331 in the mass reconstruction within the critic radius, while with \textbf{Galrotpy} the peripheral region was analyzed and although this degeneracy could not be broken completely, the estimated parameters have concordance with the obtained results of other authors \cite{Dutt,Dutt2}. This gives reliability to the  constructed routines.

On the other hand, it is important to highlight the use of mass models with spherical symmetry, the ones are used by distinct authors \cite{Petter_circular} \cite{Courbin_circular}, and which allows to get results as in mass reconstructions as in estimations of Hubble parameter.

Regarding future improvements for \textbf{Gallenspy}, the increase in the number of mass profiles used in this routine is considered, besides there is the possibility of extending this code for reconstructions of superficial brightness functions in lens galaxies, like the estimation of temporary cosmological delays for the study of the universe expansion.

Finally it is important to mention the advantages of performing visuals fitting with \textbf{Galrotpy} and \textbf{Gallenspy} for rotation curves and GLE, since through this process it is possible get the initial set of values for the MCMC in both routines.


\bibliographystyle{elsarticle-num-names}
\bibliography{References}

\end{document}